\documentclass[a4paper,fleqn,usenatbib,useAMS]{mnras}

\usepackage{graphicx}	
\usepackage{amsmath}	
\usepackage{amssymb}	
\usepackage{multicol}        
\usepackage{bm}		
\usepackage{pdflscape}	
\usepackage{comment}

\usepackage[T1]{fontenc}
\usepackage{ae,aecompl}

\usepackage{newtxtext,newtxmath}

\def\kms{{\hbox{km s$^{-1}$}}}
\def\HI{{H\,{\textsc i}}}

\title{The evolution of neutral hydrogen over the past 11 Gyr via \HI\ 21 cm absorption}

\author[K. Grasha et al.]{
Kathryn Grasha,$^{1,2,3}$\thanks{E-mail: kathryn.grasha@anu.edu.au}
Jeremy Darling,$^{3}$
Adam K. Leroy$^{4}$
Alberto D. Bolatto$^{5}$
\\
$^{1}$Research School of Astronomy and Astrophysics, Australian National University, Weston Creek, ACT 2611, Australia\\
$^{2}$ARC Centre of Excellence for All Sky Astrophysics in 3 Dimensions (ASTRO 3D), Australia\\
$^{3}$Center for Astrophysics and Space Astronomy, Department of Astrophysical and Planetary Sciences, University of Colorado, 389 UCB, Boulder, CO 80309, USA\\
$^4$Department of Astronomy, The Ohio State University, McPherson Laboratory, 140 West 18th Avenue, Columbus, OH 43210, USA\\
$^5$Department of Astronomy, University of Maryland, College Park, MD 20742, USA
}

\pubyear{2020}

\begin{document}
\label{firstpage}
\pagerange{\pageref{firstpage}--\pageref{lastpage}}
\maketitle

\begin{abstract}
We present the results of a blind search for intervening \HI\ 21~cm absorption toward 260 radio sources in the redshift range $0<z<2.74$ with the Green Bank Telescope. The survey has the sensitivity to detect sub-damped Lyman-alpha (DLA) systems for \HI\ spin temperatures $T_s/f$ = 100~K, and despite the successful re-detection of ten known 21~cm absorbers in the sample, we detect no new absorption lines in the full survey. Sources detected in 21~cm absorption were also searched for hydroxyl (OH) 18~cm absorption and we re-detect 1667~MHz OH absorption toward PKS~1830-211. We searched for intervening \HI\ 21~cm absorption along the line of sight in each source achieving a total redshift coverage of $\Delta z=88.64$ (comoving absorption path of $\Delta X=159.5$) after removing regions affected by radio frequency interference. We compute a 95\% confidence upper limit on the column density frequency distribution $f(N_{\rm HI})$ and set a statistical constraint on the spin temperature $T_s$ in the range 100--1000~K, consistent with prior redshifted optical DLA surveys and \HI\ 21~cm emission observations at the same redshifts. We infer a value for the cosmological mass density of neutral gas, $\Omega_{\rm HI}$. Through comparison with prior $\Omega_{\rm HI}$ measurements, we place a statistical constraint on the mean spin temperature of $T_s/f=175$~K. Our derived $\Omega_{\textrm {HI}}$ values support a relative mild evolution in $\Omega_{\textrm {HI}}$ over the last 11~Gyr and are consistent with other methods that measure $\Omega_{\textrm {HI}}$.
\end{abstract}




\section{Introduction}
Neutral hydrogen gas (\HI) is a primary ingredient for star formation and is a crucial component in understanding the physical processes that convert gas into stars and govern galaxy formation and evolution. At low and high redshift, the majority of the \HI\ gas (by mass) is located in the neutral gas disks in galaxies with column densities exceeding $N_{\rm HI}\ge2\times10^{20}$~cm$^{-2}$ \citep{wolfe86}. At the characteristic column densities of galactic \HI\ disks, the ultraviolet Ly$\alpha$ transition is damped and the observed absorption systems are referred to as damped Ly$\alpha$ systems \citep[DLAs;][]{wolfe05}. A complete census of DLAs and their properties allow us to investigate the neutral gas content in our Universe and shed light on the role these gas reservoirs play in the formation and evolution of galaxies over cosmic time.

Absorption against redshifted quasars allows us to study the abundance and evolution of \HI\ gas residing within intervening absorption systems located along the line of sight to the bright background source. While DLAs have been studied extensively beyond $z\gtrsim1.6$, when the Ly$\alpha$ ultraviolet transition becomes redshifted into optical wavelengths, DLA studies below $z\lesssim1.6$ requires UV observations with spectrographs on the Hubble Space Telescope. The expense of space-based observations coupled with the rarity of DLAs in random lines of sight makes it impossible to undertake large-scale surveys comparable with large ground-based (optical) surveys such as the Sloan Digital Sky Survey (SDSS). 

The hyperfine transition of hydrogen in the radio regime (21~cm; 1420.405752~MHz) opens an alternate window into studying neutral gas systems at all redshifts. Flux limitations confine 21~cm emission surveys to the local universe (i.e., $z\lesssim0.2$), making previous high redshift studies only possible with 21~cm absorption line surveys \footnote{Future \HI\ emission studies with the Square Kilometre Array (SKA) and its pathfinder telescopes, ASKAP and MeerKAT, will enable deeper \HI\ emission studies beyond $z\gtrsim0.2$ \citep[e.g.,][]{staveley-smith15}.}. The absorption signal is not distance dependent and depends only on the column, spin temperature, and background source flux. Finally, unlike the saturated lines of the Ly$\alpha$ ultraviolet transition, 21~cm absorption profiles are usually optically thin owing to a very low Einstein rate coefficient for spontaneous emission. Large observational studies of \HI\ 21-cm absorption could thus provide a comprehensive observational census of the physical conditions of cold neutral \HI\ gas. 

Despite the benefits, surveys for 21~cm absorption encounter challenges that make the discovery of new detections especially challenging. Known redshifted 21~cm absorbers are rare in the universe, with less than 50 intervening absorbers known to date above $z>0.1$ \citep{curran16}. The 21~cm line has a low observed optical depth, requiring long integration times to achieve high sensitivity for detection. The presence of radio frequency interference (RFI) at sub-GHz frequencies interferes with much of redshift space, making it difficult to confirm potential weak absorption lines of 21~cm. Small bandwidth coverage from radio receivers force small redshift intervals to be searched, causing large redshift coverage to be very time consuming. Finally, the spin temperature of the absorbing cloud and the covering factor of the background radio source can also reduce sensitivity to low-column density systems, where high spin temperatures coupled with low covering factors decrease sensitivity to low-column systems. 

The neutral hydrogen column density frequency distribution, $f(N_{\rm HI},X)$, and its moments quantifies the overall \HI\ content of the universe and constrains the evolution of neutral hydrogen absorption systems over cosmic time \citep{prochaska05}. $f(N_{\rm HI},X)$ describes the projected column density distribution of \HI\ gas in galaxies per comoving absorption distance path length $dX$ \citep{wolfe95, prochaska09}. $f(N_{\rm HI},X)$ is easily constrained from intervening absorption surveys, as has been done at high redshift with Ly$\alpha$ \citep{prochaska09, noterdaeme09, noterdaeme12} and at low redshift with 21~cm emission \citep{zwaan05} and 21~cm absorption \citep{darling11, allison20} studies. These prior studies find that the shape of $f(N_{\rm HI},X)$ is invariant whereas the overall normalization decreases with redshift. This may signify a conversion of gas into stars \citep{prochaska09}. 

The gas mass density of neutral gas $\Omega_{\rm HI}$, the first moment of $f(N_{\rm HI},X)$, can be leveraged to show the evolution of the \HI\ gas content associated in high column \HI\ absorption systems. Studies of the evolution of neutral gas in DLAs have begun to shed light on the census of the neutral gas reservoirs, revealing that there may exist twice as much hydrogen gas in the high redshift Universe as measured with Ly$\alpha$ studies \citep[e.g., ][]{storrie-lombardi00, rao00, prochaska05, prochaska09, noterdaeme09, noterdaeme12, bird17} than in the local Universe as measured with 21~cm studies \citep[e.g., ][]{zwaan05, lah07, martin10, delhaize13, hoppmann15, rhee16, rhee18, mjones18, bera19, hu19}. 

The \HI\ content is less constrained at intermediate redshifts $0.2\lesssim z\lesssim2$ where there are a dearth of measurements because Ly$\alpha$ cannot be observed at optical wavelengths and 21~cm is not yet viable for emission studies. Studies in this redshift range often show discrepant results in the measured values of $\Omega_{\rm HI}$  and introduce controversy in the exact form and evolution of $\Omega_{\rm HI}$ from high redshift to today \citep{rao06, rao17, lah07, kanekar16}. The difficulty in understanding the redshift evolution of DLAs and the possible disconnect between low redshift 21~cm studies and high redshift Ly$\alpha$ studies is complicated by the different methods used in these different redshift regimes. Observations from $0.2\lesssim z\lesssim2$ require space-based observatories to observe Ly$\alpha$, which result in smaller sample sizes and correspondingly poorer statistical constraints due to the expensive nature of observations from space. Possible selection effects between low and high redshift methods highlight the need to carry out blind surveys across all redshift ranges. 21~cm absorption surveys may be the next step necessary to bridge the gap between $z\sim0$ and $z>2$ observations and constrain the redshift-evolution of $\Omega_{\rm HI}$ and the role neutral gas systems play in galaxy and stellar evolution \citep[see e.g., recent work by][]{sadler20}. The work here will help pave the way for future large-scale 21-cm absorption line surveys to place strong statistically constraints on the cosmological evolution of \HI\ gas at intermediate redshifts with the Square Kilometre Array and its pathfinder telescopes, including the First Large Absorption Survey in \HI\ \citep[FLASH;][]{allison2016_mnras} with the Australian Square Kilometre Pathfinder \citep[ASKAP;][]{johnston07} and the South African MeerKAT Absorption Line Survey \citep[MALS;][]{gupta16}. 

Observations with Ly$\alpha$ and 21~cm absorption can in principle provide the necessary information to measure the spin temperature of the absorbing gas from the total neutral hydrogen \HI\ column density. This requires the assumption that Ly$\alpha$ and 21~cm absorption trace the same sight line, which may not always be valid \citep[see, e.g.,][]{kanekar07}. The few intervening absorbers with observations of both 21~cm absorption and Ly$\alpha$ absorption reveal a broad range in the measured spin temperature of the absorbing gas, from 100~K to nearly 10,000~K \citep[e.g., ][]{wolfe81, kanekar03, york07, nroy13, kanekar14_mnras, dutta17, curran19_mnras}. Combining our large intervening \HI\ 21~cm absorption study with prior Ly$\alpha$ and 21~cm emission studies potentially provides the ability to constrain the largely uncertain spin temperature and covering factor of the absorbers through means of comparison with prior measurements of $f(N_{\rm HI},X)$ and $\Omega_{\rm HI}$. Measurements of $f(N_{\rm HI},X)$ and $\Omega_{\rm HI}$ are temperature independent when constrained via Ly$\alpha$ and 21~cm emission surveys. The unknown spin temperature for 21~cm absorbers can be tuned to constrain the temperature range such that the observations are consistent with the Ly$\alpha$ and/or 21~cm emission measurements. Large surveys such as this one can potentially supply adequate coverage to provide a physically motivated constraint for a range of spin temperatures.

As an added benefit, 21~cm absorption lines, combined with the hydroxyl (OH) 18~cm transition line and other millimetre rotational lines, allows for accurate measurements of fundamental physical constants, such as the fine structure constant $\alpha \equiv e^2/\hbar c$, the proton-electron mass ratio $\mu \equiv m_{\rm p}/m_{\rm e}$, and the proton g-factor $g_{\rm p}$ \citep{darling04, kanekar05}. It is possible to constrain cosmological evolution of these constants with these molecular absorption systems. Redshifted OH 18~cm absorption systems are very rare, with only five systems known, all limited to $z<1$ \citep{chengalur99, kanekar03a, darling04, kanekar05}. The few known redshifted OH absorbers are always found in systems with 21~cm absorption. For this reason, studies aimed at observing new 21~cm absorbers quite often tune to the frequencies of 18~cm absorption as well, as we do this study. 

The blind 21~cm absorption line survey presented here was established to have the sensitivity to detect all DLA absorption systems in a large number of sightlines from $z=0$ to $z=2.74$ with just minutes of observing time per source. Our survey aims at detecting new intervening \HI\ 21~cm and molecular OH 18~cm absorption systems in a redshift-independent fashion with no prior knowledge of known absorption systems in order to provide adequate statistics to help constrain factors of cosmological importance. We demonstrate that despite the lack of new detections, largely due to severe RFI along much of redshift space, we are able to redetect all known absorbers present in our survey and it is possible with our large sample size and redshift coverage to place meaningful limits on the spin temperature of \HI\ gas with the column density frequency distribution function $f(N_{\rm HI},X)$ (Section \ref{sec:fN}) and the evolution of the cosmological neutral gas mass density $\Omega_{\rm HI}$ (Section \ref{sec:omega}). 

Throughout this paper, we adopt a $\Lambda$CDM cosmology of $H_\circ = 71$~\kms~Mpc$^{-1}$, $\Omega_m = 0.27$, and $\Omega_{\Lambda} =0.73$.

\section{Sample Selection}
\begin{figure}
\includegraphics[scale=0.43]{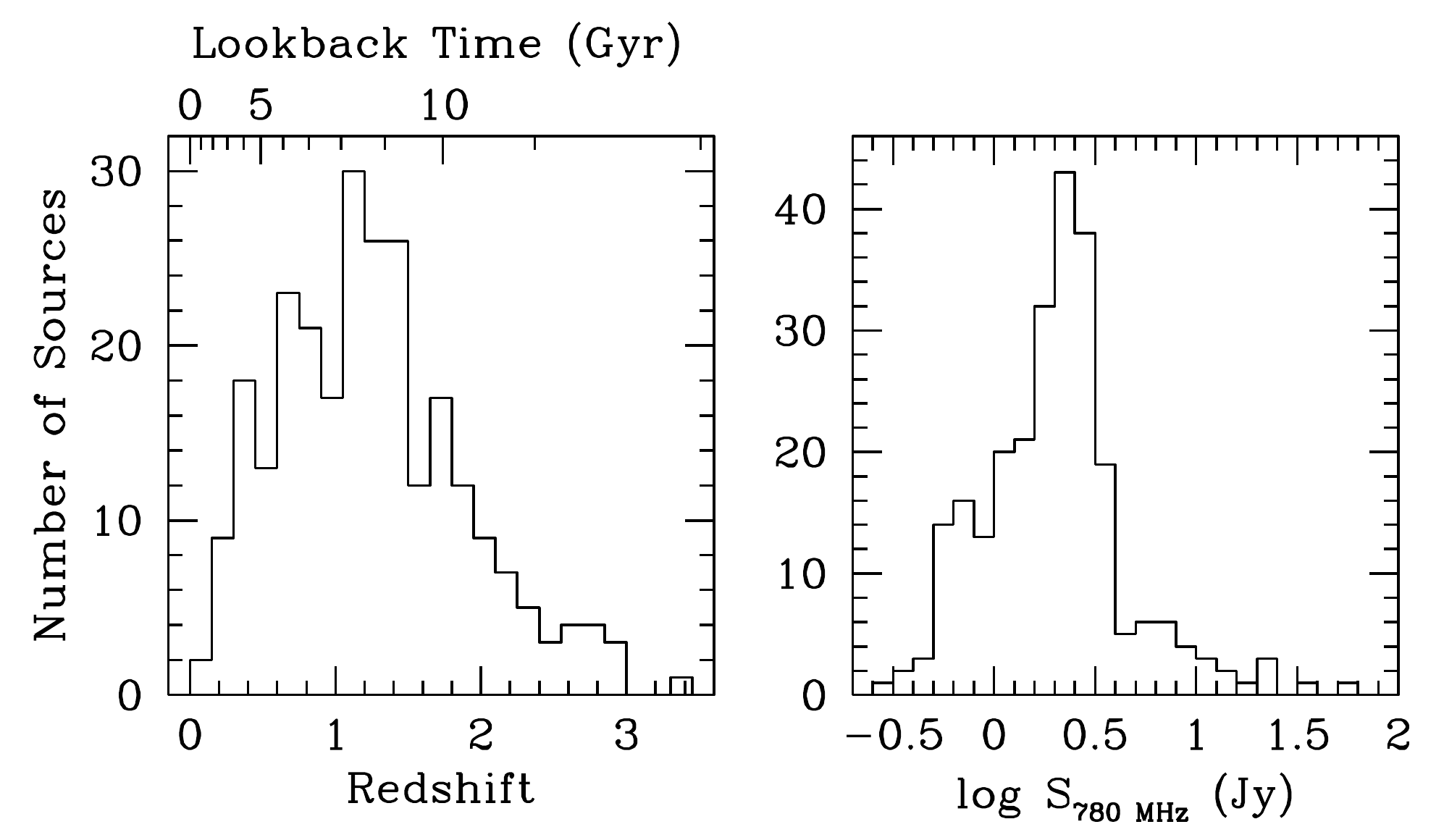}
\vspace{-15pt}
\caption{Left: Histogram of the number of sources versus redshift for the 260 sources in this survey with the lookback time listed on the top axis. The average redshift is 1.244. Right: Flux density of each source at 780~MHz.
\label{fig:sources}}
\end{figure}

We select 260 radio sources to search for new intervening 21~cm lines along the sight lines with no requirement placed on the source type. The sources are selected from the \citet{white92} catalog with requirements of a known optical redshift, north of declination $\delta \gtrsim 30$~deg, and a continuum flux density of $S>0.3$~Jy at 780~MHz (Figure~\ref{fig:sources}). No prior knowledge about possible absorption systems, intervening or intrinsic, was considered for source selection. Requiring an optical redshift does impose some bias, selecting the brightest, least obscured, and UV-bright radio sources at all redshifts. Figure~\ref{fig:sources} shows the distribution the redshift of the sources and the rest-frame radio continuum at 780 MHz for the sources in the sample. 

The majority of our sources are classified as Flat-Spectrum Radio Sources (FSRSs), radio sources characterized with a double-peaked synchrotron/Compton spectral energy distribution and possibly associated with a blazar or with the compact core of a radio galaxy \citep{fugmann88}. Our sample also includes Giga-Hertz Peaked (GPS) sources, radio sources that are believed to be intrinsically small (not foreshortened by projection effects; \citealt{fanti90}) and exhibit a low-frequency turn-over in their spectra, attributed to synchrotron self-absorption and thermal bremsstrahlung absorption \citep{jones74, menon83, odea97}. 

\section{Observations and Data Analysis}
We observed 89 sources in January 2004 (GBT program 02C--008) for intervening \HI\ 1420.405752~MHz absorption from $z=0.6$ to $z=1.1$ at the Green Bank Telescope\footnote{The National Radio Astronomy Observatory (NRAO) is a facility of the National Science Foundation operated under cooperative agreement by Associated Universities, Inc.} (GBT). We chose GBT's Prime Focus 1 receiver (PF1-800), covering 680--880~MHz ($0.6<z<1.1$) due to the relatively minimal RFI present at these frequencies. 

These observations are complemented with observations of an additional 182 radio sources between September 2004 and August 2005 (GBT program 04C--018) at the GBT for \HI\ 21~cm absorption in intervening \HI\ systems between $z = 0$ up to $z = z_{\rm source}$. The redshift coverage is not always complete for each source. 11 sources appeared in both observing programs, resulting in a total of 260 unique radio sources in this survey. 

Observations were divided among the GBT receivers as follows: 147 sources were observed with the 1.4~GHz L-band receiver ($0<z<0.235$), 126 with the 1~GHz Prime Focus (PF) 2 receiver ($0.150<z<0.570$), 204 with the 800~MHz PF1-800 receiver ($0.536<z<1.10$; including both GBT programs), two with the 600~MHz PF1-600 receiver ($1.06<z<1.79$), and 12 sources with the 450~MHz PF1-450 receiver ($1.58<z<2.74$). These receivers were selected to cover the full spectral path for each target from $z=0$ to the redshift of the source. 

Six intermediate frequencies (IFs) were employed per receiver for both PF2 and PF1-800, PF1-600 and PF1-450 employed four IFs, and L-band employed eight IFs. All IFs have a bandpass of 50~MHz, divided into 4098 channels (channel spacing of 12.21~kHz). The measured noise levels for identical integration times are comparable for the L-band, PF2 (where the bandpass is not heavily impacted by RFI), and PF1-800 receivers, but is greatly increased in the PF1-600 and PF1-450 receivers. The redshift coverage of the L-band is the cleanest with minimal RFI presence. Most of the redshift coverage of the 1~GHz receiver (PF2), 600~MHz receiver (PF1-600), and 450~MHz receiver (PF1-450) are unusable due to severe RFI. Total on-source integration times are listed for each source in Table \ref{tab:obs} for each receiver with typical noise for each receiver, and total redshift coverage $\Delta z$ searched per source. 

All observations were conducted in position-switched mode in two linear polarizations with 9-level sampling, and a range of 1.5-second to 15-second records. For all observations, a calibration diode signal was injected for half of each record. 

Each flattened and calibrated spectral record were inspected and flagged for RFI. All records were then averaged in time and polarization for each observed source and Hanning smoothed to a resolution of 12.2~kHz per channel (a resolution of 4.7~\kms\ per channel at 780~MHz). RFI features were interactively removed and the continuum at the RFI feature replaced with an average of the noise from the surrounding five channels to aid in measuring the noise across the spectral region and aid in base-line flattening the spectrum. Each spectrum was baseline-flattened using a polynomial fit, typically of fifth order. 

Final mean spectra are inspected for absorption lines in each 50~MHz bandwidth region. The actual range of each region was typically determined by the RFI conditions of each spectrum. In cases of non-detections, spectral noise measurements are recorded for the maximum range possible. All data reduction was performed using GBTIDL\footnote{GBTIDL (\url{http://gbtidl.nrao.edu/}) is the data reduction package produced by NRAO and written in the IDL language for the reduction of GBT data.}

Feed resonances at 1263, 924, 817 and 769~MHz, introduced by the GBT receivers, are removed by fitting the resonance with another observation in the same observing session, using the second observation as a template and subtracting the feed shape, flattening the baseline across the feed resonance. The 21~cm absorber toward 2351+456 occurred at the edge of the 796~MHz feed resonance, and was removed in this fashion with a template supplied by another source in the same observing run as demonstrated by \citet{darlingetal04}.

\section{Results}\label{sec:results}
\subsection{Flux Density Values}\label{sec:cntm}
Low-frequency ($\lesssim1$~GHz) spectra frequently contain negative continuum values, slopes, steps, and other unnatural features caused by instrumental effects. As a result, the flux density at the frequency of the 21~cm observations for each object cannot be measured directly from our spectra. We obtain the estimates of the flux density at our desired frequency by interpolating between extant continuum measurements in the NASA Extragalactic Database (NED) using a single spectral index power-law fit (a linear fit in log~$S_{\nu}$--log~$\nu$ space). In order to cope with the heterogeneous literature continuum measurements, we try whenever possible to select values published from the same catalog sources using the same pass bands, most often 1410, 1340, 750, 635, 408, and 365~MHz observations \citep{laing80, ficarra85, large81, white92,  douglas96, rengelink97, condon98, stanghellini98, stanghellini05, orienti07, petrov08}. This treatment neglects potentially significant time variation in radio source fluxes. The continuum flux densities are listed in Table~\ref{tab:obs}.

\subsection{Line Search Results}
\subsubsection{HI 21~cm Intervening Line Observations}\label{sec:interveningD}
Out of the 260 sources in this survey, eight sources remain indeterminate to absorption systems due to persistent and irremovable RFI for the entire observed spectral range (OJ+287, 3C308, HB89~1602$-$001, 1743+173, 1800+440, 1921$-$293, 1958$-$179, and 2007+777). 

\begin{figure*}
\includegraphics[width=\linewidth]{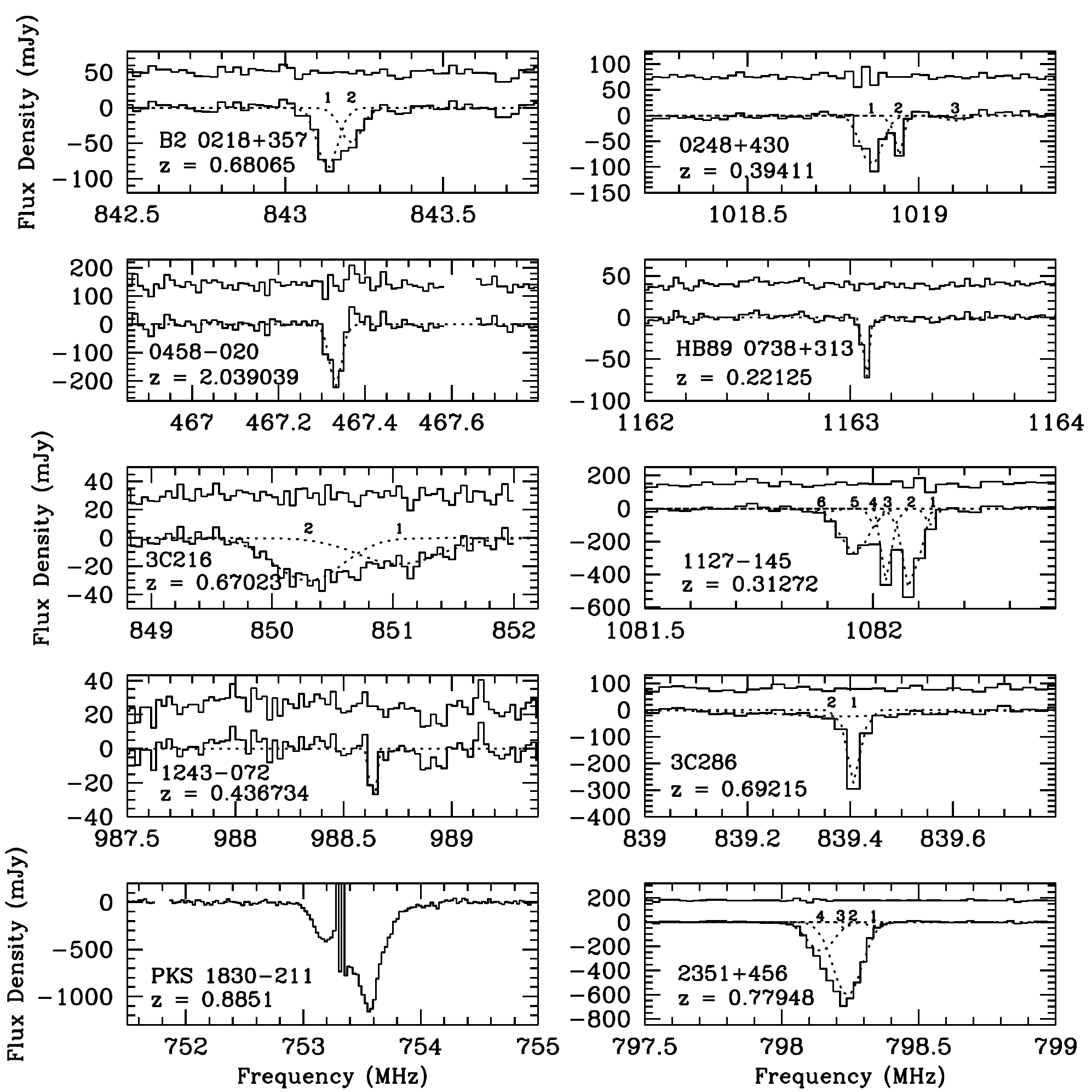}
\vspace{-15pt}
\caption{Sources detected in \HI\ 21~cm absorption. The dotted lines indicate Gaussian fits, numbered by component listed in Table \ref{tab:detections}. Upper spectra show the residual spectrum, offset for clarity. Spectral regions lost to radio frequency interference have not been shown. 3C216 is an \HI\ absorber associated with its host galaxy ($z_{\rm absorption} = z_{\rm source}$) and is excluded from all plots an analysis related to the intervening absorption and the calculations for $\Omega_{\rm HI}$ and $f(N_{\textrm {HI}},X)$. The Gaussian fit to absorption line in PKS~1830$-$211 is not shown due to the presence of RFI. }\label{fig:detects}
\end{figure*}

We have searched the remaining 252 partially RFI-free sources for intervening \HI\ 21~cm absorption and we find intervening 21~cm absorbers to be present in the spectra toward nine radio sources. All of the absorption systems in this survey are re-detections, shown in Figure~\ref{fig:detects}. Table~\ref{tab:detections} lists the properties of the detected absorption lines, fitted Gaussian components, and comparison to the prior 21~cm detections of the absorption systems. The exact value of the spin temperature $T_s$ for intervening 21~cm absorbers and the covering factor $f$ of their background radio source are unconstrained. We therefore report all column density measurements without an assumption on $T_s/f$, where possible. For our analysis in Section~\ref{sec:fN} and Section~\ref{sec:omega}, in order to compare our work to prior literature measurements that do not depend on spin temperature assumptions, we normalize all prior \HI\ column density values to a range of values that span values commonly adopted in the literature: $T_s/f=100, 250, 500$, and 1000~K \citep[e.g., ][]{curran05, curran17b, curran17c, darling11, allison20}. 

Under the assumption that  radio and ultraviolet illumination trace the same line of sight, it is possible to constrain the spin temperature of \HI\ absorbers with measurements of both Ly$\alpha$ and 21~cm absorption. \citet{kanekar14_mnras} uses these temperature constraints, combined with very long baseline interferometric observations to constrain the covering factor, to report estimates of $T_s/f$ for four of our intervening absorbers\footnote{A caveat to consider is the relative sizes of the \HI\ absorber and background radio source; at radio wavelengths, the source may be significantly larger than the spatial distribution of \HI\ absorber, influencing the inferred spin temperature for the absorber \citep{curran05, braun12}. }, listed in Table~\ref{tab:detections}. These four absorbers show a range of $T_s/f = 560- 965$~K. \citet{kanekar14_mnras} find a strong redshift dependency to the spin temperature for absorbers at high redshift ($z>2.4$); high redshifts absorbers are more likely to exhibit spin temperature above 1000~K compared to low redshift absorbers \citep[see, however,][where the spin temperature may inversely trace the star formation rate density, with $T_s$ increasing towards low redshifts]{curran19_mnras}. As all of our absorbers have  measured $T_s/f<1000$~K and are located at redshifts below $z=2.4$, $T_s/f=1000$~K represents a realistic upper limit in this study.

\subsubsection{HI 21~cm Intrinsic Line Observations}\label{sec:intrinsicD}
97 sources have spectral coverage of the redshift of the host galaxy, enabling a search for intrinsic 21~cm absorption. This paper focuses only on the search for intervening systems and we disregard all spectral measurements within $\pm$3000~\kms\ of the systemic redshift of the host galaxy as any absorbers found here may be influenced by the AGN of the host galaxy and can potentially impact cosmological calculations (Section \ref{sec:fN} and \ref{sec:omega}). We re-detect a previously known 21~cm intrinsic absorption system of 3C216, shown in Figure~\ref{fig:detects}, along with the nine re-detections of intervening 21~cm absorption systems. The intrinsic \HI\ 21~cm absorbers are treated separately (Section~\ref{sec:interveningD}) and are further discussed in \citet{grasha19_apjs} in the context of searching for neutral gas intrinsic to these radio sources. 

\begin{figure*}
\includegraphics[scale=0.65]{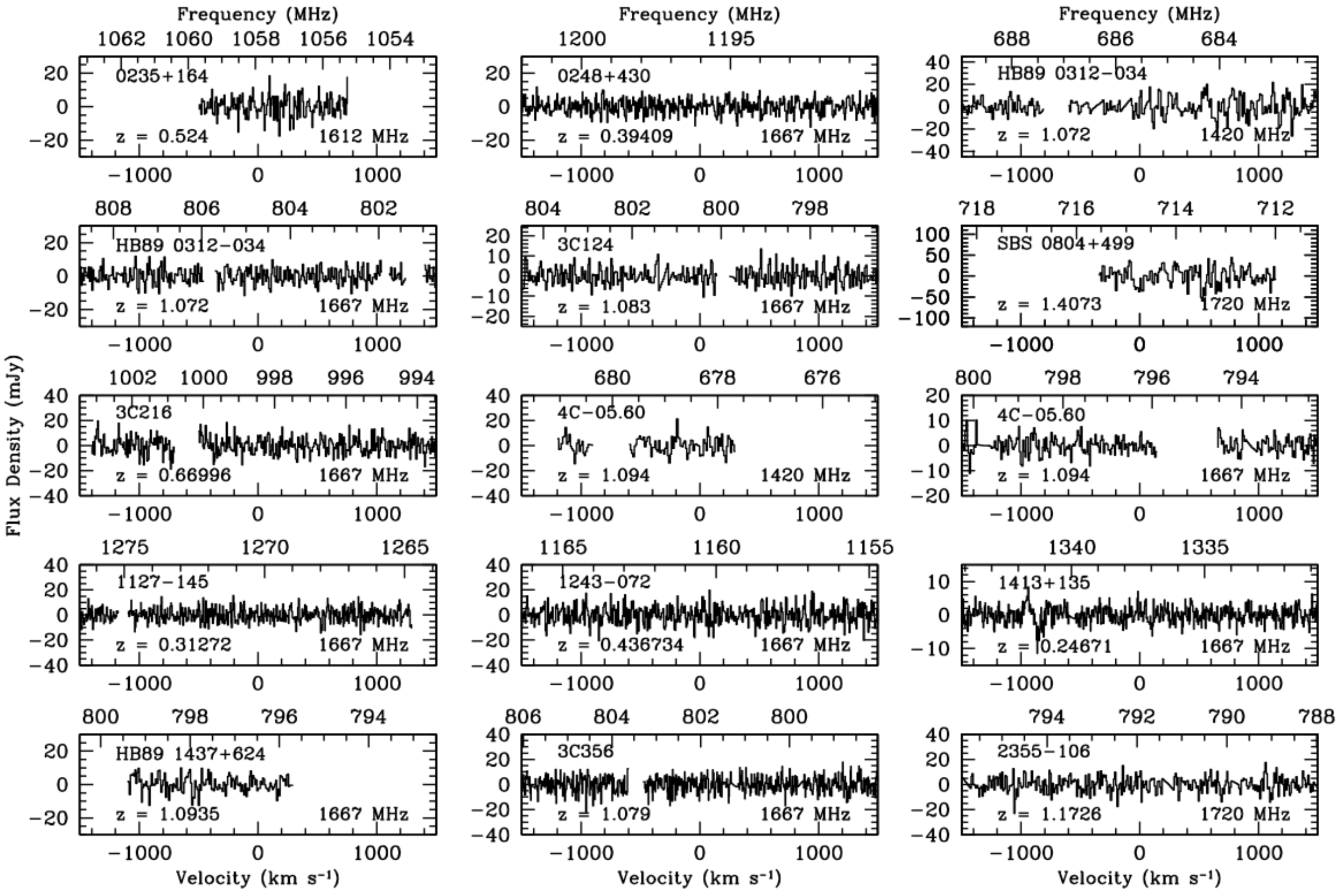}
\vspace{-15pt}
\caption{
Spectra of all non-detection observations in the two intrinsic 21~cm absorption searches (HB989 0312--034, 4C--05.60; Section~\ref{sec:intrinsicD}) and the OH line search of known intervening 21~cm absorbers. The velocity scale is in the rest-frame of each object source. All spectral regions lost to RFI have been omitted. 
While we search each system for all four OH lines (when observations are available), we only show the redshifted OH line responsible for deriving the OH column density (preferentially the 1667~MHz line, only using the 1612 or 1720~MHz lines if 1667~MHz is unusable due to RFI). 
}
\label{fig:OH_nonD}
\end{figure*}

There are an additional five sources with have spectral coverage at the redshift of the host galaxy that were not included in the associated \HI\ 21~cm absorption study \citep{grasha19_apjs}: HB89~0312$-$034, 3C124, 4C$-$05.60, HB89~1437+624, and 3C356. Of these five sources, three remain indeterminate due to RFI. We place upper limits on the intrinsic \HI\ column density for the remaining two sources, HB89~0312$-$034 and 4C$-$05.60. These two systems have no prior 21~cm absorption limits and this marks the first time these sources have been searched for intrinsic 21~cm absorption (as well as OH absorption; Section~\ref{sec:OH}) associated with the radio source. The non-detection spectra of these two intrinsic \HI\ systems are shown in Figure~\ref{fig:OH_nonD} and the $3\sigma$ column density limits are reported in Table~\ref{tab:OH}. 

\subsubsection{HI 21~cm Non-Detection Observations}
Our final sample consists of 2147 individual spectral measurement regions in 252 partially RFI-free sources. Each 50~MHz spectral region is typically limited by the presence of RFI. 99.5\% of our individual spectral regions (2137 regions) show no absorption lines. The 3$\sigma$ upper limit to the \HI\ column density in each spectral region not detected in 21~cm absorption (and not lost to RFI) is calculated with the average spectral noise $rms$ measured across the region. Figure~\ref{fig:NHI} shows the {\HI\ 21~cm line strength ($N_{\rm HI}/(T_s/f)$) for each individual spectral measurement in our survey as a function of redshift as well as the redshift and measured column densities for the re-detections.

\begin{figure}
\includegraphics[scale=0.45]{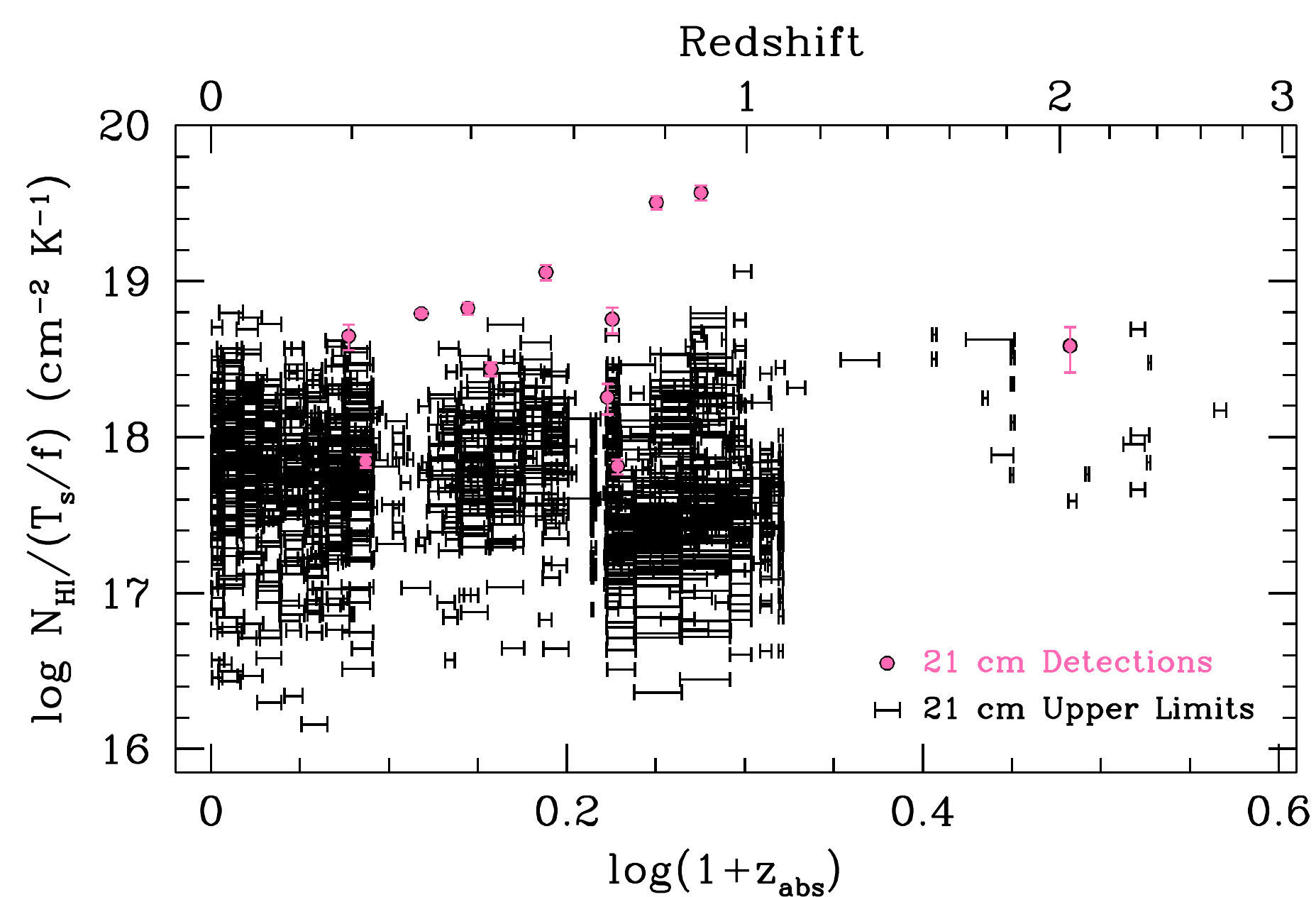}
\caption{\HI\ column density $N_{\rm HI}/(T_s/f)$ versus $\log(1+z_{\rm abs}$) for spectral regions where we search for \HI\ 21~cm absorption (redshift on the top axis). We show either the detected \HI\ column density versus systems detected at the 21~cm redshift (pink solid circles and accompanying 1$\sigma$ uncertainty in the measured column density) or its upper limit in the case of non-detections (black lines). The black horizontal bars on the 21~cm non-detections represent the 3$\sigma$ upper limit to the column density and the width of the bar indicates the redshift search region, where there is an implied downward-pointing arrow (not shown for clarity). There are a total of 2147 individual regions searched for 21~cm absorption, generally determined by RFI conditions. 58\% of the 3$\sigma$ upper limits to the \HI\ column density lie below the smallest column absorber, indicating that we reached sufficient sensitivity to detect new systems had they been present in our survey. 
\label{fig:NHI}}
\end{figure}

Although this survey was designed with the sensitivity to be able to detect DLA systems, RFI affected large redshift regions, lowering our ability to search the full potential redshift range observed toward each source (almost all spectral measurements are rendered unusable due to RFI for large regions of the 1~GHz receiver (PF2) and for most of the 600~MHz (PF1-600) and 450~MHz (PF1-450) receivers). Absorption lines that might appear in these regions therefore remain undetected. Additionally, increased rms noise levels at lower frequencies further hampered our sensitivity to low-column density systems. 

The non-detections reach an average optical depth of $\tau_{3\sigma}=0.0146$. This corresponds to 93\% of our upper limit measurements being below the DLA column density limit of $N_{\rm HI}=2\times10^{20}$~cm$^{-2}$ with the assumption $T_s/f=100$~K. While we cannot say there are not other DLAs present in our survey, we are confident we would have detected the presence of any major 21~cm absorbing feature as 58\% of our 3$\sigma$ column density upper limits lie below the weakest absorber ($N_{\rm HI}/(T_s/f)=6.5\times10^{17}$~cm$^{-2}$~K$^{-1}$) and 100\% of our 3$\sigma$ column density upper limits lie below the strongest absorber ($N_{\rm HI}/(T_s/f) = 3.2\times10^{19}$~cm$^{-2}$~K$^{-1}$), barring regions lost to RFI. Figure~\ref{fig:delz} highlights the strong absorbers ($N_{\rm HI}/(T_s/f)> 10^{19}$~cm$^{-2}$~K$^{-1}$) re-detected in the survey that lie significantly above the non-detections, demonstrating the sensitivity of our survey. 

Our low rate of detections is consistent with prior surveys of intervening \HI\ 21~cm absorption \citep{darling11, allison14, allison20}. Despite this, the large coverage of our survey allows us to attempt to place meaningful constraints on both the spin temperature of the \HI\ gas (Section~\ref{sec:fN}) and the cosmological evolution of the mass density of cold neutral gas as traced with the \HI\ 21~cm transition (Section~\ref{sec:omega}). 

\begin{figure}
\includegraphics[scale=0.45]{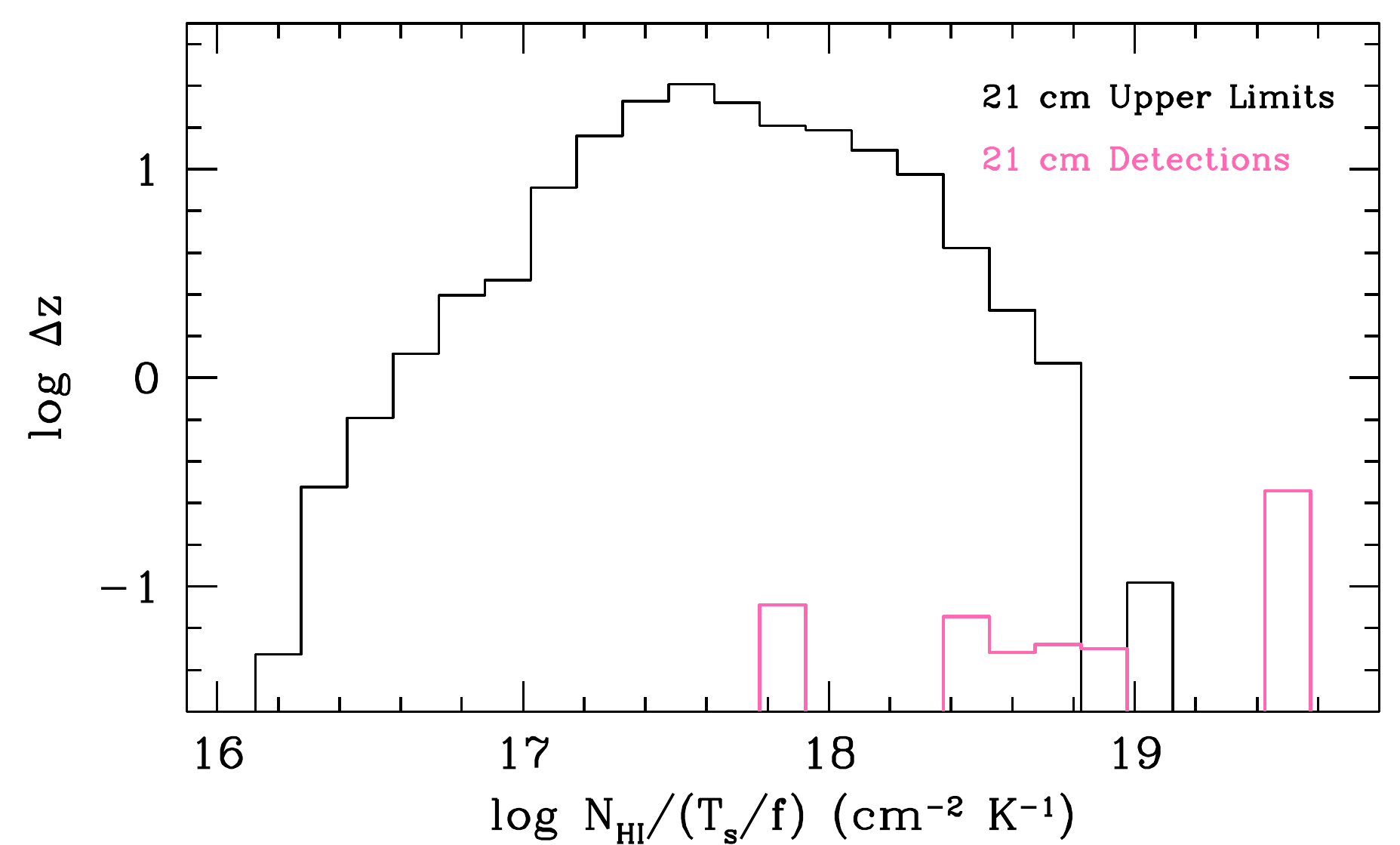}
\caption{
Redshift search path versus the 21~cm column density sensitivity for the nine intervening detections (pink) and non-detections (black). Each column density bin indicates the lower bound on the column density detectable toward the sources in that bin. Our sensitivity for non-detections peaks at $N_{\rm HI}/(T_s/f) \sim 3 \times 10^{17}$~cm$^{-2}$~K$^{-1}$, corresponding to $N_{\rm HI, 100K} \sim 5 \times 10^{19}$~cm$^{-2}$ under the assumption of  $T_s/f=100$~K, well below the sub-DLA limit of $N_{\rm HI} \geq 2\times10^{20}$~cm$^{-2}$. At $T_s/f=100$~K, 93\% of the upper limits below the DLA column density threshold. A value of $T_s/f=1000$~K corresponds to our survey peaking at a column density sensitivity of $N_{\rm HI, 1000K} \sim 3 \times 10^{20}$~cm$^{-2}$. At a value of $T_s/f=1000$~K, 15\% of the 3$\sigma$ upper limits lie below the DLA column density limit.
\label{fig:delz}}
\end{figure}

\subsubsection{OH 18~cm Line Observations}\label{sec:OH}
We search for the four OH 18~cm absorption lines at 1612.231, 1665.4018, 1667.35903, and 1720.5300~MHz in 11 sources associated with known 21~cm intervening absorbers as well as in the five systems where the redshift of the host galaxy falls in our spectral coverage that were not included in the \citet{grasha19_apjs} intrinsic absorption dataset. Redshifted OH absorption systems are very rare and are always found in systems that also exhibit 21~cm absorption and a strong correlation is observed between the full width half max (FWHM) of the OH and \HI\ absorption profiles \citep{curran07}. We thus only search for OH lines in the 21~cm absorber systems. Two sources were unusable due to RFI at the frequencies of all four OH lines. Of the 14 sources with usable spectra, two have prior OH measurements and we supply the first OH observations for the remaining 12 systems. 

We use the 1667~MHz line to derive OH column density because it places the strongest upper limit on the OH column density among all of the OH 18~cm lines. The continuum flux density refers to the redshifted 1667~MHz line frequency, except in the cases where the 1667~MHz was lost to RFI and the 1612 or 1720~MHz transition was used instead, and was obtained from power-law fits to literature radio continuum measurements. 

1667~MHz OH absorption is re-detected in the spectrum of PKS~1830$-$211, a known 21~cm and molecular absorber at $z=0.885$. Figure~\ref{fig:OH} shows this sole OH absorber in our sample with the corresponding 21~cm absorption line. We measure an OH column density of $N_{\rm OH}/T_x = 60 \pm 3 \times10^{13}$~cm$^{-2}$~K$^{-1}$, compared to the value of $N_{\rm OH}/T_x = 40 \times10^{13}$~cm$^{-2}$~K$^{-1}$ as first reported by \citet{chengalur99}. 

We do not re-detect the OH 1665 or 1667~MHz absorption lines toward 1413+135 at $z_{\rm abs}=0.24671$ despite reaching a smoothed spectral noise at a resolution similar to that of \citet{kanekar02}. They report an OH column density of $N_{\rm OH}/T_x = 5.1 \times10^{13}$~cm$^{-2}$~K$^{-1}$ and we place an upper limit of $N_{\rm OH}/T_x < 3.8 \times10^{13}$~cm$^{-2}$~K$^{-1}$. We are unable to re-detect the satellite OH lines at 1612 and 1720~MHz detected by \citet{darling04} and \citet{kanekar04_phrvl} due to noise at those frequencies. 

Excluding our re-detection of OH 1667~MHz toward PKS~1830$-$211, no new OH emission or absorption lines are detected in the remaining 13 sources with usable, RFI-free spectra in at least one of the OH transitions. The lack of detected OH absorbers is not surprising given that we also do not detect new \HI\ 21~cm absorbers; the optical selection of our DLA targets may bias against selecting obscured, molecular-rich absorbers along the sight-line to the DLA \citep{curran11b}. The RFI-free spectra for non-detections are shown in Figure~\ref{fig:OH_nonD}. Table~\ref{tab:OH} lists the results of the OH search, with the optical depths and column density limits quoted at the $3\sigma$ confidence level using the 1667~MHz line observations, when available.  

\begin{figure}
\includegraphics[scale=0.45]{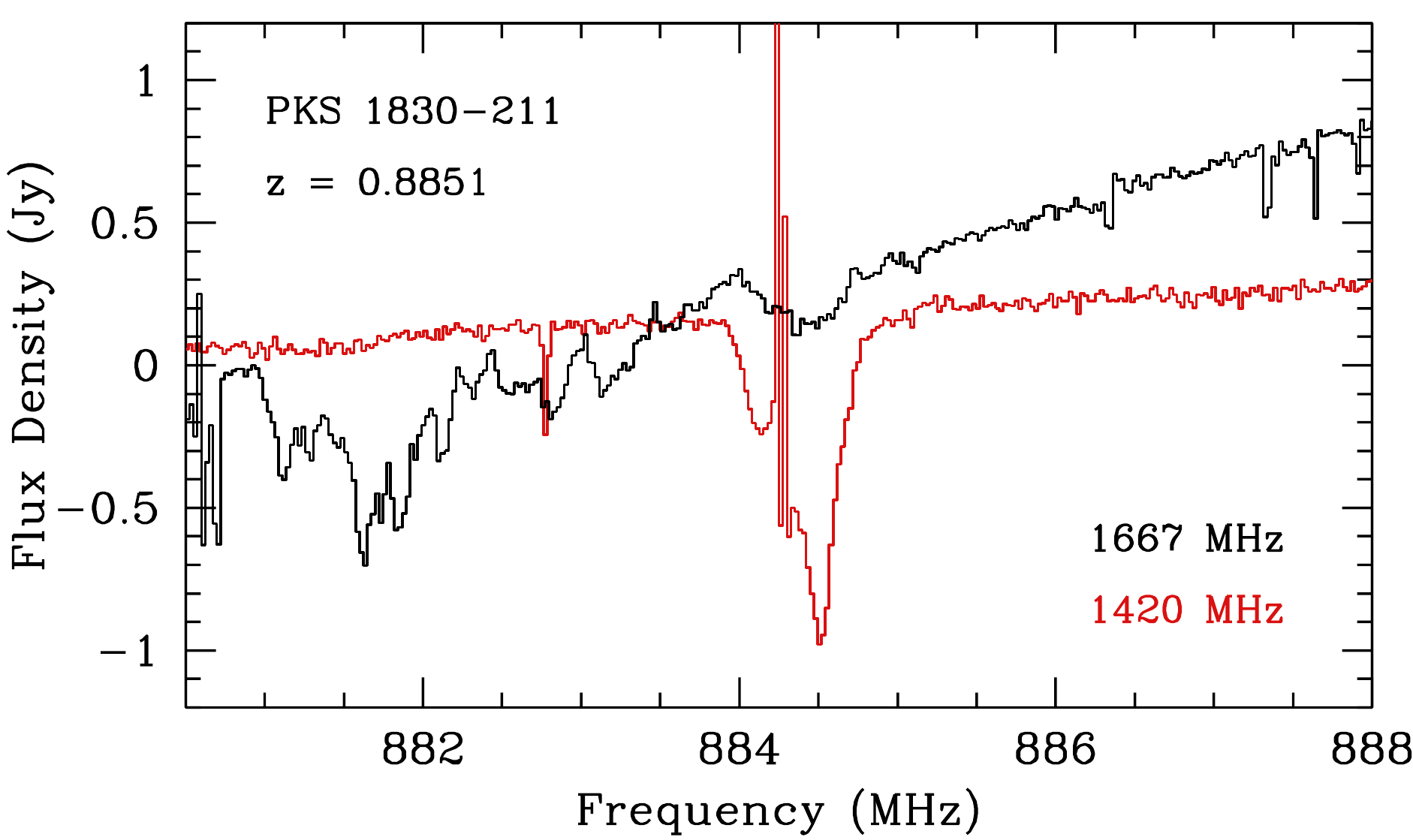}
\vspace{-15pt}
\caption{OH 18~cm 1667~MHz absorption toward PKS 1830$-$211 (black) overlaid with the corresponding 21~cm absorber (red) at the same redshift, shifted to the frequency of the 1667~MHz OH absorption line. Both continua have been arbitrarily normalized. The features to the left of the OH absorption line are RFI features as well as the spike centered in the 21~cm absorption profile.} 
\label{fig:OH}
\end{figure}

\subsection{Column Density Measurements}\label{sec:NHI}
\subsubsection{HI Column Density}\label{sec:columnHI}
Column density measurements are derived from the integrated optical depths of absorption lines. For optically thin systems ($\tau << 1$), the \HI\ column density of each absorption system is related to the optical depth $\tau$ of the line and the spin temperature $T_s$ of the absorbing system as \citep{wolfe75} 
	\begin{equation}\label{eq:NHI}
		N{\rm _{HI}} = 1.8 \times 10^{18} \ \textrm{cm}^{-2} \ \frac{T_s}{f} \int
                \tau \ {\rm d}v, 
	\end{equation}
where $f$ is the covering factor of the fractional amount of
the background continuum flux occulted by the absorber, $T_s$ is the spin temperature of the 21~cm absorption line in Kelvin, and $\int \tau \ dv$~(\kms) is the line-integrated optical depth. Lines with Gaussian absorption profiles can be integrated as $\int \tau \ dv = \sqrt{\pi/\ln 2}\ \tau_{\rm max}\,\Delta v/2$, where $\tau_{\rm max}$ is the peak optical depth of a line with a FWHM of $\Delta v$. The column densities for the ten re-detections (Section~\ref{sec:interveningD}) are listed in Table~\ref{tab:detections}. 

For non-detections in the optically thin regimes ($\tau \la 0.1$, which applies to all our sources) the $3\sigma$ upper limit column density can be approximated by assuming a Gaussian absorption profile, where the integrated optical depth reduces to $\int \tau \ dv \approx 1.0645 \ dv$, and $\tau$ is the $3\sigma$ peak optical depth for an assumed line FWHM of $dv$. Our $3\sigma$ upper limit column density measurements are approximated as 
\begin{equation}
	N{\rm _{HI, 3\sigma}}/(T_s/f) <
		1.8\times 10^{18} \  \textrm{cm}^{-2} \times \ 1.0645 \ \tau_{3\sigma} \ {\rm d}v, 
	\end{equation}
for the sources not detected in 21~cm absorption, where $\tau_{3\sigma} \approx 3\sigma/S$ and $S$ is the continuum flux value (obtained in Section \ref{sec:cntm}) of the background source at the redshifted line with measured spectral noise $rms$. We assume a FWHM width of 30~\kms\ for all sources not detected in 21~cm absorption. For all column density measurements and limits, we make no assumption on the covering factor or spin temperature of the gas.

\subsubsection{OH Column Density}
For the sources where we have spectral coverage to search for OH 18~cm absorption lines, we measure a column density using the 1667~MHz OH transition to place the strongest upper limit on the derived column densities for our OH search, given by 
	\begin{equation}
		N{\rm _{OH, 3\sigma}} < X_{1667~ \rm MHz} \  \textrm{cm}^{-2} \times 1.0645 \ \frac{T_x}{f}
                \ \tau_{3\sigma} \ {\rm d}v, 
	\end{equation}
where $X_{1667~ \rm MHz} = 2.38 \times 10^{14}$ \citep{curran08}, $T_x$ is the excitation temperature of the OH line, and the integrated optical depth $\tau$ ($\tau_{3\sigma}$ for non-detections) is calculated in the same way as the \HI\ 21~cm line detections (or non-detections). When the 1667~MHz line is unobservable due to RFI or is located outside our observing frequencies, we compute the OH column density using the 1612 or 1720~MHz transition, where $X = 2.14 \times 10^{15}$ ($X_{1720~ \rm{MHz}} = X_{1612~ \rm{MHz}}= 9 \times X_{1667~ \rm MHz}$). The observational details for the OH search toward known 21~cm absorbers are listed in Table~\ref{tab:OH}. No assumption is made about the excitation temperature $T_x$. We assume the OH absorbing systems have the same width as \HI\ absorbing profiles of 30~\kms.

\section{Analysis}\label{sec:analysis}

\subsection{The \HI\ Column Density Frequency Distribution Function, $f(N_{\textrm {HI}},X)$}\label{sec:fN} 
The parameter $f(N_{\textrm {HI}},X)$ describes the distribution of \HI\ column density in galaxies per comoving `absorption distance' path length $\Delta X$. $f(N_{\textrm {HI}},X)$ is calculated \citep{prochaska05} as 
	\begin{equation}\label{eq:fN}
		f(N_{\textrm {HI}},X) = \frac{\lambda_{absorbers}}{\Delta
                  N_{\textrm {HI}} \Delta X} , 
	\end{equation}
where $\lambda_{absorbers}$ is the total number of absorption systems within each column density interval $N$ to $N + dN$ and $\Delta X$ is the comoving redshift path over which the absorption systems can be detected for a given column density sensitivity. 

The absorption distance $\Delta
X$ is calculated \citep{wolfe05} as 
	\begin{equation}\label{eq:X}
		\Delta X = \int_{\rm z_{min}}^{\rm z_{max}}
                \frac{\mathrm{d}z \ (1+z)^2}{\sqrt{(1+z)^2~(1+z \
                    \Omega_m) - z~(z+2)~\Omega_{\Lambda}}~}, 
	\end{equation}
integrated from $z_{\rm min}$ to $z_{\rm max}$ from all steps in redshift searched over the whole range. $\Delta X$ is crucial in estimating the frequency distribution per unit column density and the cosmological mass density in neutral gas from the damped systems we are sensitive to in our survey. Our survey has a total redshift region searched $\Delta z=88.64$, corresponding to a comoving absorption distance $\Delta X=159.5$.

We calculate $f(N_{\textrm {HI}},X)$ for the nine detected intervening 21~cm absorbers with a column density $N_{\rm HI}$ in interval $\Delta N_{\rm HI}$ using Eq.~\ref{eq:fN}. We assume that $\Delta X$ does not vary with $N_{\rm HI}$. Table \ref{tab:fN} lists the values for $f(N_{\textrm {HI}},X)$ for values of $T_s/f=100, 250, 500$, and 1000~K along with the comoving absorption path length $\Delta X$ for which we are sensitive in each column density interval bin. The intrinsic detection 3C216 is not included in the $f(N_{\textrm {HI}},X)$ calculation. 

For the case of non-detections, limiting column density sensitivity estimates become an upper limit, calculated as 
	\begin{equation}
		f(N_{\textrm {HI}},X) < \frac{\lambda_{max}}{\Delta
                  N_{\textrm {HI}} \Delta X}, 
	\end{equation}
where $\lambda_{max}$ is the Poisson upper limit on the detection rate of absorption systems in a given column density limit bin. The 95\% confidence upper limit on the Poisson rate is $\lambda_{max} = 3.0$ for column density sensitivity bins with no detections \citep{gehrels86}. 

Figure~\ref{fig:fN} shows our upper limits on the column density frequency distribution function, consistent with previous studies of $f(N_{\textrm {HI}},X)$ in high redshift DLAs \citep{prochaska09, noterdaeme12}, low/intermediate redshift DLAs \citep{rao17}, and low redshift 21~cm emission \citep{zwaan05}. Due to the unconstrained nature of the hydrogen gas spin temperature, we compare the $f(N_{\textrm {HI}},X)$ distribution calculated at varying spin temperatures of 100, 250, 500, and 1000~K against the these four prior empirical measurements to enable a statistical constraint on the spin temperature and source covering fraction ratio $T_s/f$. An increase in the spin temperature for the \HI\ column density limit measurements shifts the data to both larger column densities and smaller $f(N_{\rm HI},X)$ values.

\begin{figure*}
\includegraphics[scale=0.93]{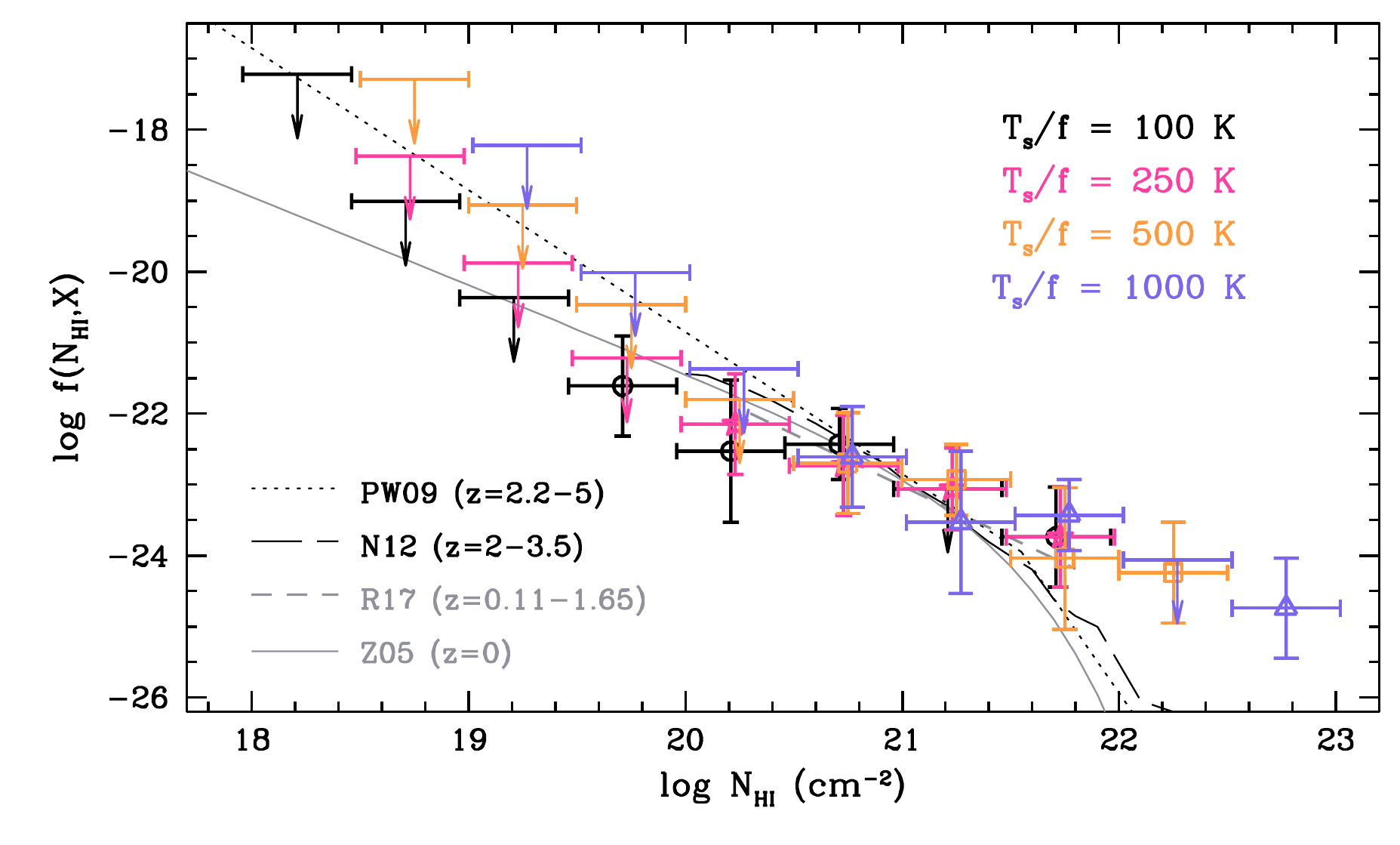} 
\vspace{-10pt}
\caption{
The column density frequency distribution function $f(N_{\textrm {HI}},X)$ versus \HI\ column density for varying values of spin temperature and covering factor $T_s/f$ for the detections and upper limits. Black points represent the measurements with an assumed spin temperature of $T_s/f=100$~K, pink represent $T_s/f=250$~K, orange represents $T_s/f=500$~K, and purple represents $T_s/f=1000$~K. The median value of $T_s/f$ for the detections with known measurements in the survey is $T_s/f=800$~K (Table~\ref{tab:detections}). The horizontal bars indicate the column density range spanned by each value. Open symbols represent column density bins with at least one 21~cm absorber detected and bins with a downward pointing arrow are non-detections with upper limits to the measured column density. The upper limits show the 95\% Poisson confidence level ($\lambda_{max}=3.0$) for bins with no absorption lines. The short dotted black line (PW09) indicates the double-power law fit by \citet{prochaska09} for absorption systems at $z =2-5$, the long dashed black line (N12) indicates the  empirical relation to $z=2-3.5$ DLAs from SDSS-III DR9 by \citet{noterdaeme12}, the gray dotted line (R17) shows the power law fit to ultraviolet measurements of Ly$\alpha$ from $0.11 < z < 1.65$ by \citet{rao17}, and the solid grey line (Z05) is the $\Gamma$-function fit by \citet{zwaan05} with 21~cm emission observations at $z=0$. Our results for $f(N_{\textrm {HI}},X)$ are consistent with earlier $f(N_{\textrm {HI}},X)$ estimates at both low and high redshifts. We use our detections and upper limit measurements to constrain the range of the spin temperature of the absorbing gas. Our column density measurements and limits are consistent with \HI\ spin temperatures up 1000~K with prior literature measurements in the same redshift range. Our observations support lower $T_s/f$ values in the highest column density bins in order to reproduce the powerlaw break in $f(N_{\textrm {HI}},X)$ at $N_{\rm HI}>10^{21}$~cm$^{-2}$ as seen in prior literature measurements. 
\label{fig:fN}}
\end{figure*}

Despite being dominated by non-detections, our 21~cm observations are able to accurately describe the $f(N_{\rm HI},X)$ distribution even with the ill-constrained nature of the spin temperature. The $f(N_{\rm HI},X)$ measurements for low column systems ($N_{\rm HI}<10^{21.5}$~cm$^{-2}$) are consistent with prior estimates of $f(N_{\textrm {HI}},X)$ with a range of temperatures from $T_s/f=100-1000$~K. The upper limits best agree with prior fits to $f(N_{\textrm {HI}},X)$ for gas with spin temperatures warmer than $T_s/f=500$~K. For sub-DLA systems ($N_{\rm HI}<2\times10^{20}$~cm$^{-2}$), our limits and measurements are consistent with the low redshift 21~cm emission measurements from \citet{zwaan05}. While overlapping with the reshift of our survey, the observations from \citet{rao17} do not extend to sub-DLA absorbers and place no additional constraint on the $T_s/f$.

In comparison, in order to reproduce the turnover in $f(N_{\textrm {HI}},X)$ above $N_{\rm HI}>10^{22}$~cm$^{-2}$, our data do not warrant these high column absorbers residing in systems with spin temperatures significantly warmer than $\sim250$~K. These high column absorbers are inconsistent with prior measurements from \citet{zwaan05, prochaska09, noterdaeme12} at all values above $T_s/f=250$~K; the same motivating constraint for adopting a spin temperature in these absorbers of $T_s/f\sim250$~K is reflected in our constraint of $\Omega_{\textrm {HI}}$ (Section~\ref{sec:omega}) as well.

While the assumption of a constant and universal spin temperature across all of our absorbers is likely too simplified, \citet{darling11} shows that a random distribution of temperatures between 100-1000~K give results very similar to a constant value of $T_s/f=550$~K. Their temperature range is consistent with the temperatures we find for agreement with prior empirical fits for absorbers and upper limits with column densities below $N_{\rm HI}<10^{21.5}$~cm$^{-2}$. 

Our results suggest that an increase in the column density sensitivity along with larger redshift coverage should place more meaningful constraints on the spin temperature of \HI\ absorbing systems, especially toward the lower column systems where there still remains little constraint on $Ts/f$ from our data.

\subsection{The Cosmological Mass Density of Neutral Gas, $\Omega_{\textrm {HI}}$}\label{sec:omega} 
The \HI\ gas mass density $\Omega_{\textrm {HI}}$ describes the total \HI\ column
density per unit absorption distance, obtained by the first moment of the
column density frequency distribution $f(N_{\textrm {HI}},X)$. The \HI\ gas mass density $\Omega_{\textrm {HI}}$ is a useful parameter to express how the \HI\ gas content of galaxies evolves as a function of cosmic time.

We estimate $\Omega_{\textrm {HI}}$ from 21~cm detections \citep{prochaska05} as 
	\begin{equation}
		\Omega_{\textrm {HI}} \equiv \frac{H_{\circ} \ m_{\rm
                    H}}{c \rho_{crit}} \int_{N_{\rm min}}^{N_{\rm max}}
                N_{\rm HI} \ f(N_{\rm HI},X) \ \rm{d}N, 
	\end{equation}
where $f(N_{\rm HI},X)$ is described and calculated in Section~\ref{sec:fN}, $m_{\textrm H}$ is the mass of the hydrogen atom, $\rho_{crit}$ is the critical mass density, $c$ is the speed of light, and $H_{\circ}$ is Hubble's constant. The integral can be evaluated discretely, which is most commonly done by setting the limits of $N_{\rm min}=2\times10^{20}$~cm$^{-2}$ and $N_{\rm max}=\infty$ to find the cosmological mass density of neutral gas associated with DLAs, $\Omega_{\rm DLA}$.  $\Omega_{\rm DLA}$ may be significantly less than $\Omega_{\textrm {HI}}$ if absorbers below the damped Ly$\alpha$ threshold contribute to the global \HI\ mass density \citep{peroux03}. Sub-DLAs ($19.0 \leq \log N_{\rm HI} < 20.3$ do contribute 10--20\% of the total neutral gas content of the universe at redshifts $2 < z < 5$, and thus these sub-DLAs are a small contribution to the total \HI\ budget of the universe over this range \citep{tberg19}. $\Omega_{\textrm {HI}}$ integrated from damped Ly$\alpha$ systems thus best defines the predominately neutral gas reservoir that are available for star formation at high redshift. 

At the smallest value of $T_s/f=100$~K that we consider in this study, two of our intervening 21~cm absorbers do not meet the classical definition of a DLA. For consistency, we refer to the neutral mass density of gas as $\Omega_{\textrm {HI}}$, the total gas mass density associated with \HI\ atoms regardless of the column density of the absorbing system from which they arise. The integral for the mass density associated with all \HI\ atoms within our 21~cm absorbing systems is integrated from the lowest column density of our nine intervening absorbers ($N_{\rm HI}=0.65\times10^{20}$~cm$^{-2}$) to $N_{\rm max}=\infty$. The discrete limit of $\Omega_{\textrm {HI}}$ in our
survey is estimated as: 
	\begin{equation}\label{eq:omega}
		\Omega_{\textrm {HI}} = \frac{H_{\circ} \ m_{\textrm H}}{c \rho_{crit}} 
 			\frac{\sum_i N_{i, \textrm {HI}}}{\Delta X},
	\end{equation}
where $\sum_i N_{i, \textrm {HI}}$ is the sum of the column density measurements for each system in a given redshift interval $\Delta X$ for which each absorber $N_{i, \textrm {HI}}$  could be detected. The intrinsic detection in 3C216 is not included in the $\Omega_{\rm HI}$ calculation. We assume no contribution from helium or molecular hydrogen in our calculations of $\Omega_{\rm HI}$. To addresses the statistical error dominated from small number of absorbers in each redshift bin, we calculate 1$\sigma$ errors in $\Omega_{\rm HI}$ from 10,000 bootstrap samples. 

Despite the evolution of $\Omega_{\rm HI}$ with redshift being rather uncertain, prior measurements of $\Omega_{\rm HI}$ at $z\sim0$ show remarkable consistency with each other and serve as anchor points for our low redshift observations. We therefore physically motivate a range of $T_s/f$ for our calculations of $\Omega_{\rm HI}$ by leveraging prior literature measurements of $\Omega_{\rm HI}$. The assumed value of $T_s/f$ will affect the overall normalization of the $\Omega_{\textrm {HI}}$ and is discussed further below. We show our measurements of $\Omega_{\rm HI}$ with select prior values of low redshift 21~cm emission surveys, 21~cm spectral stacking, and Ly$\alpha$ studies of DLAs \citep{zwaan05, rao06, lah07, noterdaeme12, hoppmann15, neeleman16, bird17, rhee18, hu19} in Figure~\ref{fig:omega}. We remove the mean molecular weight of $\mu=1.3$ (correcting for 25\% contribution of neutral gas by He II) from literature values if applicable. 

Despite only nine intervening absorption systems, we supply the very first measurement of $\Omega_{\rm HI}$ from a blind intervening 21~cm absorption survey with continuous redshift coverage over a lookback time of 11~Gyr (Figure~\ref{fig:omega}). Our sensitivity to low-column systems (Figure~\ref{fig:delz}) coupled with a large total redshift region searched $\Delta z=88.64$ ($\Delta X=159.5$; Eqn.~\ref{eq:X}) allows our measurement of $\Omega_{\rm HI}$ to be competitive with previous studies even with our low detection rate. To trace the evolution of neutral gas over cosmic time, we split our sample into two redshift bins at $z=0.69$. Five detections fall into the lower redshift bin and four detections lie beyond $z=0.69$. 

In the low-redshift interval of $0<z<0.69$ we calculate $\Omega_{\rm HI/(T_s/f) = 0.21 \pm 0.10 \times 10^{-5}~K^{-1}}$. We then compare our $\Omega_{\rm HI/(T_s/f)}$ values with the average of prior $\Omega_{\textrm {HI}}$ measurements obtained from previous low redshift \HI\ studies to anchor our $\Omega_{\rm HI/(T_s/f)}$ values and constrain the unknown quantity $T_s/f$. Previous 21~cm emission studies and \HI\ spectral stacking \citep[e.g.,][]{zwaan05, hoppmann15, rhee18, hu19} show consistent values of $\Omega_{\rm HI} \sim 0.3-0.4 \times 10^{-3}$. Using the average of these previous low redshift $\Omega_{\rm HI}$ studies, we compare our derive values of $\Omega_{\rm HI/(T_s/f)}$ and constrain a spin temperature -- covering factor of $T_s/f \sim$ 175~K. At a value of $T_s/f=$ 175~K, we measure $\Omega_{\rm HI} = 0.37 \pm 0.13 \times 10^{-3}$ for the low redshift interval $0<z<0.69$. 

In the high-redshift interval of $0.69<z<2.74$ we calculate $\Omega_{\rm HI/(T_s/f) = 0.69 \pm 0.45 \times 10^{-5}~K^{-1}}$. We use the same $T_s/f=175$~K constrained from prior low redshift $\Omega_{\rm HI}$ studies to calculate $\Omega_{\rm HI}$ in our high redshift bin. We report $\Omega_{\rm HI} = 1.2 \pm 0.6 \times 10^{-3}$ for $0.69<z<2.74$. This constrained value of $T_s/f=$ 175~K falls in the range of $T_s/f=100-1000$~K we considered for the the column density frequency distribution $f(N_{\textrm {HI}},X)$ analysis (Section~\ref{sec:fN}) in this study.

The small number of absorbers results in poor statistical constraints on our measured $\Omega_{\rm HI}$ and we have large uncertainties in comparison with prior observations, especially within the higher redshift interval of $0.69<z<2.74$ where our two strongest absorbers reside. We additionally measure $\Omega_{\textrm {HI}}$ at the same four spin temperatures of $T_s/f=100, 250, 500,$ and 1000~K that used for the  $f(N_{\rm HI},X)$ calculations, listed in Table~\ref{tab:fN} in each redshift bin. This allows us to estimate upper limits to the average $T_s/f$ value of our survey consistent with prior studies in the same redshift range. In our low redshift interval, we remain consistent within the errors with prior UV-selected DLAs studies \citep{rao06, rao17} for a spin temperature up to $T_s/f=500$~K ($\Omega_{\rm HI} = 1.05 \pm 0.33 \times 10^{-3}$; Table~\ref{tab:fN}). 

In the high redshift interval of $0.69<z<2.74$, compared against prior measurements of $\Omega_{\rm HI}$ near $z\sim1$, the median redshift of the absorbers in this interval, we constrain the spin temperature up to a value of $T_s/f\sim250$~K while still consistent with prior measurements, calculating $\Omega_{\rm HI} = 1.7 \pm 0.9 \times 10^{-3}$. While our value of $\Omega_{\rm HI}$ at $T_s/f=250$~K is larger than prior $\Omega_{\rm HI}$ measurements at the same redshift, it remains consistent given the large uncertainties resulting from few absorbers. The two largest column absorbers reside in this high redshift bin and the upper constraint for $T_s/f=250$~K from prior DLA $\Omega_{\rm HI}$ measurements agrees with the constraint we placed on $T_s/f$ from prior the $f(N_{\textrm {HI}},X)$ measurements in Section~\ref{sec:fN}.

\begin{figure*}
\includegraphics[scale=0.93]{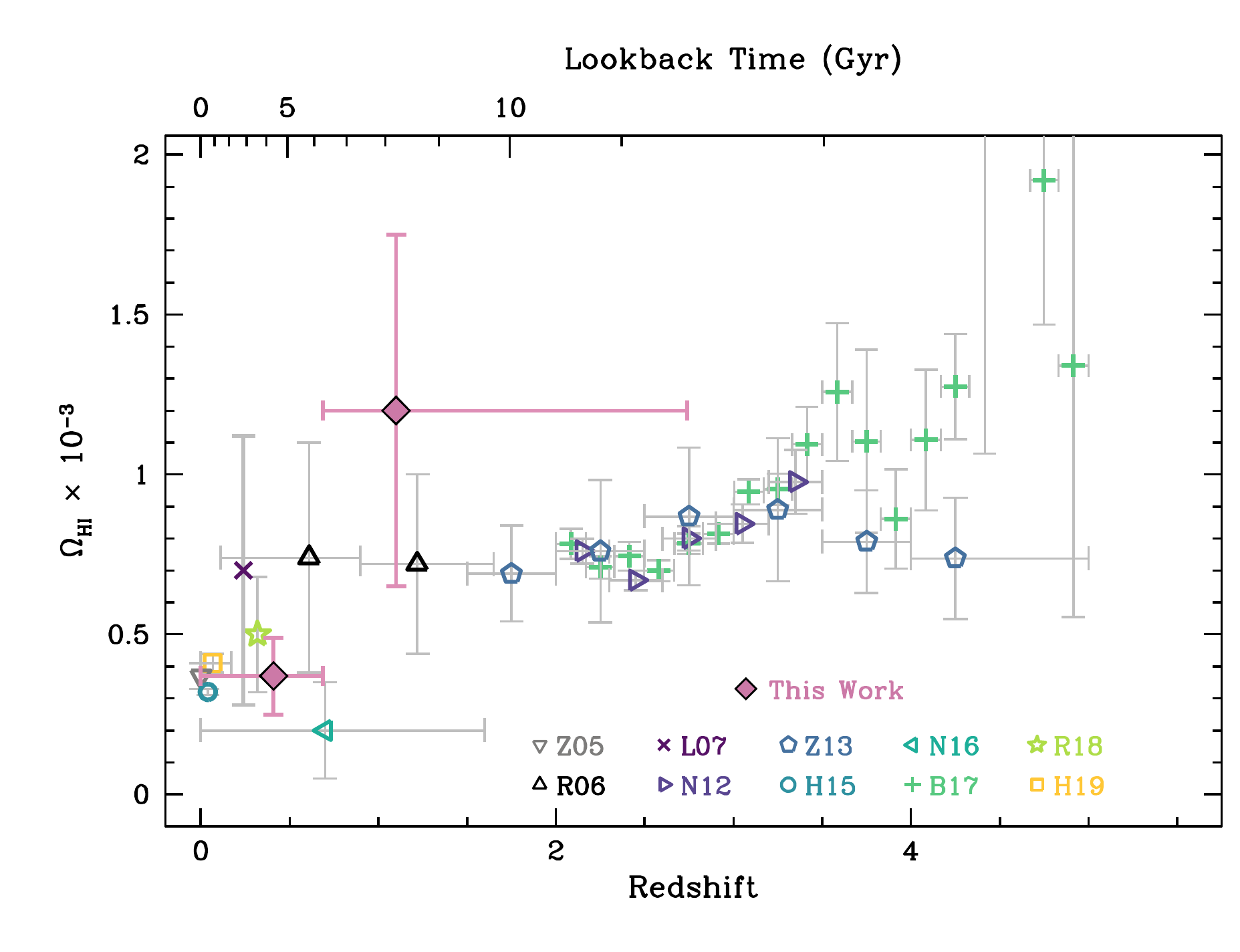}
\vspace{-20pt}
\caption{
The cosmological mass density of neutral gas $\Omega_{\textrm {HI}}$ over the past 11 Gyr as a function of redshift (lookback time on the top axis) shown as pink diamonds in two redshift bins from $0<z<2.74$ for $T_s/f=175$~K; this temperature is consistent with $\Omega_{\textrm {HI}}$ obtained from prior low redshift measurements. At $T_s/f=$ 175~K, we measure $\Omega_{\rm HI} = 0.37 \pm 0.13 \times 10^{-3}$ for the low redshift interval $0<z<0.69$ and $\Omega_{\rm HI} = 1.2 \pm 0.6 \times 10^{-3}$ in the high-redshift interval of $0.69<z<2.74$. 
The vertical error bars represent the 1$\sigma$ uncertainties of $\Omega_{\textrm {HI}}$ from 10,000 bootstrap samples and the horizontal error bars indicate the redshift bin sizes. We compare estimates of the cosmological mass density of neutral \HI\ to prior surveys: 
the down-pointing triangle (Z05) at $z=0$ represents the HIPASS \HI\ 21~cm emission survey \citep{zwaan05}, 
the triangles (R06) at $0<z<1.7$ represent a survey for DLAs selected as strong Mg II absorbers with the HST \citep{rao06}, 
the X-symbol (L07) represents a survey of \HI\ emission from star-forming galaxies at $z=0.24$ \citep{lah07}, 
the right-pointing triangles (N09) at $2.2<z<5$ represent a DLA survey from the SDSS-DR9 \citep{noterdaeme12}, 
the hexagons (Z13) at $1.5<z<5$ are VLT/UVES measurements of DLAs \citep{zafar13}, 
the circle (H15) represents HI spectral stacking from the Arecibo Ultra Deep Survey at $z=0$ \citep{hoppmann15},  
the left-pointing triangle (N16) represent a DLA survey with the HST from $0<z<1.5$ \citep{neeleman16}, 
the $+$ symbols (B17) from $2<z<5$ represent a DLA survey from the SDSS-DR12 \citep{bird17}, 
the star (R18) at $z=0.32$ is from \HI\ stacking with the WSRT \citep{rhee18}, and
the square (H19) at $0<z<0.11$ represents a survey to identify new DLA and Lyman-limit systems with spectral stacking from the WSRT \citep{hu19}, 
We have corrected all measurements to a consistent definition of $\Omega_{\textrm {HI}}$. Table~\ref{tab:fN} reports the calculations of $\Omega_{\textrm {HI}}$ for a range of $T_s/f$.
\label{fig:omega}}
\end{figure*}

\section{Discussion}\label{sec:discussion}
21~cm emission line measurements at low redshift all provide consistent constraints with each other for the \HI\ mass density in the local Universe \citep{zwaan05, martin10, hoppmann15}. DLA surveys at redshifts $z>2$ suggest an increase in the \HI\ gas mass density, suggesting significant evolution of \HI\ gas from $z\sim2$ to $z=0$ \citep{storrie-lombardi00, peroux03, prochaska05, noterdaeme09, noterdaeme12, crighton15, bird17}. Despite the fact that measurements at $z \gtrsim 2$ consistently tend to show larger values of $\Omega_{\rm HI}$ than those at low redshifts, $\Omega_{\rm HI}$ measurements at intermediate redshift $0.4\lesssim z \lesssim2$ are inconsistent and there is still much controversy in the exact form and evolution of $\Omega_{\rm HI}$ over cosmic time \citep[see, e.g.,][and references therein]{crighton15, rhee18}. Some of the discrepancies in this redshift range likely come from the different techniques used for target acquisition and subsequent measurement of $\Omega_{\textrm {HI}}$ in high redshift DLA surveys, such as potential biases arising from target pre-selection. Pre-selecting sightlines based on the strength of the metal absorption is commonly done at intermediate redshifts to improve the DLA detection rate with space-based observatories, however this may bias the detection rate of \HI\ absorbers, especially against those with low column densities and metallicities \citep[][and references therein, but see Rao et al. 2017]{dessauges-zavadsky09, neeleman16, berg17}. 

Despite uncertainty in the spin temperature of intervening \HI\ 21~cm absorbers, our measurements allow us to bridge the redshifted DLA surveys with low redshift 21~cm survey results and allow us to place constrains on $\Omega_{\textrm {HI}}$ when anchored to $z\sim0$ \HI\ 21~cm surveys \citep[e.g., ][]{zwaan05, hoppmann15, mjones18}. We find agreement between our low redshift $\Omega_{\textrm {HI}}$ value from $0<z<0.69$ with previous $\Omega_{\textrm {HI}}$ measurements in the same redshift range for spin temperatures in the range 100~K to 500~K -- where prior low redshift $\Omega_{\textrm {HI}}$ measurements constrain our survey to $T_s/f=175$~K -- supporting a mild evolution in $\Omega_{\rm HI}$ in the redshift range from $z=0$ to $z\sim2$. Several of our absorbers have a constraint on $T_s/f$ from Ly$\alpha$ observations (Table~\ref{tab:detections}), with a median value of $T_s/f\sim 800$~K. For the lower redshift absorbers ($0<z<0.69$), this corresponds to $\Omega_{\rm HI} = 1.7 \pm 0.5 \times 10^{-3}$, which remain consistent with larger values of $\Omega_{\textrm {HI}}$ at the low redshift range \citep{rao06, lah07}. A spin temperature of $T_s/f\sim 800$~K for the high redshift bin gives a measurement of $\Omega_{\rm HI} = 5.4 \pm 2.7 \times 10^{-3}$, significantly larger than prior $\Omega_{\textrm {HI}}$ observations for absorbers at $z>1$. Theoretical understanding of the evolutionary trend of $\Omega_{\textrm {HI}}$ remain incomplete and our survey demonstrates the success of 21~cm absorption surveys to assist in solving tension and shaping future theoretical evolutionary models $\Omega_{\textrm {HI}}$ \citep[e.g.,][]{kimhs15, dave17, lagos18, diemer19, hu20}. 

Despite DLAs tracing the bulk of neutral gas at $2<z<3$ \citep[e.g.,][]{prochaska05,noterdaeme09, crighton15} and expectations that the abundance of hydrogen is higher during the era of cosmic noon than in the present-day universe \citep{peroux01}, 21~cm absorption line surveys at high redshift do not have the detection rate seen in low redshift surveys. There are several factors that negatively impact the detection rates of 21~cm absorbers. The biggest detriment to the detection of new 21~cm absorbing systems in our survey is irremovable RFI in our single-dish observations. RFI generally worsens toward lower frequencies and renders large areas of the radio spectrum unusable, especially toward high redshift. Spectral noise measurements, coupled with radio-faint sources, result in weak constraints to our measured integrated optical depths, lowering our sensitivity to detect absorption systems. While there is no easy way to alleviate RFI in low-frequency spectral observations with single-dish observations such as from GBT, increased redshift coverage and sample size, as well as targeting regions with minimal RFI impact, will help future surveys to reach the sensitivity necessary to detect new 21~cm absorption systems. Interferometric data are in general less susceptible to RFI than single-dish observatories. ASKAP Band 1 (700--1000~MHz) is essentially free of RFI and will make large blind surveys, such as FLASH \citep{allison16, allison20}, more effective than previous searches.

The spin temperature -- covering factor degeneracy $T_s/f$ (Eqn.~\ref{eq:NHI}) may further impact the sensitivity of observations, worsening toward high redshifts. Poor covering of the background radio source will decrease sensitivity of the absorber, negatively impacting the detection of new absorbers. Absorbers at higher redshifts will suffer from geometric effects more and lower the incidence of the absorber covering the background source \citep{curran12a}. Likewise, spin temperature may also be systematically higher at higher redshifts, decreasing the measured absorber optical depth with increasing redshift \citep{kanekar03, srianand12, nroy13, curran19_mnras}. Neither the spin temperature of the absorber nor the covering factor of the background source is constrained for most \HI\ 21~cm absorption systems. As the spin temperature cannot be constrained with observations of \HI\ 21~cm absorption alone, studies typically assume a canonical value ranging from $T_{s}/f=100-1000$~K and these appear to be reasonable estimates for intervening \HI\ absorption systems at redshifts $z\lesssim2$ \citep{kanekar14_mnras}. Our intervening 21~cm absorbers support this range in $T_{s}/f$ values through comparing our $f(N_{\textrm {HI}},X)$ (Figure~\ref{fig:fN}) results with expectations from prior temperature-independent methods. Our two largest absorbers do not have prior constraints on $T_{s}/f$ and are more consistent with prior $f(N_{\textrm {HI}},X)$ measurements for lower spin temperature values of $T_s/f=\sim250$~K. This suggests that the $T_s/f$ distribution may depends on the \HI\ column density of the absorber where absorbers with the largest \HI\ column densities contain a larger contribution of cooler gas. Deviations toward higher temperatures and/or lower covering factors will decrease sensitivity and bias against detection of lower column systems and is an important factor in our lack of new detections. 

Lastly, source selection criteria may bear some of the responsibility, worsening at high redshifts, where optical biases select galaxies that are less likely to be obscured by gas and dust. Future ``blind'' 21~cm absorption surveys along dusty sightlines may alleviate the bias of selecting against UV-faint and dusty sources where it is more likely to have cold gas along those lines of sight, otherwise missed by optical/UV surveys \citep[e.g., ][]{curran11b, yan12, yan16, glowacki19}. Such surveys may be more fruitful in selecting cold gas systems readily available for detection in 21~cm absorption as well as rarer molecular absorption systems in order to provide adequate statistics to help constrain factors of cosmological importance.

\section{Summary and Conclusions}\label{sec:conclusions}
We present the results of a blind survey for intervening \HI\ 21~cm absorption along the line of sight to 260 radio sources in the redshift range $0<z<2.74$. We make 2147 individual RFI-free spectral measurements and we re-detect ten previously known 21~cm absorbers in our sample (9 intervening, 1 intrinsic) and observe no new detections of intervening 21~cm absorption systems. We place a $3\sigma$ upper limit on the \HI\ column density for all searchable spectra not detected in 21~cm absorption in 2137 individual spectral regions across the sample for a total redshift searched path of $\Delta z=88.64$ (comoving path of $\Delta X=159.5$). We record a mean 21~cm optical depth for the non-detections of $\tau_{3\sigma}=0.0146$, corresponding to 93\% of upper limit spectral measurements lying below the DLA column density threshold under the assumption of a spin temperature and covering factor for the absorbers of $T_s/f=100$~K. At $T_s/f=100$~K, 15\% of upper limit spectral measurements are below the DLA column density limit. 99\% of our spectral measurements are constrained to $z<1$.

Sixteen of our sources are searched for OH 18~cm absorption at the redshift of the known 21~cm absorbing systems. We re-detect the 1667~MHz line toward PKS~1830$-$211. We do not re-detect the 1665 or 1667~MHz absorption features toward 1413+135 as detected by \citet{kanekar02} despite smoothing our spectral measurements to the same resolution. We do not have adequate signal and sensitivity to re-detect the two satellite OH lines toward 1413+135 at 1612 and 1720~MHz at the same redshift by \citet{darling04, kanekar04_phrvl}. We place upper limits on the OH column density for the remaining 12 sources with usable OH data. While not all of our OH column density limits are significant due to RFI and/or bad spectral measurements, it provides a reference point for observations and the sensitivity necessary in the future to detect possible OH absorbing systems. 

We place measurements and upper limit estimates for the \HI\ column density frequency distribution function $f(N_{\rm HI},X)$. We estimate $f(N_{\rm HI},X)$ at spin temperatures of $T_s/f = 100, 250, 500$, and 1000~K. Through comparison with prior empirical measurements of 21~cm emission line surveys at $z\sim0$ \citep{zwaan05}, ultraviolet DLA surveys at $z<1.65$ \citep{rao17}, and optical DLA surveys at $z>2$ \citep{prochaska09, noterdaeme12}, we constrain the spin temperature of the \HI\ absorption systems and place limits on the spin temperature of our non-detection systems. We find that our absorbers and upper limit measurements are consistent with prior measurements in the range $T_s/f = 100-1000$~K for the redshift range of our survey. We find that the largest \HI\ column absorbers $N_{\rm HI} > 10^{22}$~cm$^{-2}$ require temperatures lower than $T_s/f\lesssim250$~K in order to reproduce the turnover in $f(N_{\rm HI},X)$ distribution at high \HI\ column densities observed in prior temperature-independent observations. This suggests that the spin temperature values for our non-detection and low \HI\ column density absorbers have larger fractional contributions of warmer gas compared to the high \HI\ column density absorbers. 

We estimate the cosmological mass density of neutral gas $\Omega_{\rm HI}$, the largest blind survey of redshifted \HI\ 21~cm absorbers to constrain $\Omega_{\rm HI}$ continuously in the redshift range $0<z<2.74$. We estimate $\Omega_{\rm HI}$ at spin temperatures of $T_s/f = 100, 250, 500$, and 1000~K in two redshift bins: $0<z<0.69$ and $0.69<z<2.74$ to constrain the relative difference between $\Omega_{\rm HI}$ with redshift and infer a possible redshift evolution between the low and high redshift absorbers. 

We are able to place statistical constraints on $T_s/f$ through comparing our low redshift $\Omega_{\rm HI}$ value with prior 21~cm emission and 21~cm absorption stacking studies. With previous $z<0.69$ studies of $\Omega_{\rm HI}$ serving as anchor points, we estimate a mean $T_s/f=175$~K for our calculations of $\Omega_{\rm HI}$. At $T_s/f=$ 175~K, we measure $\Omega_{\rm HI} = 0.37 \pm 0.13 \times 10^{-3}$ for the low redshift interval $0<z<0.69$ and $\Omega_{\rm HI} = 1.2 \pm 0.6 \times 10^{-3}$ in the high-redshift interval of $0.69<z<2.74$. Our values remain consistent with studies up to $T_s/f = 250-500$~K. 
Despite our estimate of the \HI\ mass density at high redshift ($0.69<z<2.74$) accompanied with large uncertainties, we show agreement with previous estimates owing to the large redshift search path and sensitivity to sub-DLA absorption systems. Our results support an overall relative decrease in the neutral gas density over the past $\sim$11~Gyr between the two redshift bins we use to constrain $\Omega_{\rm HI}$. This demonstrates that it is possible for a redshifted blind 21~cm absorption survey to complement and connect high redshift Ly$\alpha$ studies to the 21~cm emission studies that anchor $\Omega_{\rm HI}$ at $z\sim0$ and will be insightful in forthcoming SKA pathfinder surveys.

\section*{Acknowledgements}
We are grateful for the valuable comments on this work by an anonymous referee that improved the scientific outcome and quality of the paper. 
The authors thank the staff members at the Green Bank Telescope for their assistance and support.
This research has made use of the NASA/IPAC Extragalactic Database (NED) which is operated by the Jet Propulsion Laboratory, California Institute of Technology, under contract with NASA. 
This research has made use of NASA’s Astrophysics Data System Bibliographic Services.
KG gratefully acknowledges the support of Lisa Kewley’s ARC Laureate Fellowship (FL150100113).
Parts of this research were supported by the Australian Research Council Centre of Excellence for All Sky Astrophysics in 3 Dimensions (ASTRO 3D), through project number CE170100013. 
This research made use of Astropy, a community-developed core Python package for Astronomy \citep{astropy13, astropy18}. The authors thank the invaluable labor of the maintenance and clerical staff at their institutions, whose contributions make scientific discoveries a reality. 

Facility: {GBT.}\\
Software: {ASTROPY \citep{astropy13, astropy18}, GBTIDL \citep{marganian06}.}\\
Data availability: The data underlying this article are available in National Radio Astronomy Observatory Archive at https://archive.nrao.edu/archive/advquery.jsp, and can be accessed with project code 02C-008, 04C-018.

\begin{landscape}
 \begin{table}
 \tiny
  \caption{Journal of Observations. Columns list the 
source name, 
right ascension, 
declination (J2000 coordinates), 
optical redshift of the background radio object, total integration time spent on the reduced spectrum for each receiver bandpass, the average measured spectral noise level for each receiver, the flux and column density limits $N_{\rm HI} / (T_s/f)$ measured as closely to 1420, 1000, 800, and 450~MHz (exact frequency determined by RFI conditions) and the  total redshift searched per source in all the receivers. 
The L-band covers the range of $0<z<0.235$ (1150-1420~MHz), PF2 covers the range of $0.150<z<0.570$ (905-1235~MHz), PF1-800 covers $0.536<z<1.10$ (675-925~MHz), PF1-600 covers $1.06<z<1.79$ (510-690~MHz), and PF1-450 covers $1.58<z<2.74$ (380-550~MHz). \newline
$a$ -- Only two sources were observed with the GBT's PF1-600 receiver, 0016+731 and SBS~0804+499 with 516 and 596 seconds, respectively. The entire spectral range for 0016+731 was affected by RFI and the average measured spectral noise level for SBS~0804+499 was 16~mJy and a column density limit of $N_{\rm} < 3.6\times10^{20}$~cm$^{-2}$ at 600~MHz. \newline
$b$ -- This source was observed in the 800~MHz band in both GBT programs. The observations from the 02C--008 program suffer less RFI and are used for this band throughout this paper. \newline
$c$ -- This source was observed in the 800~MHz band in both GBT programs. The observations from the 04C--018 program suffer less RFI and are used for this band throughout this paper. \newline
This table is available in its entirety in machine-readable form.
}
  \label{tab:obs}
  \hskip-1.5cm\begin{tabular}{lllccccccccccccccccccccc}
    \hline
& & & & \multicolumn{4}{c}{1.4 GHz} & & \multicolumn{4}{c}{1.0 GHz} & & \multicolumn{4}{c}{800 MHz} & & \multicolumn{4}{c}{450 MHz} & \\
\cline{5-8} \cline{10-13} \cline{15-18} \cline{20-23} \noalign{\smallskip}
& & & & \multicolumn{4}{c}{$z=0-0.235$} & & \multicolumn{4}{c}{$0.150-0.570$} & & \multicolumn{4}{c}{$0.536-1.10$} & & \multicolumn{4}{c}{$1.58-2.74$} & \vspace{8pt}\\
    Source & $\alpha$ & $\delta$ & $z$ & Time & $\overline{rms}$ & S$_{\rm 1.4 GHz}$ & $N_{\rm HI}$ & & Time & $\overline{rms}$ & S$_{\rm 1.0 GHz}$ & $N_{\rm HI}$ & & Time & $\overline{rms}$ & S$_{\rm 0.8 GHz}$ & $N_{\rm HI}$ & & Time & $\overline{rms}$ & S$_{\rm 0.45 GHz}$ & $N_{\rm HI}$ & $\Delta z_{\rm total}$ \\
     & (J2000) & (J2000) & & (sec) & (mJy) & (Jy) & (10$^{18}$ ($Ts/f$) cm$^{-2}$) & & (sec) & (mJy) & (Jy) & (10$^{18}$ ($Ts/f$) cm$^{-2}$) & & (sec) & (mJy) & (Jy) & (10$^{18}$ ($Ts/f$) cm$^{-2}$) & & (sec) & (mJy) & (Jy) & (10$^{18}$ ($Ts/f$) cm$^{-2}$) & \\
    \hline
0016+731$^a$ 	&	 00 19 45.79 	&	 +73 27 30.0 	&	1.781	&	0	&	 ...     	&	...	&	...	&	&	392	&	5.4	&	1.162	&	$<$0.65	&	&	268	&	12	&	1.058	&	$<$2.1	&	&	0	&	...		&	...	&	...	&	 0.31		\\
3C013              		&		00 34 14.5		&		 +39 24 17		&		1.351			&	0	&	...  	&	...	&	...	&	&	0	&	...  	&	...	&	...	&	&		1490		&	4.8	&	3.476	&	$<$0.22	&	&	0	&	...  	&	...	&	...	&	 0.39		\\
3C014              		&		00 36 06.5		&		 +18 37 59		&		1.4690(10)		&	0	&	...  	&	...	&	...	&	&	0	&	...  	&	...	&	...	&	&		1494		&	4.6	&	3.483	&	$<$0.21	&	&	0	&	...  	&	...	&	...	&	 0.4147		\\
B3 0035+413 		&	 00 38 24.84 	&	 +41 37 06.0 	&	 1.353(3) 		&		  329	&	5.1	&	0.692	&	$<$1.6	&	&	398	&	4.3	&	0.614	&	$<$1.2	&	&	278	&	6.6	&	0.589	&	$<$1.7	&	&	0	&	...		&	...	&	...	&	 0.40		\\
LBQS 0106+0119$^b$ 	&	 01 08 38.77 	&	 +01 35 00.3 	&	 2.099(5) 		&	 	  279	&	6.2	&	2.314	&	$<$0.59	&	&	0	&	...  	&	...	&	...	&	&		1464		&	4.8	&	2.715	&	$<$0.27	&	&	564	&	 9.6		&	2.829	&	$<$0.58	&	0.8304	\\
UM310 			&	 01 15 17.10 	&	 $-$01 27 04.6 	&	 1.365 		&		  260	&	4.7	&	1.052	&	$<$0.85	&	&	388	&	4.3	&	1.031	&	$<$0.66	&	&	448	&	8.2	&	1.016	&	$<$1.3	&	&	0	&	...		&	...	&	...	&	 0.61		\\
0113$-$118		&	 01 16 12.52 	&	 $-$11 36 15.4 	&	 0.67 		&		  258	&	5.5	&	1.787	&	$<$0.64	&	&	360	&	 RFI   	&	...	&	...	&	&	398	&	4.5	&	1.977	&	$<$0.37	&	&	0	&	...		&	...	&	...	&	 0.27		\\
3C036              		&		01 17 59.5		&		 +45 36 22		&		1.301			&	0	&	...  	&	...	&	...	&	&	0	&	...  	&	...  	&	...  	&	&		1487		&	5.4	&	2.342	&	$<$0.36	&	&	0	&	...  	&	...	&	...	&	 0.3654		\\
0119+041 		&	 01 21 56.86 	&	 +04 22 24.7 	&	 0.637 		&		  277	&	5.3	&	1.44	&	$<$0.72	&	&	382	&	 RFI   	&	...	&	...	&	&	0	&	...   	&	...	&	...	&	&	0	&	...		&	...	&	...	&	 0.22		\\
HB89 0119$-$046			&		01 22 27.9		&		$-$04 21 27		&		1.9250(10)		&	0	&	...  	&	...	&	...	&	&	0	&	...  	&	...	&	...	&	&		1482		&	5.1	&	1.99	&	$<$0.41	&	&	0	&	...  	&	...	&	...	&	 0.3934		\\
UM321$^b$		&	 01 25 28.84 	&	 $-$00 05 55.9 	&	 1.07481(15)	&		  259	&	4.4	&	1.543	&	$<$0.59	&	&	0	&	...  	&	...	&	...	&	&	1490	&	4.5	&	1.334	&	$<$0.54	&	&	0	&	...		&	...	&	...	&	0.6189	\\
0133+476 		&	 01 36 58.59 	&	 +47 51 29.1 	&	 0.859 		&		  398	&	15	&	2.294	&	$<$0.93	&	&	398	&	6.8	&	2.061	&	$<$0.52	&	&	293	&	6.5	&	1.883	&	$<$0.54	&	&	0	&	...		&	...	&	...	&	 0.71		\\
3C048 			&	 01 37 41.30 	&	 +33 09 35.1 	&	 0.367 		&		  557	&	6.7	&	17.4	&	$<$0.058	&	&	378	&	4.9	&	21.4	&	$<$0.037	&	&	0	&	...   	&	...	&	...	&	&	0	&	...		&	...	&	...	&	 0.18		\\
0138$-$097 		&	 01 41 25.83 	&	 $-$09 28 43.7 	&	 0.5 		&		  298	&	5.1	&	0.661	&	$<$1.5	&	&	0	&	...  	&	...	&	...	&	&	298	&	9.5	&	0.765	&	$<$1.9	&	&	0	&	...		&	...	&	...	&	 0.35		\\
0149+218 		&	 01 52 18.06 	&	 +22 07 07.7 	&	 1.32 		&		  687	&	5.7	&	1.151	&	$<$1.2	&	&	687	&	4.8	&	1.454	&	$<$0.58	&	&	283	&	6.5	&	1.528	&	$<$0.94	&	&	0	&	...		&	...	&	...	&	 0.41		\\
4C+15.05 		&	 02 04 50.41 	&	 +15 14 11.0 	&	 0.405 		&		  359	&	5.2	&	4.182	&	$<$0.17	&	&	0	&	...  	&	...	&	...	&	&	0	&	...   	&	...	&	...	&	&	0	&	...		&	...	&	...	&	 0.26		\\
0202+319 		&	 02 05 04.92 	&	 +32 12 30.1 	&	 1.466 		&		  398	&	9.8	&	0.656	&	$<$3.3	&	&	586	&	5.8	&	0.695	&	$<$1.3	&	&	283	&	RFI	&	...	&	...	&	&	0	&	...		&	...	&	...	&	 0.32		\\
0212+735 		&	 02 17 30.81 	&	 +73 49 32.6 	&	 2.367 		&		  0	&	...   	&	...	&	...	&	&	270	&	6	&	2.04	&	$<$0.46	&	&	283	&	RFI	&	...	&	...	&	&	528	&	 13		&	1.65	&	$<$1.2	&	 0.13		\\
3C066A 			&	 02 22 39.61 	&	 +43 02 07.8 	&	 0.444 		&		  380	&	8.7	&	2.226	&	$<$0.72	&	&	299	&	8.3	&	2.83	&	$<$0.58	&	&	0	&	...   	&	...	&	...	&	&	0	&	...		&	...	&	...	&	 0.16		\\
B2 0218+357     	&	 02 21 05.47 	&	 +35 56 13.7 	&	 0.68466(4)		&	0	&	...   	&	...	&	...	&	&	0	&	... 	&	...	&	...	&	&	299	&	7.1	&	1.453	&	$<$0.98	&	&	0	&	...		&	...	&	...	&	 0.025		\\
3C065				&		02 23 43.2		&		 +40 00 52		&		1.176			&	0	&	...  	&	...	&	...	&	&	0	&	...  	&	...	&	...	&	&		1498		&	1.8	&	5.445	&	$<$0.52	&	&	0	&	...  	&	...	&	...	&	 0.4015		\\
0221+067 		&	 02 24 28.43 	&	 +06 59 23.3 	&	 0.511 		&		  298	&	8	&	0.827	&	$<$2.3	&	&	296	&	5.9	&	0.961	&	$<$1.0	&	&	0	&	...   	&	...	&	...	&	&	0	&	...		&	...	&	...	&	 0.22		\\
4C+34.07			&		02 26 10.3		&		 +34 21 30		&		2.910(2)		&	0	&	...  	&	...	&	...	&	&	0	&	...  	&	...	&	...	&	&		1493		&	5.5	&	2.891	&	$<$0.30	&	&	0	&	...  	&	...	&	...	&	 0.3853		\\
0229+131$^b$	&	 02 31 45.89 	&	 +13 22 54.7 	&	 2.0590(10)	&		  0	&	...   	&	...   	&	...   	&	&	299	&	9.5	&	1.515	&	$<$0.93	&	&		1461		&	3.8	&	1.713	&	$<$0.35	&	&	0	&	...		&	...	&	...	&	0.5757	\\
3C068.1				&		02 32 28.9		&		 +34 23 47		&		1.238			&	0	&	...  	&	...	&	...	&	&	0	&	...  	&	...	&	...	&	&		1478		&	4.4	&	4.351	&	$<$0.1.6	&	&	0	&	...  	&	...	&	...	&	 0.3926		\\
3C068.2				&		02 34 23.8		&		 +31 34 17		&		1.575			&	0	&	...  	&	...	&	...	&	&	0	&	...  	&	...	&	...	&	&		1485		&	5	&	2.45	&	$<$0.32	&	&	0	&	...  	&	...	&	...	&	 0.3925		\\
0234+285 		&	 02 37 52.40 	&	 +28 48 09.0 	&	 1.213 		&		  268	&	10	&	1.282	&	$<$1.8	&	&	299	&	11	&	1.37	&	$<$1.9	&	&	258	&	11	&	1.45	&	$<$1.2	&	&	0	&	...		&	...	&	...	&	 0.32		\\
0235+164 		&	 02 38 38.93 	&	 +16 36 59.3 	&	 0.94 		&		  557	&	6.5	&	0.979	&	$<$1.1	&	&	298	&	7.6	&	1.028	&	$<$1.2	&	&	281	&	7.5	&	1.058	&	$<$1.0	&	&	0	&	...		&	...	&	...	&	 0.48		\\
0237$-$027 		&	 02 39 45.47 	&	 $-$02 34 40.9 	&	 1.116 		&	 	  298	&	8.8	&	0.315	&	$<$5.1	&	&	277	&	 RFI   	&	...	&	...	&	&	383	&	5.6	&	0.287	&	$<$2.9	&	&	0	&	...		&	...	&	...	&	 0.41		\\
0237$-$233 		&	 02 40 08.17 	&	 $-$23 09 15.7 	&	 2.223 		&		  259	&	5.1	&	6.94	&	$<$0.36	&	&	0	&	...  	&	...	&	...	&	&	398	&	6.2	&	5.14	&	$<$0.19	&	&	716	&	 RFI		&	...	&	...	&	 0.63		\\
PKS 0239+108 		&	 02 42 29.17 	&	 +11 01 00.7 	&	 2.68 		&		  398	&	8.5	&	1.531	&	$<$1.2	&	&	263	&	 RFI   	&	...	&	...	&	&	286	&	9.6	&	1.437	&	$<$0.95	&	&	322	&	 15		&	1.406	&	$<$1.0	&	 0.35		\\
0248+430 		&	 02 51 34.54 	&	 +43 15 15.8 	&	 1.31 		&		  0	&	...   	&	...	&	...	&	&	256	&	5.8	&	0.795	&	$<$1.1	&	&	272	&	6.6	&	0.784	&	$<$1.9	&	&	0	&	...		&	...	&	...	&	 0.29		\\
0306+102 		&	 03 09 03.62 	&	 +10 29 16.3 	&	 0.863 		&		  254	&	7.8	&	0.512	&	$<$2.0	&	&	273	&	8.4	&	0.502	&	$<$2.7	&	&	297	&	6.7	&	0.48	&	$<$2.6	&	&	0	&	...		&	...	&	...	&	 0.44		\\
HB89 0307+444			&		03 10 31.0		&		 +44 35 48		&		1.165			&	0	&	...  	&	...	&	...	&	&	0	&	...  	&	...	&	...	&	&		1490		&	4.4	&	2.5	&	$<$0.28	&	&	0	&	...  	&	...	&	...	&	 0.3738		\\
TXS 0311+430			&		03 14 43.6		&		 +43 14 05		&		2.87			&	0	&	...  	&	...	&	...	&	&	0	&	...  	&	...	&	...	&	&		1498		&	6.3	&	2.712	&	$<$0.37	&	&	0	&	...  	&	...	&	...	&	 0.3928		\\
HB89 0312$-$034			&		03 15 22.7		&		$-$03 16 46		&		1.072			&	0	&	...  	&	...	&	...	&	&	0	&	...  	&	...	&	...	&	&		1688		&	5.8	&	2.505	&	$<$0.37	&	&	0	&	...  	&	...	&	...	&	 0.4233		\\
NGC 1275        	&	 03 19 48.16 	&	 +41 30 42.1 	&	 0.01765(4)	&	278	&	4.9	&	23.3	&	$<$0.037	&	&	0	&	...  	&	...	&	...	&	&	0	&	...   	&	...	&	...	&	&	0	&	...		&	...	&	...	&	 0.033		\\
TXS 0319+131			&		03 21 53.1		&		 +12 21 14		&		2.662			&	0	&	...  	&	...	&	...	&	&	0	&	...  	&	...	&	...	&	&		1543		&	3.2	&	2.242	&	$<$0.22	&	&	0	&	...  	&	...	&	...	&	 0.3686		\\
PKS 0327$-$241 		&	 03 29 54.07 	&	 $-$23 57 08.8 	&	 0.895 		&		  267	&	5.2	&	0.682	&	$<$0.93	&	&	252	&	 RFI   	&	...	&	...	&	&	278	&	6.4	&	0.902	&	$<$1.3	&	&	0	&	...   		&	...	&	...	&	 0.44		\\
0333+321 		&	 03 36 30.11 	&	 +32 18 29.3 	&	 1.258 		&		  279	&	5.8	&	2.68	&	$<$0.61	&	&	299	&	9.2	&	2.8	&	$<$0.57	&	&	299	&	9.2	&	2.9	&	$<$0.46	&	&	0	&	...		&	...	&	...	&	 0.83		\\
0336$-$019 		&	 03 39 30.94 	&	 $-$01 46 35.8 	&	 0.852 		&		  597	&	7.7	&	2.098	&	$<$0.61	&	&	268	&	8.1	&	2.146	&	$<$0.64	&	&	290	&	6.8	&	2.264	&	$<$0.47	&	&	0	&	...		&	...	&	...	&	 0.35		\\
0346$-$279 		&	 03 48 38.14 	&	 $-$27 49 13.6 	&	 0.991 		&		  268	&	5.3	&	0.837	&	$<$1.1	&	&	298	&	4.8	&	0.957	&	$<$0.95	&	&	297	&	9.7	&	1.032	&	$<$1.6	&	&	0	&	...		&	...	&	...	&	 0.31		\\
0405$-$123 		&	 04 07 48.43 	&	 $-$12 11 36.7 	&	 0.5726(2)	&		  298	&	6.8	&	2.912	&	$<$0.63	&	&	286	&	20	&	3.99	&	$<$0.89	&	&	0	&	...   	&	...	&	...	&	&	0	&	...		&	...	&	...	&	 0.23		\\
3C108				&		04 12 43.6		&		 +23 05 05		&		1.215			&	0	&	...  	&	...	&	...	&	&	0	&	...  	&	...	&	...	&	&		1490		&	6.5	&	1.83(10)	&	$<$0.56	&	&	0	&	...  	&	...	&	...	&	 0.3005		\\
0414$-$189 		&	 04 16 36.54 	&	 $-$18 51 08.3 	&	 1.536(2)	&		  259	&	8	&	1.258	&	$<$1.1	&	&	299	&	5.2	&	1.109	&	$<$0.69	&	&	263	&	6.4	&	1.064	&	$<$1.5	&	&	0	&	...		&	...	&	...	&	 0.49		\\
0420$-$014 		&	 04 23 15.80 	&	 $-$01 20 33.1 	&	 0.91609(18)	&		  298	&	8.9	&	2.587	&	$<$0.48	&	&	278	&	8.6	&	2.097	&	$<$0.68	&	&	298	&	13	&	1.901	&	$<$1.6	&	&	0	&	...		&	...	&	...	&	 0.33		\\
0422+004 		&	 04 24 46.84 	&	 +00 36 06.3 	&	 0.31 		&		  398	&	9.2	&	0.494	&	$<$3.7	&	&	298	&	7.1	&	0.516	&	$<$2.4	&	&	0	&	...   	&	...	&	...	&	&	0	&	...		&	...	&	...	&	 0.11		\\
3C124				&		04 41 59.1		&		 +01 21 02		&		1.083			&	0	&	...  	&	...	&	...	&	&	0	&	...  	&	...	&	...	&	&		1494		&	4.7	&	2.058(8)	&	$<$0.36	&	&	0	&	...  	&	...	&	...	&	 0.4163		\\
0440$-$003 		&	 04 42 38.66 	&	 $-$00 17 43.4 	&	 0.844 		&		  255	&	5.8	&	3.342	&	$<$0.28	&	&	254	&	5.5	&	2.763	&	$<$0.31	&	&	274	&	6.1	&	2.372	&	$<$0.53	&	&	0	&	...		&	...	&	...	&	 0.38		\\
4C+02.16        		&		04 44 40.5		&		 +02 47 53		&		1.43			&	0	&	...  	&	...	&	...	&	&	0	&	...  	&	...	&	...	&	&		1494		&	5.5	&	1.979(5)	&	$<$0.44	&	&	0	&	...  	&	...	&	...	&	 0.4146		\\
PKS 0454$-$234 		&	 04 57 03.18 	&	 $-$23 24 52.0 	&	 1.003 		&		  268	&	8.5	&	1.654	&	$<$0.97	&	&	265	&	11	&	1.589	&	$<$0.85	&	&	298	&	12	&	1.564	&	$<$3.8	&	&	0	&	...		&	...	&	...	&	 0.41		\\
0458$-$020$^b$ 	&	 05 01 12.81 	&	 $-$01 59 14.3 	&	 2.286 		&		  298	&	5.1	&	2.274	&	$<$0.34	&	&	299	&	4	&	2.302	&	$<$0.30	&	&	1499	&	3.9	&	2.37(4)	&	$<$0.26	&	&	680	&	 33			&	2.436	&	$<$2.2	&	0.5969	\\
PKS 0500+019 		&	 05 03 21.20 	&	 +02 03 04.7 	&	 0.58457(2)	&		  398	&	7.3	&	2.335	&	$<$0.71	&	&	299	&	6.8	&	1.736	&	$<$0.62	&	&	0	&	...   	&	...	&	...	&	&	0	&	...		&	...	&	...	&	 0.33		\\
PKS 0511$-$220 		&	 05 13 49.11 	&	 $-$21 59 16.1 	&	 1.296 		&		  269	&	7.2	&	0.648	&	$<$2.6	&	&	273	&	5.8	&	0.626	&	$<$1.6	&	&	297	&	7.5	&	0.619	&	$<$1.9	&	&	0	&	...		&	...	&	...	&	 0.36		\\
3C138 			&	 05 21 09.88 	&	 +16 38 22.1 	&	 0.759 		&		  278	&	5.7	&	10.14	&	$<$0.078	&	&	279	&	 RFI   	&	...	&	...	&	&	291	&	8.9	&	12.86	&	$<$0.14	&	&	0	&	...		&	...	&	...	&	 0.35		\\
0528$-$250 		&	 05 30 07.96 	&	 $-$25 03 29.3 	&	 2.778(7) 		&		  298	&	5.6	&	1.19	&	$<$0.70	&	&	298	&	18	&	1.284	&	$<$3.3	&	&	280	&	7.9	&	1.139	&	$<$1.2	&	&	687	&	 26		&	1.454	&	$<$4.2	&	 0.82		\\
PKS 0528+134 		&	 05 30 56.42 	&	 +13 31 55.1 	&	 2.06 		&		  361	&	4.8	&	1.829	&	$<$0.23	&	&	277	&	6.7	&	1.579	&	$<$0.55	&	&	298	&	8.8	&	1.414	&	$<$1.0	&	&	0	&	...		&	...	&	...	&	 0.50		\\
TXS 0529+483 		&	 05 33 15.86 	&	 +48 22 52.8 	&	 1.162		&	 	  0	&	...   	&	...	&	...	&	&	298	&	7.6	&	0.514	&	$<$1.9	&	&	298	&	8.8	&	0.549	&	$<$2.3	&	&	0	&	...		&	...	&	...	&	 0.14		\\
PKS 0539$-$057 		&	 05 41 38.08 	&	 $-$05 41 49.4 	&	 0.839(3)	&		  398	&	5.5	&	0.882	&	$<$1.0	&	&	250	&	 RFI   	&	...	&	...	&	&	298	&	7.6	&	0.915	&	$<$1.5	&	&	0	&	...		&	...	&	...	&	 0.36		\\
3C147 			&	 05 42 36.14 	&	 +49 51 07.2 	&	 0.545 		&		  361	&	7.3	&	21.44	&	$<$0.059	&	&	298	&	26	&	29.35	&	$<$0.16	&	&	0	&	...   	&	...	&	...	&	&	0	&	...		&	...	&	...	&	 0.17		\\
0552+398 		&	 05 55 30.80 	&	 +39 48 49.2 	&	 2.365(5) 		&		  0	&	...   	&	...	&	...	&	&	299	&	8.3	&	0.997	&	$<$1.6	&	&	298	&	9.7	&	0.89	&	$<$2.1	&	&	716	&	 6.8		&	0.447	&	$<$3.0	&	 0.15		\\
0605$-$085 		&	 06 07 59.70 	&	 $-$08 34 50.0 	&	 0.872 		&		  398	&	6.6	&	2.942	&	$<$0.58	&	&	299	&	9.8	&	2.862	&	$<$0.62	&	&	516	&	7.8	&	2.828	&	$<$0.51	&	&	0	&	...		&	...	&	...	&	 0.53		\\
0607$-$157 		&	 06 09 40.95 	&	 $-$15 42 40.7 	&	 0.3226(2)	&		  299	&	9.5	&	2.261	&	$<$0.61	&	&	299	&	 RFI   	&	...	&	...	&	&	0	&	...   	&	...	&	...	&	&	0	&	...		&	...	&	...	&	 0.21	\\
    \hline
  \end{tabular}
 \end{table}
\end{landscape}

\begin{landscape}
 \begin{table}
 \tiny
  \contcaption{Journal of Observations. }
  \hskip-1.5cm\begin{tabular}{lllccccccccccccccccccccc}
    \hline
& & & & \multicolumn{4}{c}{1.4 GHz} & & \multicolumn{4}{c}{1.0 GHz} & & \multicolumn{4}{c}{800 MHz} & & \multicolumn{4}{c}{450 MHz} & \\
\cline{5-8} \cline{10-13} \cline{15-18} \cline{20-23} \noalign{\smallskip}
& & & & \multicolumn{4}{c}{$z=0-0.235$} & & \multicolumn{4}{c}{$0.150-0.570$} & & \multicolumn{4}{c}{$0.536-1.10$} & & \multicolumn{4}{c}{$1.58-2.74$} & \vspace{8pt}\\
    Source & $\alpha$ & $\delta$ & $z$ & Time & $\overline{rms}$ & S$_{\rm 1.4 GHz}$ & $N_{\rm HI}$ & & Time & $\overline{rms}$ & S$_{\rm 1.0 GHz}$ & $N_{\rm HI}$ & & Time & $\overline{rms}$ & S$_{\rm 0.8 GHz}$ & $N_{\rm HI}$ & & Time & $\overline{rms}$ & S$_{\rm 0.45 GHz}$ & $N_{\rm HI}$ & $\Delta z_{\rm total}$ \\
     & (J2000) & (J2000) & & (sec) & (mJy) & (Jy) & (10$^{18}$ ($Ts/f$) cm$^{-2}$) & & (sec) & (mJy) & (Jy) & (10$^{18}$ ($Ts/f$) cm$^{-2}$) & & (sec) & (mJy) & (Jy) & (10$^{18}$ ($Ts/f$) cm$^{-2}$) & & (sec) & (mJy) & (Jy) & (10$^{18}$ ($Ts/f$) cm$^{-2}$) & \\
    \hline
0642+449 		&	 06 46 32.02 	&	 +44 51 16.6 	&	 3.3960(10)	&		  0	&	...   	&	...	&	...	&	&	286	&	4.7	&	0.52	&	$<$1.5	&	&	298	&	6.7	&	0.51	&	$<$1.6	&	&	693	&	 12			&	0.475	&	$<$4.9	&	 0.33		\\
4C+68.08           		&		07 13 14.0		&		 +68 52 09		&		1.141(2)		&	0	&	...  	&	...	&	...	&	&	0	&	...  	&	...	&	...	&	&		1498		&	4.7	&	2.788(9)	&	$<$0.27	&	&	0	&	...  	&	...	&	...	&	 0.4038		\\
PKS 0727$-$11 		&	 07 30 19.11 	&	 $-$11 41 12.6 	&	 1.591(3)	&		  299	&	6.1	&	2.765	&	$<$0.29	&	&	299	&	6.2	&	2.457	&	$<$0.33	&	&	365	&	6.4	&	2.359	&	$<$0.36	&	&	0	&	...		&	...	&	...	&	 0.49		\\
0735+178 		&	 07 38 07.39 	&	 +17 42 19.0 	&	 0.424 		&		  362	&	4.4	&	2.143	&	$<$0.31	&	&	298	&	7.3	&	2.16	&	$<$0.72	&	&	0	&	...   	&	...	&	...	&	&	0	&	...		&	...	&	...	&	 0.25		\\
0736+017 		&	 07 39 18.03 	&	 +01 37 04.6 	&	 0.189410(9)	&	 	  349	&	6.6	&	2.53	&	$<$0.54	&	&	0	&	...  	&	...	&	...	&	&	0	&	...   	&	...	&	...	&	&	0	&	...		&	...	&	...	&	 0.17		\\
HB89 0738+313	 	&	 07 41 10.70 	&	 +31 12 00.2 	&	0.6320(4)   	&		  0	&	...   	&	...	&	...	&	&	289	&	5.4	&	1.861	&	$<$0.37	&	&	0	&	...   	&	...	&	...	&	&	0	&	...		&	...	&	...	&	 0.18		\\
3C186				&		07 44 17.4		&		 +37 53 17		&		1.0670(19)		&	0	&	...  	&	...	&	...	&	&	0	&	...  	&	...	&	...	&	&		1477		&	5.9	&	2.698(2)	&	$<$0.35	&	&	0	&	...  	&	...	&	...	&	 0.3938		\\
0743$-$006 		&	 07 45 54.08 	&	 $-$00 44 17.5 	&	 0.994 		&	 	  398	&	5.6	&	0.688	&	$<$1.7	&	&	597	&	4	&	0.643	&	$<$1.2	&	&	278	&	5.3	&	0.608	&	$<$2.8	&	&	0	&	...		&	...	&	...	&	 0.74		\\
PKS 0745+241 		&	 07 48 36.11 	&	 +24 00 24.1 	&	 0.4099(4)	&		  0	&	...   	&	...	&	...	&	&	299	&	4.9	&	1.039	&	$<$1.4	&	&	0	&	...   	&	...	&	...	&	&	0	&	... 		&	...	&	...	&	 0.055		\\
0746+483 		&	 07 50 20.43 	&	 +48 14 53.6 	&	 1.9562(6)	&		  0	&	...   	&	...	&	...	&	&	298	&	4.8	&	0.699	&	$<$1.1	&	&	298	&	6.5	&	0.686	&	$<$1.5	&	&	0	&	...		&	...	&	...	&	 0.40		\\
0748+126 		&	 07 50 52.04 	&	 +12 31 04.8 	&	 0.889 		&		  398	&	4.8	&	1.495	&	$<$0.38	&	&	283	&	4.3	&	1.49	&	$<$0.41	&	&	283	&	8.7	&	1.48	&	$<$0.75	&	&	0	&	...		&	...	&	...	&	 0.64		\\
0754+100 		&	 07 57 06.64 	&	 +09 56 34.9 	&	 0.2660(10)	&		  398	&	4.9	&	1.121	&	$<$0.91	&	&	299	&	6.7	&	1.03	&	$<$0.76	&	&	298	&	7.5	&	0.998	&	$<$1.2	&	&	0	&	...		&	...	&	...	&	 0.51		\\
FBQS J0804+3012			&		08 04 42.1		&		 +30 12 38		&		1.4513(16)		&	0	&	...  	&	...	&	...	&	&	0	&	...  	&	...	&	...	&	&		1487		&	5.8	&	2.09	&	$<$0.44	&	&	0	&	...  	&	...	&	...	&	 0.4174		\\
3C191				&		08 04 47.9		&		 +10 15 23		&		1.956			&	0	&	...  	&	...	&	...	&	&	0	&	...  	&	...	&	...	&	&		1482		&	6.3	&	3.459	&	$<$0.29	&	&	0	&	...  	&	...	&	...	&	 0.3977		\\
0805$-$077 		&	 08 08 15.53 	&	 $-$07 51 09.9 	&	 1.837 		&		  385	&	6.8	&	1.523	&	$<$0.48	&	&	281	&	5.5	&	1.726	&	$<$0.48	&	&	398	&	9.7	&	1.905	&	$<$0.83	&	&	0	&	...		&	...	&	...	&	 0.75		\\
SBS 0804+499$^a$ 		&	 08 08 39.66 	&	 +49 50 36.5 	&	 1.4352(5) 		&	0	&	...   	&	...	&	...	&	&	266	&	6	&	0.952	&	$<$0.74	&	&	277	&	7.4	&	0.847	&	$<$1.2	&	&	0	&	...  	&	...	&	...	&	0.48	\\
0808+019 		&	 08 11 26.71 	&	 +01 46 52.2 	&	 1.148 		&		  398	&	4.7	&	0.612	&	$<$1.0	&	&	299	&	5.7	&	0.589	&	$<$1.3	&	&	298	&	6.3	&	0.566	&	$<$4.2	&	&	0	&	...		&	...	&	...	&	 0.57		\\
FBQS J082455.4+391641		&		08 24 55.5		&		 +39 16 42		&		1.2157(11)		&	0	&	...  	&	...	&	...	&	&	0	&	...  	&	...	&	...	&	&		1475		&	5.9	&	1.951	&	$<$0.48	&	&	0	&	...  	&	...	&	...	&	 0.4211		\\
PKS 0823+033 		&	 08 25 50.34 	&	 +03 09 24.5 	&	 0.506 		&		  298	&	5.6	&	1.368	&	$<$0.68	&	&	262	&	6.2	&	1.293	&	$<$0.75	&	&	0	&	...   	&	...	&	...	&	&	0	&	...		&	...	&	...	&	 0.39		\\
PKS 0823$-$223 		&	 08 26 01.57 	&	 $-$22 30 27.2 	&	 0.91 		&		  994	&	5	&	0.519	&	$<$1.2	&	&	277	&	6.3	&	0.573	&	$<$2.2	&	&	298	&	8.5	&	0.63	&	$<$2.1	&	&	0	&	...		&	...	&	...	&	 0.50		\\
0828+493 		&	 08 32 23.22 	&	 +49 13 21.0 	&	 0.548 		&	 	  361	&	5.1	&	0.798	&	$<$1.1	&	&	256	&	7.7	&	0.939	&	$<$1.1	&	&	0	&	...   	&	...	&	...	&	&	0	&	...		&	...	&	...	&	 0.23		\\
0834$-$201 		&	 08 36 39.21 	&	 $-$20 16 59.5 	&	 2.752 		&		  298	&	7.5	&	2.052	&	$<$0.57	&	&	269	&	7.9	&	2.473	&	$<$0.92	&	&	298	&	6.4	&	2.672	&	$<$0.96	&	&	252	&	 9.9		&	3.457	&	$<$0.46	&	 0.50		\\
SBS 0833+585 		&	 08 37 22.41 	&	 +58 25 01.8 	&	 2.101 		&	 	  0	&	...   	&	...	&	...	&	&	298	&	5.4	&	0.723	&	$<$1.3	&	&	299	&	7.6	&	0.743	&	$<$1.8	&	&	716	&	 RFI		&	...	&	...	&	 0.36		\\
3C205              		&		08 39 06.4		&		 +57 54 17		&		1.534			&	0	&	...  	&	...	&	...	&	&	0	&	...  	&	...	&	...	&	&		1498		&	6.4	&	3.95	&	$<$0.26	&	&	0	&	...  	&	...	&	...	&	 0.3929		\\
0836+710 		&	 08 41 24.36 	&	 +70 53 42.2 	&	 2.1720(10)	&	 	  0	&	...   	&	...	&	...	&	&	298	&	6.9	&	4.248	&	$<$0.21	&	&	256	&	6.5	&	4.41	&	$<$0.22	&	&	520	&	 11			&	4.794	&	$<$0.39	&	 0.40		\\
OJ+287 			&	 08 54 48.87 	&	 +20 06 30.6 	&	 0.30560(10) 		&		  0	&	...   	&	...	&	...	&	&	270	&	 RFI   	&	...	&	...	&	&	0	&	...   	&	...	&	...	&	&	0	&	...		&	...	&	...	&	 RFI		\\
3C210              		&		08 58 09.9		&		 +27 50 52		&		1.169			&	0	&	...  	&	...	&	...	&	&	0	&	...  	&	...	&	...	&	&		1493		&	7.9	&	3.2	&	$<$0.39	&	&	0	&	...  	&	...	&	...	&	 0.3649		\\
0859$-$140 		&	 09 02 16.83 	&	 $-$14 15 30.9 	&	 1.3324(2) 		&		  398	&	5.3	&	3.343	&	$<$0.29	&	&	228	&	6.5	&	3.574	&	$<$0.26	&	&	398	&	9.8	&	3.743	&	$<$0.35	&	&	0	&	...		&	...	&	...	&	 0.39		\\
0859+470$^b$ 	&	 09 03 03.99 	&	 +46 51 04.1 	&	 1.4650(5) 		&		  0	&	...   	&	...	&	...	&	&	299	&	5.1	&	2.34	&	$<$0.39	&	&	1491	&	4.8	&	2.2	&	$<$0.34	&	&	0	&	...		&	...	&	...	&	0.6165	\\
0906+015 		&	 09 09 10.09 	&	 +01 21 35.6 	&	 1.0249(4) 		&	 	  0	&	...   	&	...	&	...	&	&	259	&	5.5	&	1.36	&	$<$0.64	&	&	298	&	7.1	&	1.32	&	$<$0.89	&	&	0	&	...		&	...	&	...	&	 0.34		\\
3C216 			&	 09 09 33.49 	&	 +42 53 46.1 	&	 0.6699(4)	&		  259	&	4.8	&	4.05	&	$<$0.14	&	&	248	&	6.6	&	5.4	&	$<$0.20	&	&	271	&	8.1	&	6.4	&	$<$0.18	&	&	0	&	...		&	...	&	...	&	 0.32		\\
0917+449 		&	 09 20 58.46 	&	 +44 41 54.0 	&	 2.1864(7) 	&	 	  0	&	...   	&	...	&	...	&	&	289	&	6	&	1.02	&	$<$0.96	&	&	398	&	8.3	&	0.99	&	$<$2.6	&	&	577	&	 RFI		&	...	&	...	&	 0.32		\\
0919$-$260 		&	 09 21 29.35 	&	 $-$26 18 43.4 	&	 2.3 		&		  299	&	4.3	&	1.296	&	$<$0.42	&	&	299	&	 RFI   	&	...	&	...	&	&	597	&	8.8	&	1.06	&	$<$1.3	&	&	597	&	 4.2		&	0.864	&	$<$0.91	&	 0.47		\\
TXS 0917+624		&	 09 21 36.23 	&	 +62 15 52.2 	&	 1.446(3)		&		  0	&	...   	&	...	&	...	&	&	299	&	5.4	&	1.1	&	$<$0.77	&	&	298	&	7	&	1.2	&	$<$1.0	&	&	0	&	...		&	...	&	...	&	 0.41		\\
0923+392 		&	 09 27 03.01 	&	 +39 02 20.9 	&	 0.6953(3) 	&		  893	&	4.8	&	2.728	&	$<$0.15	&	&	299	&	7.1	&	3.013	&	$<$0.39	&	&	298	&	9.4	&	3.131	&	$<$0.39	&	&	0	&	...		&	...	&	...	&	 0.37		\\
0925$-$203 		&	 09 27 51.82 	&	 $-$20 34 51.2 	&	 0.34741(15)	&		  526	&	5.1	&	0.778	&	$<$1.1	&	&	298	&	7.8	&	0.887	&	$<$1.3	&	&	0	&	...   	&	...	&	...	&	&	0	&	...		&	...	&	...	&	 0.19		\\
3C220.2				&		09 30 33.5		&		 +36 01 24		&		1.1577(14)		&	0	&	...  	&	...	&	...	&	&	0	&	...  	&	...	&	...	&	&		1496		&	5	&	2.363	&	$<$0.33	&	&	0	&	...  	&	...	&	...	&	 0.3452		\\
FBQS J094855.3+403944		&		09 48 55.3		&		 +40 39 45		&		1.2491(13)		&	0	&	...  	&	...	&	...	&	&	0	&	...  	&	...	&	...	&	&		1493		&	1.9	&	2.07	&	$<$0.14	&	&	0	&	...  	&	...	&	...	&	 0.39	        	\\
3C230              		&		09 51 58.8		&		 +00 01 27		&		1.487			&	0	&	...  	&	...	&	...	&	&	0	&	...  	&	...	&	...	&	&		1490		&	2.6	&	6.23	&	$<$0.066	&	&	0	&	...  	&	...	&	...	&	 0.3951		\\
0954+556 		&	 09 57 38.18 	&	 +55 22 57.8 	&	 0.8996(4)		&		  0	&	...   	&	...	&	...	&	&	291	&	5.9	&	4.102	&	$<$0.24	&	&	268	&	8.6	&	4.354	&	$<$0.32	&	&	0	&	...	&	...	&	...	&	 0.20		\\
0954+658 	&	 09 58 47.24 	&	 +65 33 54.8	&	0.368 		&		  0	&	...   	&	...	&	...	&	&	292	&	6.8	&	0.764	&	$<$1.4	&	&	0	&	...   	&	...	&	...	&	&	0	&	...	&	...	&	...	&	 0.12		\\
1004+141 		&	 10 07 41.50 	&	 +13 56 29.6 	&	 2.707(5) 		&		  297	&	 RFI    	&	...	&	...	&	&	282	&	8.2	&	1.027	&	$<$0.99	&	&	228	&	8.4	&	1.087	&	$<$1.2	&	&	519	&	 9.4		&	1.305	&	$<$1.5	&	 0.30		\\
3C239				&		10 11 45.4		&		 +46 28 20		&		1.781			&	0	&	...  	&	...	&	...	&	&	0	&	...  	&	...	&	...	&	&		1690		&	5.1	&	3.049	&	$<$0.26	&	&	0	&	...  	&	...	&	...	&	 0.3706		\\
PKS 1012+232 	&	 10 14 47.06 	&	 +23 01 16.6 	&	0.5664(4) 		&		  258	&	3.4	&	1.208	&	$<$0.48	&	&	256	&	6	&	1.366	&	$<$0.63	&	&	0	&	...   	&	...	&	...	&	&	0	&	...	&	...	&	...	&	 0.22		\\
3C241				&		10 21 54.5		&		 +21 59 30		&		1.617			&	0	&	...  	&	...	&	...	&	&	0	&	...  	&	...	&	...	&	&		1497		&	7.9	&	3.125	&	$<$0.40	&	&	0	&	...  	&	...	&	...	&	 0.4205		\\
1034$-$293 		&	 10 37 16.08 	&	 $-$29 34 02.8 	&	 0.312 		&		  596	&	6.8	&	1.108	&	$<$1.1	&	&	298	&	4.4	&	1.06	&	$<$0.62	&	&	0	&	...   	&	...	&	...	&	&	0	&	...		&	...	&	...	&	 0.15		\\
1039+811 		&	 10 44 23.06 	&	 +80 54 39.4 	&	 1.26 		&		  263	&	4.5	&	0.829	&	$<$0.91	&	&	285	&	5.2	&	0.78	&	$<$1.1	&	&	299	&	6.8	&	0.744	&	$<$1.5	&	&	0	&	...		&	...	&	...	&	 0.52		\\
1045$-$188 		&	 10 48 06.62 	&	 $-$19 09 35.7 	&	 0.595 		&		  596	&	4.7	&	1.154	&	$<$0.59	&	&	299	&	7	&	1.394	&	$<$0.76	&	&	0	&	...   	&	...	&	...	&	&	0	&	...		&	...	&	...	&	 0.35		\\
WMAP J1047+7143 	&	 10 48 27.62 	&	 +71 43 35.9 	&	 1.15		&		  298	&	6	&	0.743	&	$<$1.4	&	&	259	&	6.7	&	0.756	&	$<$1.3	&	&	298	&	5.8	&	0.761	&	$<$1.4	&	&	0	&	...		&	...	&	...	&	 0.36		\\
1049+215 		&	 10 51 48.79 	&	 +21 19 52.3 	&	 1.300(5) 		&		  297	&	4.4	&	1.237	&	$<$0.61	&	&	248	&	5.9	&	1.359	&	$<$0.69	&	&	299	&	6.3	&	1.466	&	$<$0.59	&	&	0	&	...		&	...	&	...	&	 0.47		\\
1055+018 		&	 10 58 29.60 	&	 +01 33 58.8 	&	 0.890(5) 		&	 	  546	&	5.9	&	3.315	&	$<$0.20	&	&	597	&	5.9	&	3.544	&	$<$0.26	&	&	243	&	13	&	3.759	&	$<$0.54	&	&	0	&	...		&	...	&	...	&	 0.33		\\
3C252				&		11 11 33.1		&		 +35 40 42		&		1.1			&	0	&	...  	&	...	&	...	&	&	0	&	...  	&	...	&	...	&	&		1490		&	4.9	&	2.67	&	$<$0.29	&	&	0	&	...  	&	...	&	...	&	 0.3992		\\
HB89 1109+437			&		11 12 39.4		&		 +43 25 47		&		1.6819(19)		&	0	&	...  	&	...	&	...	&	&	0	&	...  	&	...	&	...	&	&		1492		&	2.3	&	2.737	&	$<$0.13	&	&	0	&	...  	&	...	&	...	&	 0.3863		\\
1116+128$^b$ 	&	 11 18 57.30 	&	 +12 34 41.7 	&	 2.1257(4) 		&	 	  595	&	5	&	2.3	&	$<$0.40	&	&	291	&	8.3	&	2.7	&	$<$0.57	&	&	1386	&	2.9	&	3	&	$<$0.13	&	&	567	&	 RFI		&	...	&	...	&	0.8691	\\
3C255              		&		11 19 25.2		&		$-$03 02 52		&		1.355			&	0	&	...  	&	...	&	...	&	&	0	&	...  	&	...	&	...	&	&		1489		&	7.3	&	3.813	&	$<$0.30	&	&	0	&	...  	&	...	&	...	&	  0.4185		\\
3C256  				&		11 20 43.0		&		 +23 27 55		&		1.819			&	0	&	...  	&	...	&	...	&	&	0	&	...  	&	...	&	...	&	&		1490		&	5.7	&	2.7	&	$<$0.33	&	&	0	&	...  	&	...	&	...	&	 0.3945		\\
3C257              		&		11 23 09.2		&		 +05 30 19		&		2.474			&	0	&	...  	&	...	&	...	&	&	0	&	...  	&	...	&	...	&	&		1486		&	6.9	&	2.997	&	$<$0.37	&	&	0	&	...  	&	...	&	...	&	 0.3919		\\
4C+33.26          		&		11 26 23.7		&		 +33 45 27		&		1.23			&	0	&	...  	&	...	&	...	&	&	0	&	...  	&	...	&	...	&	&		1482		&	2	&	2.1	&	$<$0.15	&	&	0	&	...  	&	...	&	...	&	 0.3819		\\
PKS 1124$-$186 		&	 11 27 04.39 	&	 $-$18 57 17.4 	&	 1.05 		&		  526	&	5	&	0.538	&	$<$1.6	&	&	272	&	11	&	0.5	&	$<$3.0	&	&	249	&	8.4	&	0.475	&	$<$2.6	&	&	0	&	...		&	...	&	...	&	 0.31		\\
1127$-$145 		&	 11 30 07.05 	&	 $-$14 49 27.4 	&	 1.184 		&		  596	&	6.3	&	5.59	&	$<$0.17	&	&	245	&	9.9	&	5.538	&	$<$0.27	&	&	262	&	7.5	&	5.496	&	$<$0.20	&	&	0	&	...		&	...	&	...	&	 0.49		\\
B2 1128+38 		&	 11 30 53.28 	&	 +38 15 18.5 	&	 1.7405(5)		&		  556	&	5.2	&	0.703	&	$<$1.2	&	&	299	&	8	&	0.692	&	$<$1.2	&	&	299	&	8.3	&	0.679	&	$<$1.6	&	&	0	&	... 		&	...	&	...	&	 0.48		\\
3C266              		&		11 45 43.4		&		 +49 46 08		&		1.275			&	0	&	...  	&	...	&	...	&	&	0	&	...  	&	...	&	...	&	&		1488		&	5.1	&	2.465	&	$<$0.29	&	&	0	&	...  	&	...	&	...	&	 0.0589		\\
B3 1144+402 		&	 11 46 58.30 	&	 +39 58 34.3 	&	 1.0901(5)	&		  529	&	4.5	&	0.325	&	$<$2.3	&	&	299	&	6.9	&	0.431	&	$<$3.3	&	&	262	&	7.5	&	0.539	&	$<$1.9	&	&	0	&	...		&	...	&	...	&	 0.46		\\
3C267				&		11 49 56.5		&		 +12 47 19		&		1.14			&	0	&	...  	&	...	&	...	&	&	0	&	...  	&	...	&	...	&	&		1498		&	7.1	&	4.137	&	$<$0.27	&	&	0	&	...  	&	...	&	...	&	 0.3985		\\
LBQS 1148$-$0007$^b$ 	&	 11 50 43.87 	&	 $-$00 23 54.2 	&	 1.9801(5)	&		  596	&	7.2	&	3.28	&	$<$0.37	&	&	289	&	5.5	&	3.32	&	$<$0.25	&	&		1481		&	2	&	3.36	&	$<$0.093	&	&	0	&	...		&	...	&	...	&	0.6398	\\
1150+812 		&	 11 53 12.50 	&	 +80 58 29.2 	&	 1.25 		&		  298	&	6.1	&	1.34	&	$<$0.81	&	&	299	&	9.5	&	1.35	&	$<$1.2	&	&	268	&	7.9	&	1.36	&	$<$1.1	&	&	0	&	...		&	...	&	...	&	 0.48		\\
FBQS J1159+2914 	&	 11 59 31.83 	&	 +29 14 43.8 	&	 0.7247(4)	&		  596	&	4.7	&	1.85	&	$<$0.39	&	&	263	&	7.5	&	2.086	&	$<$0.57	&	&	298	&	9.1	&	2.25	&	$<$0.59	&	&	0	&	...		&	...	&	...	&	 0.38		\\
1202$-$262 		&	 12 05 33.21 	&	 $-$26 34 04.5 	&	 0.789 		&		  556	&	10	&	1.455	&	$<$1.5	&	&	261	&	8.4	&	1.856	&	$<$0.79	&	&	298	&	8.4	&	2.209	&	$<$0.36	&	&	0	&	...		&	...	&	...	&	 0.26		\\
PKS B1206$-$238 	&	 12 09 02.44 	&	 $-$24 06 20.8 	&	 1.299 		&		  544	&	6.3	&	0.423	&	$<$1.7	&	&	253	&	9.1	&	0.361	&	$<$5.3	&	&	299	&	9	&	0.328	&	$<$6.2	&	&	0	&	...		&	...	&	...	&	 0.42		\\
3C268.4            		&		12 09 13.5		&		 +43 39 18		&		1.3978(14)		&	0	&	...  	&	...	&	...	&	&	0	&	...  	&	...	&	...	&	&		1498		&	2.3	&	3.52	&	$<$0.10	&	&	0	&	...  	&	...	&	...	&	 0.3795		\\
B2 1219+28		&	 12 21 31.69 	&	 +28 13 58.5 	&	 0.102 		&		  596	&	7.7	&	0.875	&	$<$1.7	&	&	0	&	...  	&	...	&	...	&	&	0	&	...   	&	...	&	...	&	&	0	&	...		&	...	&	...	&	 0.11		\\
1222+037 		&	 12 24 52.42 	&	 +03 30 50.3 	&	 0.9550(4)	&		  596	&	6.6	&	1.292	&	$<$0.53	&	&	299	&	7.8	&	1.046	&	$<$0.80	&	&	261	&	6.7	&	0.882	&	$<$1.2	&	&	0	&	...		&	...	&	...	&	 0.54		\\
PG 1222+216 		&	 12 24 54.46 	&	 +21 22 46.4 	&	 0.4320(10)	&		  546	&	5.9	&	1.715	&	$<$0.49	&	&	299	&	7.7	&	2.048	&	$<$0.59	&	&	0	&	...   	&	...	&	...	&	&	0	&	...		&	...	&	...	&	 0.30		\\
3C273 			&	 12 29 06.70 	&	 +02 03 08.6 	&	 0.15834(7)	&		  595	&	7.7	&	45.1	&	$<$0.028	&	&	0	&	...  	&	...	&	...	&	&	0	&	...   	&	...	&	...	&	&	0	&	...		&	...	&	...	&	 0.18		\\
1243$-$072 		&	 12 46 04.23 	&	 $-$07 30 46.6 	&	 1.286 		&		  298	&	7.3	&	0.55	&	$<$1.3	&	&	266	&	7.5	&	0.637	&	$<$1.8	&	&	299	&	9.3	&	0.724	&	$<$2.1	&	&	0	&	...		&	...	&	...	&	 0.57		\\
1244$-$255 		&	 12 46 46.80 	&	 $-$25 47 49.3 	&	 0.633 		&		  447	&	6.8	&	1.17	&	$<$0.89	&	&	261	&	7.1	&	1.185	&	$<$0.92	&	&	0	&	...   	&	...	&	...	&	&	0	&	...		&	...	&	...	&	 0.23		\\
PKS 1245$-$19 		&	 12 48 23.90 	&	 $-$19 59 18.6 	&	 1.275 		&	893	&	6.7	&	5.324	&	$<$0.15	&	&	299	&	7.5	&	6.052	&	$<$0.20	&	&	277	&	9.3	&	6.793	&	$<$0.22	&	&	0	&	...		&	...	&	...	&	 0.52		\\
\hline
  \end{tabular}
 \end{table}
\end{landscape}

\begin{landscape}
 \begin{table}
 \tiny
  \contcaption{Journal of Observations. }
  \hskip-1.5cm\begin{tabular}{lllccccccccccccccccccccc}
    \hline
& & & & \multicolumn{4}{c}{1.4 GHz} & & \multicolumn{4}{c}{1.0 GHz} & & \multicolumn{4}{c}{800 MHz} & & \multicolumn{4}{c}{450 MHz} & \\
\cline{5-8} \cline{10-13} \cline{15-18} \cline{20-23} \noalign{\smallskip}
& & & & \multicolumn{4}{c}{$z=0-0.235$} & & \multicolumn{4}{c}{$0.150-0.570$} & & \multicolumn{4}{c}{$0.536-1.10$} & & \multicolumn{4}{c}{$1.58-2.74$} & \vspace{8pt}\\
    Source & $\alpha$ & $\delta$ & $z$ & Time & $\overline{rms}$ & S$_{\rm 1.4 GHz}$ & $N_{\rm HI}$ & & Time & $\overline{rms}$ & S$_{\rm 1.0 GHz}$ & $N_{\rm HI}$ & & Time & $\overline{rms}$ & S$_{\rm 0.8 GHz}$ & $N_{\rm HI}$ & & Time & $\overline{rms}$ & S$_{\rm 0.45 GHz}$ & $N_{\rm HI}$ & $\Delta z_{\rm total}$ \\
     & (J2000) & (J2000) & & (sec) & (mJy) & (Jy) & (10$^{18}$ ($Ts/f$) cm$^{-2}$) & & (sec) & (mJy) & (Jy) & (10$^{18}$ ($Ts/f$) cm$^{-2}$) & & (sec) & (mJy) & (Jy) & (10$^{18}$ ($Ts/f$) cm$^{-2}$) & & (sec) & (mJy) & (Jy) & (10$^{18}$ ($Ts/f$) cm$^{-2}$) & \\
    \hline
3C279 			&	 12 56 11.16 	&	 $-$05 47 21.5 	&	 0.5362(14) 	&	298	&	13	&	9.71	&	$<$0.23	&	&	299	&	5.2	&	10.7	&	$<$0.097	&	&	0	&	...   	&	...	&	...	&	&	0	&	...		&	...	&	...	&	 0.27		\\
PG 1302$-$102 		&	 13 05 33.01 	&	 $-$10 33 19.4 	&	 0.2784(4)	&	298	&	7.4	&	0.713	&	$<$1.3	&	&	0	&	...  	&	...	&	...	&	&	0	&	...   	&	...	&	...	&	&	0	&	...		&	...	&	...	&	 0.25		\\
4C+09.45         		&		13 05 36.0		&		 +08 55 14		&		1.409(2)		&	0	&	...  	&	...	&	...	&	&	0	&	...  	&	...	&	...	&	&		1491		&	7.2	&	2.804	&	$<$0.40	&	&	0	&	...  	&	...	&	...	&	 0.3416		\\
1308+326 		&	 13 10 28.66 	&	 +32 20 43.8 	&	 0.9980(5) 		&		  565	&	4.7	&	1.158	&	$<$0.84	&	&	0	&	...  	&	...	&	...	&	&	228	&	12	&	1.168	&	$<$1.3	&	&	0	&	...		&	...	&	...	&	 0.31		\\
MG J132701+2210 	&	 13 27 00.86 	&	 +22 10 50.2 	&	 1.4 		&		  596	&	4.4	&	0.852	&	$<$1.0	&	&	0	&	...  	&	...	&	...	&	&	266	&	10	&	0.78	&	$<$1.7	&	&	0	&	...		&	...	&	...	&	 0.32		\\
3C286           	&	 13 31 08.29 	&	 +30 30 33.0 	&	 0.8499(4)	&	557	&	8.8	&	14.7	&	$<$0.10	&	&	0	&	...  	&	...	&	...	&	&	788	&	9.2	&	19.3	&	$<$0.068	&	&	0	&	...		&	...	&	...	&	 0.21		\\
1334$-$127 		&	 13 37 39.78 	&	 $-$12 57 24.7 	&	 0.539 		&	247	&	8.7	&	2.676	&	$<$0.44	&	&	0	&	...  	&	...	&	...	&	&	0	&	...   	&	...	&	...	&	&	0	&	...		&	...	&	...	&	 0.19		\\
4C+24.28          		&		13 48 14.8		&		 +24 15 52		&		2.879(6)		&	0	&	...  	&	...	&	...	&	&	0	&	...  	&	...	&	...	&	&		1496		&	5.8	&	2.21	&	$<$0.42	&	&	0	&	...  	&	...	&	...	&	 0.3994		\\
1354+195 		&	 13 57 04.43 	&	 +19 19 07.4 	&	 0.72 		&		  892	&	4.3	&	2.226	&	$<$0.36	&	&	0	&	...  	&	...	&	...	&	&	298	&	7.1	&	3.133	&	$<$0.38	&	&	0	&	...		&	...	&	...	&	 0.23		\\
1354$-$152 		&	 13 57 11.24 	&	 $-$15 27 28.8 	&	 1.89 		&		  287	&	5.6	&	0.689	&	$<$1.1	&	&	272	&	 RFI   	&	...	&	...	&	&	275	&	5.6	&	0.635	&	$<$1.7	&	&	0	&	...		&	...	&	...	&	 0.31		\\
3C294              		&		14 06 44.0		&		 +34 11 25		&		1.779			&	0	&	...  	&	...	&	...	&	&	0	&	...  	&	...	&	...	&	&		1462		&	6.3	&	2.441	&	$<$0.41	&	&	0	&	...  	&	...	&	...	&	 0.4197		\\
4C$-$05.60          		&		14 13 48.5		&		$-$05 59 56		&		1.094(2)		&	0	&	...  	&	...	&	...	&	&	0	&	...  	&	...	&	...	&	&		1478		&	2.3	&	2.55	&	$<$0.14	&	&	0	&	...  	&	...	&	...	&	 0.3969		\\
1413+135 		&	 14 15 58.82 	&	 +13 20 23.7 	&	 0.24671(10) 	&		  596	&	4.2	&	1.085	&	$<$0.58	&	&	0	&	...  	&	...	&	...	&	&	0	&	...   	&	...	&	...	&	&	0	&	...		&	...	&	...	&	 0.25		\\
3C297              		&		14 17 24.0		&		$-$04 00 48		&		1.406			&	0	&	...  	&	...	&	...	&	&	0	&	...  	&	...	&	...	&	&		1488		&	5.5	&	2.575	&	$<$0.34	&	&	0	&	...  	&	...	&	...	&	 0.3903		\\
3C298             		&		14 19 08.2		&		 +06 28 35		&		1.4373(16)		&	0	&	...  	&	...	&	...	&	&	0	&	...  	&	...	&	...	&	&		1498		&	2.2	&	12.45	&	$<$0.028	&	&	0	&	...  	&	...	&	...	&	 0.3951		\\
1418+546 		&	 14 19 46.60 	&	 +54 23 14.8 	&	 0.1526(3)		&		  298	&	5.1	&	0.919	&	$<$0.82	&	&	0	&	...  	&	...	&	...	&	&	0	&	...   	&	...   	&	...   	&	&	0	&	...		&	...	&	...	&	 0.17		\\
3C300.1            		&		14 28 31.3		&		$-$01 24 08		&		1.159			&	0	&	...  	&	...	&	...	&	&	0	&	...  	&	...	&	...	&	&		1490		&	2	&	5.32	&	$<$0.059	&	&	0	&	...  	&	...	&	...	&	 0.3407		\\
HB89 1437+624     		&		14 38 44.9		&		 +62 11 55		&		1.0935(17)		&	0	&	...  	&	...	&	...	&	&	0	&	...  	&	...	&	...	&	&		1498		&	8.9	&	2.61	&	$<$0.54	&	&	0	&	...  	&	...	&	...	&	 0.2824		\\
3C305.1            		&		14 47 09.5		&		 +76 56 22		&		1.132			&	0	&	...  	&	...	&	...	&	&	0	&	...  	&	...	&	...	&	&		1489		&	2.7	&	2.51	&	$<$0.17	&	&	0	&	...  	&	...	&	...	&	 0.3849		\\
3C308              		&		14 54 08.3		&		 +50 03 31		&		2.8490(10)		&	0	&	...  	&	...	&	...	&	&	0	&	...  	&	...	&	...	&	&		1497		&	RFI	&	...	&	...	&	&	0	&	...  	&	...	&	...	&	 RFI			\\
4C+04.49 		&	 14 58 59.35 	&	 +04 16 13.8 	&	 0.3915(4) 	&		  297	&	6.3	&	1.064	&	$<$1.3	&	&	0	&	...  	&	...	&	...	&	&	0	&	...   	&	...	&	...	&	&	0	&	...		&	...	&	...	&	 0.17		\\
3C309.1 		&	 14 59 07.58 	&	 +71 40 19.9 	&	 0.905 		&		  298	&	6.5	&	8.72	&	$<$0.11	&	&	0	&	...  	&	...	&	...	&	&	285	&	13	&	12.2	&	$<$0.16	&	&	0	&	...		&	...	&	...	&	 0.31		\\
1502+106 		&	 15 04 24.98 	&	 +10 29 39 2 	&	 1.8383(15)	&		  272	&	5.3	&	1.484	&	$<$0.63	&	&	0	&	...  	&	...	&	...	&	&	287	&	17	&	1.448	&	$<$1.9	&	&	0	&	...		&	...	&	...	&	 0.37		\\
1504$-$166 		&	 15 07 04.79 	&	 $-$16 52 30.3 	&	 0.876(4)	&		  227	&	4.7	&	3.45	&	$<$0.17	&	&	0	&	...  	&	...	&	...	&	&	298	&	9.7	&	2.82	&	$<$0.55	&	&	0	&	...		&	...	&	...	&	 0.26		\\
HB89 1508$-$055   		&		15 10 53.6		&		$-$05 43 08		&		1.185			&	0	&	...  	&	...	&	...	&	&	0	&	...  	&	...	&	...	&	&		1491		&	2	&	5.53	&	$<$0.057	&	&	0	&	...  	&	...	&	...	&	 0.3909		\\
1510$-$089 		&	 15 12 50.53 	&	 $-$09 05 59.8 	&	 0.36 		&		  297	&	5.6	&	3.56	&	$<$0.22	&	&	0	&	...  	&	...	&	...	&	&	0	&	...   	&	...	&	...	&	&	0	&	...		&	...	&	...	&	 0.18		\\
1511$-$100 		&	 15 13 44.89 	&	 $-$10 12 00.3 	&	 1.513 		&		  282	&	5.8	&	0.624	&	$<$1.4	&	&	0	&	...  	&	...	&	...	&	&	277	&	9.9	&	0.761	&	$<$3.4	&	&	0	&	...		&	...	&	...	&	 0.29		\\
3C322   			&		15 35 01.2		&		 +55 36 53		&		1.681			&	0	&	...  	&	...	&	...	&	&	0	&	...  	&	...	&	...	&	&		1498		&	2.7	&	3.418	&	$<$0.12	&	&	0	&	...  	&	...	&	...	&	 0.3574		\\
1538+149 		&	 15 40 49.49 	&	 +14 47 45.9 	&	 0.605 		&		  297	&	6.7	&	1.43	&	$<$0.77	&	&	0	&	...  	&	...	&	...	&	&	0	&	...   	&	...	&	...	&	&	0	&	...		&	...	&	...	&	 0.22		\\
1546+027 		&	 15 49 29.44 	&	 +02 37 01.2 	&	 0.4144(4) 		&		  276	&	5.1	&	0.853	&	$<$0.84	&	&	0	&	...  	&	...	&	...	&	&	0	&	...   	&	...	&	...	&	&	0	&	...		&	...	&	...	&	 0.18		\\
1548+056 		&	 15 50 35.27 	&	 +05 27 10.5 	&	 1.422 		&		  285	&	7.8	&	2.85	&	$<$0.56	&	&	0	&	...  	&	...	&	...	&	&	290	&	7.5	&	2.75	&	$<$0.44	&	&	0	&	...		&	...	&	...	&	 0.39		\\
1555+001 		&	 15 57 51.43 	&	 $-$00 01 50.4 	&	 1.77 		&		  297	&	6.3	&	2.12	&	$<$0.49	&	&	0	&	...  	&	...	&	...	&	&	272	&	10	&	1.49	&	$<$1.3	&	&	0	&	...		&	...	&	...	&	 0.29		\\
TXS 1600+335			&		16 02 07.2		&		 +33 26 53		&		1.1			&	0	&	...  	&	...	&	...	&	&	0	&	...  	&	...	&	...	&	&		1562		&	3	&	2.941	&	$<$0.16	&	&	0	&	...  	&	...	&	...	&	 0.3939			\\
HB89 1602$-$001   		&		16 04 56.1		&		 +00 19 07		&		1.629(2)		&	0	&	...  	&	...	&	...	&	&	0	&	...  	&	...	&	...	&	&		1489		&	RFI	&	...	&	...	&	&	0	&	...  	&	...	&	...	&	 RFI			\\
4C+10.45 		&	 16 08 46.20 	&	 +10 29 07.8 	&	 1.2260(10)	&		  285	&	5.9	&	1.484	&	$<$0.44	&	&	0	&	...  	&	...	&	...	&	&	271	&	7.5	&	2.058	&	$<$0.65	&	&	0	&	...		&	...	&	...	&	 0.29		\\
1611+343 		&	 16 13 41.06 	&	 +34 12 47.9 	&	 1.3991(5)	&		  297	&	5.2	&	4.05	&	$<$0.16	&	&	0	&	...  	&	...	&	...	&	&	266	&	8.6	&	3.53	&	$<$0.35	&	&	0	&	...		&	...	&	...	&	 0.23		\\
HB89 1622+158     		&		16 25 14.4		&		 +15 45 22		&		1.406			&	0	&	...  	&	...	&	...	&	&	0	&	...  	&	...	&	...	&	&		1485		&	2.6	&	1.783	&	$<$0.23	&	&	0	&	...  	&	...	&	...	&	 0.3465		\\
PKS 1622$-$253 		&	 16 25 46.89 	&	 $-$25 27 38.3 	&	 0.786 		&		  297	&	5.7	&	2.521	&	$<$0.30	&	&	0	&	...  	&	...	&	...	&	&	277	&	9	&	2.447	&	$<$0.55	&	&	0	&	...		&	...	&	...	&	 0.26		\\
HB89 1624+416     		&		16 25 57.7		&		 +41 34 41		&		2.55			&	0	&	...  	&	...	&	...	&	&	0	&	...  	&	...	&	...	&	&		1491		&	2.7	&	1.84	&	$<$0.23	&	&	0	&	...  	&	...	&	...	&	 0.3918		\\
PKS 1622$-$29 		&	 16 26 06.02 	&	 $-$29 51 27.0 	&	 0.815 		&		  288	&	8.2	&	2.278	&	$<$0.67	&	&	0	&	...  	&	...	&	...	&	&	266	&	9.3	&	2.83	&	$<$0.50	&	&	0	&	...		&	...	&	...	&	 0.18		\\
HB89 1629+120			&		16 31 45.2		&		 +11 56 03		&		1.795			&	0	&	...  	&	...	&	...	&	&	0	&	...  	&	...	&	...	&	&		1390		&	4	&	2.53	&	$<$0.25	&	&	0	&	...  	&	...	&	...	&	 0.1978		\\
1633+382 		&	 16 35 15.49 	&	 +38 08 04.5 	&	 1.8131(5) 		&		  287	&	4.5	&	2.332	&	$<$0.24	&	&	0	&	...  	&	...	&	...	&	&	286	&	9.9	&	2.349	&	$<$0.64	&	&	0	&	...		&	...	&	...	&	 0.14		\\
1637+574 		&	 16 38 13.45 	&	 +57 20 24.0 	&	 0.7506(10)	&		  296	&	4.7	&	0.792	&	$<$0.96	&	&	0	&	...  	&	...	&	...	&	&	298	&	7.3	&	0.925	&	$<$0.99	&	&	0	&	...		&	...	&	...	&	 0.34		\\
1638+398 		&	 16 40 29.63 	&	 +39 46 46.0 	&	 1.66 		&		  292	&	5	&	1.111	&	$<$0.82	&	&	0	&	...  	&	...	&	...	&	&	298	&	7	&	0.954	&	$<$1.1	&	&	0	&	...		&	...	&	...	&	 0.36		\\
1642+690 		&	 16 44 07.85 	&	 +68 56 39.8 	&	 0.751 		&		  295	&	6.3	&	1.433	&	$<$0.60	&	&	0	&	...  	&	...	&	...	&	&	297	&	9.6	&	1.675	&	$<$0.72	&	&	0	&	...		&	...	&	...	&	 0.29		\\
3C345 			&	 16 42 58.81 	&	 +39 48 37.0 	&	 0.5928(4) 	&		  297	&	5	&	7.3	&	$<$0.11	&	&	0	&	...  	&	...	&	...	&	&	0	&	...   	&	...	&	...	&	&	0	&	...		&	...	&	...	&	 0.11		\\
1656+477 		&	 16 58 02.78 	&	 +47 37 49.2 	&	 1.622 		&		  297	&	4.7	&	0.858	&	$<$0.82	&	&	0	&	...  	&	...	&	...	&	&	299	&	6.5	&	0.884	&	$<$1.2	&	&	0	&	...		&	...	&	...	&	 0.48		\\
1655+077 		&	 16 58 09.01 	&	 +07 41 27.5 	&	 0.621 		&		  284	&	6.1	&	1.688	&	$<$0.42	&	&	0	&	...  	&	...	&	...	&	&	0	&	...   	&	...	&	...	&	&	0	&	...		&	...	&	...	&	 0.23		\\
1656+053 		&	 16 58 33.45 	&	 +05 15 16.4 	&	 0.879 		&		  255	&	5.4	&	1.525	&	$<$0.45	&	&	0	&	...  	&	...	&	...	&	&	282	&	9.4	&	1.926	&	$<$0.77	&	&	0	&	...		&	...	&	...	&	 0.26		\\
HB89 1702+298			&		17 04 07.2		&		 +29 46 59		&		1.927			&	0	&	...  	&	...	&	...	&	&	0	&	...  	&	...	&	...	&	&		1498		&	3.2	&	2.17	&	$<$0.23	&	&	0	&	...  	&	...	&	...	&	 0.2828		\\
3C356				&		17 24 19.0		&		 +50 57 40		&		1.079			&	0	&	...  	&	...	&	...	&	&	0	&	...  	&	...	&	...	&	&		1587		&	7.9	&	2.842	&	$<$0.44	&	&	0	&	...  	&	...	&	...	&	 0.2849		\\
HB89 1729+501			&		17 31 03.6		&		 +50 07 34		&		1.107			&	0	&	...  	&	...	&	...	&	&	0	&	...  	&	...	&	...	&	&		1590		&	7.4	&	1.82	&	$<$0.64	&	&	0	&	...  	&	...	&	...	&	 0.3721		\\
1730$-$130 		&	 17 33 02.70 	&	 $-$13 04 49.5 	&	0.902 		&		  289	&	4.8	&	5.97	&	$<$0.15	&	&	0	&	...  	&	...	&	...	&	&	284	&	9.8	&	6.66	&	$<$0.33	&	&	0	&	...		&	...	&	...	&	 0.33		\\
HB89 1732+160			&		17 34 42.6		&		 +16 00 31		&		1.296			&	0	&	...  	&	...	&	...	&	&	0	&	...  	&	...	&	...	&	&		1599		&	4.7	&	2.63	&	$<$0.27	&	&	0	&	...  	&	...	&	...	&	 0.2656		\\
1739+522 		&	 17 40 36.98 	&	 +52 11 43.4 	&	 1.375 		&		  596	&	7.1	&	0.836	&	$<$1.5	&	&	0	&	...  	&	...	&	...	&	&	298	&	7.9	&	0.989	&	$<$1.2	&	&	0	&	...		&	...	&	...	&	 0.37		\\
1741$-$038 		&	 17 43 58.85 	&	 $-$03 50 04.6 	&	 1.054 		&		  297	&	5.2	&	1.41	&	$<$0.47	&	&	0	&	...  	&	...	&	...	&	&	271	&	14	&	1.4	&	$<$1.6	&	&	0	&	...		&	...	&	...	&	 0.28		\\
1743+173 		&	 17 45 35.21 	&	 +17 20 01.4 	&	 1.702 		&		  0	&	...   	&	...	&	...	&	&	0	&	...  	&	...	&	...	&	&	263	&	 RFI    	&	...	&	...	&	&	0	&	...		&	...	&	...	&	 RFI		\\
3C362				&		17 47 07.0		&		 +18 21 10		&		2.281			&	0	&	...  	&	...	&	...	&	&	0	&	...  	&	...	&	...	&	&		1590		&	7.2	&	1.804	&	$<$0.63	&	&	0	&	...  	&	...	&	...	&	 0.369			\\
1749+701 		&	 17 48 32.84 	&	 +70 05 50.8 	&	 0.77 		&		  269	&	5	&	1.209	&	$<$0.55	&	&	0	&	...  	&	...	&	...	&	&	278	&	7.8	&	1.307	&	$<$0.78	&	&	0	&	...		&	...	&	...	&	 0.36		\\
1803+784 		&	 18 00 45.68 	&	 +78 28 04.0 	&	 0.68 		&		  298	&	4.8	&	2.221	&	$<$0.37	&	&	0	&	...  	&	...	&	...	&	&	298	&	6	&	2.105	&	$<$0.45	&	&	0	&	...		&	...	&	...	&	 0.26		\\
1800+440 		&	 18 01 32.31 	&	 +44 04 21.9 	&	 0.663 		&		  0	&	...   	&	...	&	...	&	&	0	&	...  	&	...	&	...	&	&	298	&	 RFI    	&	...	&	...	&	&	0	&	...		&	...	&	...	&	 RFI		\\
4C+13.66           		&		18 01 38.9		&		 +13 51 24		&		1.450(5)		&	0	&	...  	&	...	&	...	&	&	0	&	...  	&	...	&	...	&	&		1510		&	9.2	&	3.05	&	$<$0.47	&	&	0	&	...  	&	...	&	...	&	 0.2025		\\
3C368				&		18 05 06.3		&		 +11 01 33		&		1.131(2)		&	0	&	...  	&	...	&	...	&	&	0	&	...  	&	...	&	...	&	&		1490		&	8.2	&	2.47	&	$<$0.51	&	&	0	&	...  	&	...	&	...	&	 0.1187		\\
1823+568 		&	 18 24 07.07 	&	 +56 51 01.5 	&	 0.6640(10)	&		  271	&	6	&	1.457	&	$<$0.80	&	&	384	&	6.2	&	1.613	&	$<$0.86	&	&	288	&	5.7	&	1.668	&	$<$0.15	&	&	0	&	...		&	...	&	...	&	 0.36		\\
PKS 1830$-$211  	&	 18 33 39.89 	&	 $-$21 03 39.8 	&	 2.507(2) 		&	0	&	...   	&	...	&	...	&	&	0	&	...  	&	...	&	...	&	&	458	&	13	&	9.9	&	$<$0.24	&	&	0	&	...		&	...	&	...	&	 0.16		\\
3C380 			&	 18 29 31.80 	&	 +48 44 46.6 	&	 0.692(2) 		&		  298	&	7.3	&	13.8	&	$<$0.089	&	&	384	&	5.3	&	19.3	&	$<$0.044	&	&	597	&	15	&	21.7	&	$<$0.098	&	&	0	&	...		&	...	&	...	&	 0.39		\\
TXS 1849+670 		&	 18 49 16.07 	&	 +67 05 41.7 	&	 0.6570(10)		&		  0	&	...   	&	...	&	...	&	&	384	&	 RFI   	&	...	&	...	&	&	283	&	6.2	&	0.624	&	$<$1.6	&	&	0	&	...		&	...	&	...	&	 0.15		\\
HB89 1857+566			&		18 58 26.7		&		 +56 45 57		&		1.595			&	0	&	...  	&	...	&	...	&	&	0	&	...  	&	...	&	...	&	&		1498		&	9.9	&	1.82	&	$<$0.86	&	&	0	&	...  	&	...	&	...	&	 0.2469		\\
PKS B1908$-$201		&	 19 11 09.65 	&	 $-$20 06 55.1 	&	 1.119 		&		  0	&	...   	&	...	&	...	&	&	384	&	5.2	&	2.703	&	$<$0.31	&	&	298	&	8.5	&	2.701	&	$<$0.59	&	&	0	&	...		&	...	&	...	&	 0.32		\\
1921$-$293 		&	 19 24 51.05 	&	 $-$29 14 30.1 	&	 0.35263(18)	&		  0	&	...   	&	...	&	...	&	&	398	&	 RFI   	&	...	&	...	&	&	0	&	...   	&	...	&	...	&	&	0	&	...		&	...	&	...	&	 RFI		\\
1928+738 		&	 19 27 48.49 	&	 +73 58 01.6 	&	 0.3021(3) 	&		  260	&	7.2	&	3.925	&	$<$0.31	&	&	398	&	 RFI   	&	...	&	...	&	&	0	&	...   	&	...	&	...	&	&	0	&	...		&	...	&	...	&	 0.23		\\
1936$-$155 		&	 19 39 26.66 	&	 $-$15 25 34.1 	&	 1.657 		&		  0	&	...   	&	...	&	...	&	&	384	&	5.7	&	0.624	&	$<$1.4	&	&	285	&	6.3	&	0.636	&	$<$2.0	&	&	0	&	...		&	...	&	...	&	 0.35		\\
1954+513$^b$ 	&	 19 55 42.74 	&	 +51 31 48.5 	&	 1.22 		&		  298	&	7	&	1.79	&	$<$0.62	&	&	0	&	...  	&	...	&	...	&	&	1490	&	9.5	&	3.24	&	$<$0.46	&	&	0	&	...		&	...	&	...	&	0.6637	\\
1958$-$179 		&	 20 00 57.09 	&	 $-$17 48 57.7 	&	 0.65 		&		  0	&	...   	&	...	&	...	&	&	384	&	 RFI   	&	...	&	...	&	&	0	&	...   	&	...	&	...	&	&	0	&	...		&	...	&	...	&	 RFI		\\
2007+777 		&	 20 05 30.93 	&	 +77 52 43.1 	&	 0.342 		&		  0	&	...   	&	...	&	...	&	&	394	&	 RFI   	&	...	&	...	&	&	0	&	...   	&	...	&	...	&	&	0	&	...		&	...	&	...	&	 RFI		\\
HB89 2003$-$025			&		20 06 08.5		&		$-$02 23 35		&		1.457			&	0	&	...  	&	...	&	...	&	&	0	&	...  	&	...	&	...	&	&		1495		&	6.7	&	2.89	&	$<$0.35	&	&	0	&	...  	&	...	&	...	&	 0.4081		\\
2005+403 		&	 20 07 44.94 	&	 +40 29 48.6 	&	 1.736 		&		  245	&	6.3	&	2.473	&	$<$0.41	&	&	377	&	5.9	&	2.26	&	$<$0.42	&	&	277	&	7.1	&	2.091	&	$<$0.74	&	&	0	&	...		&	...	&	...	&	 0.28	  \\
\hline
  \end{tabular}
 \end{table}
\end{landscape}

\begin{landscape}
 \begin{table}
 \tiny
  \contcaption{Journal of Observations. }
  \hskip-1.5cm\begin{tabular}{lllccccccccccccccccccccc}
    \hline
& & & & \multicolumn{4}{c}{1.4 GHz} & & \multicolumn{4}{c}{1.0 GHz} & & \multicolumn{4}{c}{800 MHz} & & \multicolumn{4}{c}{450 MHz} & \\
\cline{5-8} \cline{10-13} \cline{15-18} \cline{20-23} \noalign{\smallskip}
& & & & \multicolumn{4}{c}{$z=0-0.235$} & & \multicolumn{4}{c}{$0.150-0.570$} & & \multicolumn{4}{c}{$0.536-1.10$} & & \multicolumn{4}{c}{$1.58-2.74$} & \vspace{8pt}\\
    Source & $\alpha$ & $\delta$ & $z$ & Time & $\overline{rms}$ & S$_{\rm 1.4 GHz}$ & $N_{\rm HI}$ & & Time & $\overline{rms}$ & S$_{\rm 1.0 GHz}$ & $N_{\rm HI}$ & & Time & $\overline{rms}$ & S$_{\rm 0.8 GHz}$ & $N_{\rm HI}$ & & Time & $\overline{rms}$ & S$_{\rm 0.45 GHz}$ & $N_{\rm HI}$ & $\Delta z_{\rm total}$ \\
     & (J2000) & (J2000) & & (sec) & (mJy) & (Jy) & (10$^{18}$ ($Ts/f$) cm$^{-2}$) & & (sec) & (mJy) & (Jy) & (10$^{18}$ ($Ts/f$) cm$^{-2}$) & & (sec) & (mJy) & (Jy) & (10$^{18}$ ($Ts/f$) cm$^{-2}$) & & (sec) & (mJy) & (Jy) & (10$^{18}$ ($Ts/f$) cm$^{-2}$) & \\
    \hline
2008$-$159 		&	 20 11 15.71 	&	 $-$15 46 40.3 	&	 1.18 		&		  266	&	6.2	&	0.568	&	$<$1.4	&	&	384	&	5.4	&	0.569	&	$<$1.4	&	&	280	&	7	&	0.569	&	$<$2.4	&	&	0	&	...		&	...	&	...	&	 0.39		\\
COINS J2022+6136 	&	 20 22 09.68 	&	 +61 36 58.8 	&	 0.227 		&		  258	&	7	&	2.234	&	$<$0.44	&	&	0	&	...  	&	...	&	...	&	&	0	&	...   	&	...	&	...	&	&	0	&	...		&	...	&	...	&	 0.23		\\
PKS 2022+031 		&	 20 25 09.63 	&	 +03 16 44.5 	&	 2.21 		&		  0	&	...   	&	...	&	...	&	&	384	&	5.7	&	0.454	&	$<$2.1	&	&	281	&	7.2	&	0.481	&	$<$2.3	&	&	0	&	...		&	...	&	...	&	 0.29		\\
3C418 			&	 20 38 37.03 	&	 +51 19 12.7 	&	 1.686 		&		  298	&	6.8	&	6.042	&	$<$0.14	&	&	396	&	7.2	&	6.836	&	$<$0.21	&	&	398	&	15	&	7.48	&	$<$0.30	&	&	0	&	...		&	...	&	...	&	 0.34		\\
PKS 2055+054			&		20 58 28.8		&		 +05 42 51		&		1.381			&	0	&	...  	&	...	&	...	&	&	0	&	...  	&	...	&	...	&	&		1566		&	7.7	&	1.94	&	$<$0.63	&	&	0	&	...  	&	...	&	...	&	 0.4114		\\
2059+034 		&	 21 01 38.83 	&	 +03 41 31.3 	&	 1.013 		&		  0	&	...   	&	...	&	...	&	&	384	&	5.6	&	0.401	&	$<$2.3	&	&	242	&	7	&	0.326	&	$<$3.4	&	&	0	&	...		&	...	&	...	&	 0.26		\\
3C432				&		21 22 46.2		&		 +17 04 38		&		1.7850(10)		&	0	&	...  	&	...	&	...	&	&	0	&	...  	&	...	&	...	&	&		1488		&	7.5	&	2.919	&	$<$0.41	&	&	0	&	...  	&	...	&	...	&	 0.3232		\\
2121+053 		&	 21 23 44.52 	&	 +05 35 22.1 	&	 1.941 		&		  298	&	10	&	0.943	&	$<$1.8	&	&	365	&	5.5	&	0.851	&	$<$0.87	&	&	298	&	6.9	&	0.789	&	$<$1.4	&	&	0	&	...		&	...	&	...	&	 0.59		\\
2128$-$123 		&	 21 31 35.26 	&	 $-$12 07 04.8 	&	 0.501 		&		  297	&	5.2	&	1.98	&	$<$0.46	&	&	384	&	6.1	&	1.87	&	$<$0.52	&	&	0	&	...   	&	...	&	...	&	&	0	&	...		&	...	&	...	&	 0.39		\\
2131$-$021$^c$ 	&	 21 34 10.31 	&	 $-$01 53 17.2 	&	 1.285 		&		  0	&	...   	&	...	&	...	&	&	0	&	...  	&	...	&	...	&	&	252	&	5.6	&	1.94	&	$<$0.54	&	&	0	&	...		&	...	&	...	&	 0.25		\\
2134+004 		&	 21 36 38.58 	&	 +00 41 54.2 	&	 1.9446(4)	&		  298	&	11	&	3.293	&	$<$0.57	&	&	398	&	 RFI   	&	...	&	...	&	&	0	&	...   	&	...	&	...	&	&	0	&	...		&	...	&	...	&	 0.25		\\
2145+067 		&	 21 48 05.46 	&	 +06 57 38.6 	&	 0.99 		&		  298	&	8.2	&	3.145	&	$<$0.45	&	&	0	&	...  	&	...	&	...	&	&	299	&	14	&	3.355	&	$<$0.69	&	&	0	&	...		&	...	&	...	&	 0.23		\\
HB89 2149+212			&		21 51 45.9		&		 +21 30 14		&		1.5385(8)		&	0	&	...  	&	...	&	...	&	&	0	&	...  	&	...	&	...	&	&		1499		&	5.6	&	1.78	&	$<$0.50	&	&	0	&	...  	&	...	&	...	&	 0.313			\\
HB89 2150+053			&		21 53 24.7		&		 +05 36 19		&		1.9670(10)		&	0	&	...  	&	...	&	...	&	&	0	&	...  	&	...	&	...	&	&		1486		&	8.6	&	1.877	&	$<$0.72	&	&	0	&	...  	&	...	&	...	&	 0.3258		\\
2155$-$152 		&	 21 58 06.28 	&	 $-$15 01 09.3 	&	 0.672 		&		  0	&	...   	&	...	&	...	&	&	370	&	6.2	&	2.799	&	$<$0.33	&	&	298	&	5.9	&	2.681	&	$<$0.35	&	&	0	&	...		&	...	&	...	&	 0.22		\\
HB89 2156+297			&		21 58 41.9		&		 +29 59 08		&		1.753			&	0	&	...  	&	...	&	...	&	&	0	&	...  	&	...	&	...	&	&		1495		&	8.8	&	2.123	&	$<$0.66	&	&	0	&	...  	&	...	&	...	&	 0.4085		\\
2201+315 		&	 22 03 14.97 	&	 +31 45 38.3 	&	 0.29474(9) 	&		  297	&	6.5	&	1.869	&	$<$0.59	&	&	398	&	9.5	&	1.927	&	$<$0.72	&	&	0	&	...   	&	...	&	...	&	&	0	&	...		&	...	&	...	&	 0.22		\\
2203$-$188 		&	 22 06 10.42 	&	 $-$18 35 38.7 	&	 0.6185(14) 		&		  594	&	8	&	6.818	&	$<$0.22	&	&	398	&	5.4	&	7.476	&	$<$0.089	&	&	0	&	...   	&	...	&	...	&	&	0	&	...		&	...	&	...	&	 0.24		\\
HB89 2207+374			&		22 09 21.4		&		 +37 42 18		&		1.493(4)		&	0	&	...  	&	...	&	...	&	&	0	&	...  	&	...	&	...	&	&		1602		&	8.7	&	2.57	&	$<$0.54	&	&	0	&	...  	&	...	&	...	&	 0.3936		\\
2210$-$257 		&	 22 13 02.50 	&	 $-$25 29 30.1 	&	 1.833 		&		  258	&	6.6	&	1.201	&	$<$0.68	&	&	343	&	 RFI   	&	...	&	...	&	&	256	&	 RFI    	&	...	&	...	&	&	0	&	...		&	...	&	...	&	 0.24		\\
2216$-$038 		&	 22 18 52.04 	&	 $-$03 35 36.9 	&	 0.901 		&		  277	&	7	&	1.065	&	$<$1.1	&	&	398	&	4.8	&	1.212	&	$<$0.60	&	&	281	&	6.9	&	1.348	&	$<$0.79	&	&	0	&	...		&	...	&	...	&	 0.63		\\
HB89 2223+210			&		22 25 38.0		&		 +21 18 06		&		1.959			&	0	&	...  	&	...	&	...	&	&	0	&	...  	&	...	&	...	&	&		1499		&	6.5	&	2.7	&	$<$0.40	&	&	0	&	...  	&	...	&	...	&	 0.3904		\\
3C446$^b$ 		&	 22 25 47.26 	&	 $-$04 57 01.4 	&	 1.404 		&		  298	&	4.2	&	6.51	&	$<$0.061	&	&	0	&	...  	&	...	&	...	&	&	1491	&	8.8	&	9.11	&	$<$0.15	&	&	0	&	...		&	...	&	...	&	0.746	\\
2227$-$088 		&	 22 29	40.08 	&	 $-$08 32 54.4 	&	 1.5605(4)	&	298	&	12	&	0.964	&	$<$1.9	&	&	373	&	8.3	&	1.026	&	$<$1.3	&	&	262	&	12	&	1.084	&	$<$4.1	&	&	0	&	...		&	...	&	...	&	 0.48		\\
2230+114 		&	 22 32 36.41 	&	 +11 43 50.9 	&	 1.037 		&		  298	&	8.6	&	7.379	&	$<$0.23	&	&	0	&	...  	&	...	&	...	&	&	298	&	7.2	&	7.867	&	$<$0.16	&	&	0	&	...		&	...	&	...	&	 0.22		\\
2234+282 		&	 22 36 22.47 	&	 +28 28 57.4 	&	 0.795 		&		  257	&	9.1	&	0.709	&	$<$2.4	&	&	344	&	 RFI   	&	...	&	...	&	&	298	&	8.2	&	0.597	&	$<$2.1	&	&	0	&	...		&	...	&	...	&	 0.30		\\
HB89 2243$-$032			&		22 46 11.3		&		$-$03 00 38		&		1.347(5)		&	0	&	...  	&	...	&	...	&	&	0	&	...  	&	...	&	...	&	&		1472		&	5.8	&	1.998	&	$<$0.46	&	&	0	&	...  	&	...	&	...	&	 0.4659		\\
2243$-$123 		&	 22 46 18.23 	&	 $-$12 06 51.3 	&	 0.632 		&		  298	&	5.7	&	1.884	&	$<$0.51	&	&	398	&	 RFI   	&	...	&	...	&	&	0	&	...   	&	...	&	...	&	&	0	&	...		&	...	&	...	&	 0.22		\\
3C454.1            		&		22 50 32.9		&		 +71 29 19		&		1.841			&	0	&	...  	&	...	&	...	&	&	0	&	...  	&	...	&	...	&	&		1488		&	3.7	&	3.13	&	$<$0.19	&	&	0	&	...  	&	...	&	...	&	 0.3713		\\
3C454.3 		&	 22 53 57.75 	&	 +16 08 53.6 	&	 0.859 		&		  262	&	6.2	&	13.27	&	$<$0.097	&	&	0	&	...  	&	...	&	...	&	&	252	&	9.2	&	13.87	&	$<$1.1	&	&	0	&	...		&	...	&	...	&	 0.26		\\
HB89 2251+244			&		22 54 09.3		&		 +24 45 24		&		2.328(5)		&	0	&	...  	&	...	&	...	&	&	0	&	...  	&	...	&	...	&	&		1493		&	5.9	&	2.385	&	$<$0.39	&	&	0	&	...  	&	...	&	...	&	 0.2953		\\
4C+41.45         		&		22 57 22.1		&		 +41 54 17		&		2.150(10)		&	0	&	...  	&	...	&	...	&	&	0	&	...  	&	...	&	...	&	&		1482		&	8.6	&	2.85	&	$<$0.848	&	&	0	&	...  	&	...	&	...	&	 0.391			\\
2255$-$282 		&	 22 58 05.96 	&	 $-$27 58 21.3 	&	 0.92584(15)	&		  258	&	5	&	1.251	&	$<$0.72	&	&	372	&	4.2	&	1.174	&	$<$0.56	&	&	298	&	6.1	&	1.114	&	$<$0.8	&	&	0	&	...		&	...	&	...	&	 0.34		\\
3C459			&	 23 16 35.23 	&		+04 05 18.1 	&	 0.22012(3)	&	292	&	5.8	&	4.858	&	$<$0.17	&	&	0	&	...  	&	...	&	...	&	&	0	&	...   	&	...	&	...	&	&	0	&	...		&	...	&	...	&	 0.23		\\
2318+049 		&	 23 20 44.85 	&	 +05 13 50.0 	&	 0.622 		&		  245	&	5.1	&	0.543	&	$<$1.3	&	&	381	&	4.1	&	0.582	&	$<$1.2	&	&	0	&	...   	&	...	&	...	&	&	0	&	...		&	...	&	...	&	 0.28		\\
4C$-$05.96          		&		23 25 19.6		&		$-$04 57 37		&		1.188(2)		&	0	&	...  	&	...	&	...	&	&	0	&	...  	&	...	&	...	&	&		1426		&	4.6	&	2.86	&	$<$0.25	&	&	0	&	...  	&	...	&	...	&	 0.3583		\\
HB89 2338+042			&		23 40 57.9		&		 +04 31 16		&		2.594			&	0	&	...  	&	...	&	...	&	&	0	&	...  	&	...	&	...	&	&		1490		&	3.4	&	3.05	&	$<$0.18	&	&	0	&	...  	&	...	&	...	&	 0.3721		\\
2344+092 		&	 23 46 36.84 	&	 +09 30 45.5 	&	 0.677 		&		  277	&	7.3	&	2.122	&	$<$0.60	&	&	391	&	6.8	&	2.186	&	$<$0.57	&	&	298	&	RFI	&	...	&	...	&	&	0	&	...		&	...	&	...	&	 0.48		\\
2345$-$167 		&	 23 48 02.61 	&	 $-$16 31 12.0 	&	 0.576 		&		  260	&	5.1	&	1.912	&	$<$0.36	&	&	392	&	5.7	&	2.049	&	$<$0.44	&	&	0	&	...   	&	...	&	...	&	&	0	&	...		&	...	&	...	&	 0.42		\\
2351+456$^b$	&	 23 54 21.68 	&	 +45 53 04.2 	&	 1.992(3) 		&	0	&	...   	&	...	&	...	&	&	0	&	...  	&	...	&	...	&	&		1726		&	5.2	&	2.016	&	$<$0.40	&	&	0	&	...		&	...	&	...	&	0.3763	\\
3C469.1     			&		23 55 23.3		&		 +79 55 20		&		1.336			&	0	&	...  	&	...	&	...	&	&	0	&	...  	&	...	&	...	&	&		1182		&	2.8	&	3.021	&	$<$0.15	&	&	0	&	...  	&	...	&	...	&	 0.3957		\\
HB89 2354+144			&		23 57 18.6		&		 +14 46 07		&		1.8142(17)		&	0	&	...  	&	...	&	...	&	&	0	&	...  	&	...	&	...	&	&		1498		&	4	&	1.8	&	$<$0.35	&	&	0	&	...  	&	...	&	...	&	 0.3902		\\
2355$-$106 		&	 23 58 10.88 	&	 $-$10 20 08.6 	&	 1.6363(5)	&		  259	&	5.1	&	0.772	&	$<$1.1	&	&	351	&	5.6	&	0.596	&	$<$1.5	&	&	277	&	8.1	&	0.518	&	$<$2.3	&	&	0	&	...		&	...	&	...	&	0.33	\\
3C470				&		23 58 23.3		&		 +44 04 39		&		1.653			&	0	&	...  	&	...	&	...	&	&	0	&	...  	&	...	&	...	&	&		1372 		&	3.8	&	4.14	&	$<$0.15	&	&	0	&	...  	&	...	&	...	&	 0.4085  \\
\hline
  \end{tabular}
 \end{table}
\end{landscape}

 \begin{table*}
 \scriptsize
  \caption{Sources with \HI\ 21~cm Absorption. Sources in which \HI\ absorption has been detected. Columns list the
(1) source name, 
(2) numbered Gaussian component (from Figure \ref{fig:detects}), 
(3) central frequency of each component, 
(4) redshift of each absorption feature, 
(5) continuum flux density, 
(6) observed width of the line (kHz), 
(7) rest-frame line width (\kms), 
(8) depth of the absorption line, 
(9) peak optical depth, 
(10) derived HI column density $N_{\rm HI} / (T_s/f)$,  
(11) measured column density $N_{\rm HI} / (T_s/f)$ from previous 21~cm observations with respective references; and 
(12) measured $T_s/f$ values constrained from Ly$\alpha$ and VLA observations from \citet{kanekar14_mnras}. Numbers in parentheses show uncertainties in the final digit(s) of listed quantities (when available). 
References for original detection of the 21~cm absorber and measured column densities:   
1 -- \citet{carilli93}; 2 -- \citet{briggs89}; 3 -- \citet{lane01}; 4 -- \citet{wolfe85}; 5 -- \citet{lane98}; 6 -- \citet{vermeulen03}; 7 -- \citet{brown73}; 8 -- \citet{chengalur99};  9 -- \citet{darlingetal04}.  \newline
$a$ -- Intrinsic 21~cm absorption line. This system is not included in our cosmological measurements of the $f(N_{\rm HI},X)$ distribution or $\Omega_{\rm HI}$.   }
  \label{tab:detections}
  \begin{tabular}{lccccccccccc}
    \hline
    Source & Comp. & $\nu$ & $z$ & Cntm. & FWHM & FWHM & Depth & $\tau$ & $N_{\rm {HI}}$ & Prev. $N_{\rm {HI}}$ & $T_s/f$ \\
    & & (MHz) & & (Jy) & (kHz) & (\kms) & (mJy) & (10$^{-2}$) & \multicolumn{2}{c}{($10^{18} \ (T_s/f)$~cm$^{-2}$~K$^{-1}$)} & (K) \\
    \hline
B2 0218+357 & 1   & 843.134(7)   & 0.684675(14)& 1.484(4) & 74(11)  & 26(4)     & 85(9)    & 5.9(7)  & 3.0(6)&... & ... \\ %
            & 2   & 843.209(12)  & 0.68452(2)  & 1.484(4) & 76(18)  & 27(6)     & 51(9)    & 3.5(6)  & 1.8(5)&... & ... \\
            & Tot & 843.16(8)    & 0.68468(16) & 1.484(4) & 116(20) & 41(7)     & 95(4)    & 7.2(7)  & 5.7(1.1)	& 4.0$^1$ & ... \\ 
0248+430    & 1   & 1018.8618(18)& 0.39411(3)  & 0.795(3) & 71(6)   & 21.0(1.8) & 97(6)    & 13.0(3)  & 5.2(6)&... & ... \\
            & 2   & 1018.940(2)  & 0.394003(4) & 0.795(3) & 29(4)   & 8.5(1.4)  & 10(4)    & 10.5(5)  & 1.7(4)&... & ... \\
            & 3   & 1019.107(17) & 0.39378(4)  & 0.795(3) & 80(5)   & 23.5(1.2) & 79(9)    & 13(5)  & 0.6(2)&... & ... \\
            & Tot & 1018.88(3)   & 0.39409(2)  & 0.795(3) & 150(10) & 44(3)     & 109(6)   & 7.9(5)  & 6.7(6)	& 5.4(6)$^3$ & ... \\ 
0458$-$020  & 1   & 467.332(3)   & 2.03939(2)  & 2.45(5)  & 34(8)   & 22(5)     & 218(46)  & 9.0(2)   & 3.8(1.3)	& 10$^{2,4}$& 560 \\
HB89 0738+313& 1  & 1163.0764(13)& 0.22125(14) & 1.9453(10)& 35(2)  & 9.1(7)    & 76(5)    & 4.0(2)  & 0.70(7)	& 0.64(14)$^5$ & 870 \\
3C216$^a$       & 1   & 851.05(12)    & 0.6690(3)  & 6.39(4)  & 811(90) & 285(32)   & 18(2)& 0.29(3) & 1.6(2)&... & ... \\
            & 2   & 850.30(6)   & 0.67047(12)  & 6.37(4)  & 671(95) & 237(33)   & 30(4)    & 0.47(6)& 2.1(4)&... & ... \\
            & Tot & 850.56(5)    & 0.66996(14) & 6.39(4)  & 682(98) & 240(35)   & 27(5)    & 0.40(6) & 1.8(4)	& 1.2$^6$ & ... \\
1127$-$145  & 1   & 1082.130(2)  & 0.312602(2) & 5.548(3) & 20(2)   & 5.5(6)    & 115(40)  & 2.1(7)  & 0.22(8)&... & ... \\
            & 2   & 1082.08(3)   & 0.31266(4)  & 5.548(3) & 41(3)   & 11.4(8)   & 512(42)  & 9.7(8)  & 2.1(2)&... & ... \\
            & 3   & 1082.032(6)  & 0.312721(7) & 5.548(3) & 24(4)   & 6.71(1.1) & 472(63)  & 8.90(12) & 1.1(2)&... & ... \\
            & 4   & 1082.001(5)  & 0.312758(6) & 5.548(3) & 9.3(1.1)& 2.6(3)    & 136(22)  & 2.5(4)  & 0.12(2)&... & ... \\
            & 5   & 1081.96(5)   & 0.31281(6)  & 5.548(3) & 71(5)   & 19.7(1.4) & 276(11)  & 5.1(2)  & 1.92(16)&... & ... \\
            & 6   & 1081.89(7)   & 0.31289(8)  & 5.548(3) & 73(11)  & 20(3)     & 22(3)    & 0.40(5) & 0.15(3)&... & ... \\
            & Tot & 1082.04(5)   & 0.31272(6)  & 5.548(3) & 170(17) & 47(5)     & 543(8)   & 6.8(7)  & 6.2(2)	& 5.1(9)$^5$ & 820 \\
1243$-$072  & 1   & 988.635(5)   & 0.436734(7) & 0.64(2)  & 14.4(1.0)& 4.4(3)   & 179(10)  & 33(2)& 2.7(3)	& 1.7(6)$^3$ & ... \\ 
3C286       & 1   & 839.4074(9)  & 0.6921530(18)& 18.64(9)& 32.3(1.4)& 11.5(5)  & 274(9)   & 1.48(5) & 0.327(18)&... & ... \\
            & 2   & 839.39(2)    & 0.69219(5)  & 18.64(9) & 348(64) & 124(23)   & 23(4)    & 0.12(2) & 0.29(7)&... & ... \\
            & Tot & 839.41(3)    & 0.69215(6)  & 18.64(9) & 24(2)   & 8.6(7)    & 296(6)   & 3.9(3)  & 0.65(7)	& 2.6$^7$ & 965 \\
PKS 1830$-$211  & 1   & 753.5(3)     & 0.8851(8)   & 9.95(4)  & 450(40) & 179(16)   &  1030(40) & 10.9(5)  & 37(4)	& 30$^8$ &... \\ 
2351+456    & 1   & 798.334(9)   & 0.77921(2)  & 2.006(2) & 14.4(8) & 5.4(3)   & 253(8)    & 13.5(5) & 1.4(9)&... & ... \\
            & 2   & 798.240(6)   & 0.77942(13) & 2.006(2) & 123(8)  & 46(3)    & 545(34)   & 32(2)   & 28(3)&... & ... \\
	    & 3   & 798.23(3)	 & 0.77944(6)  & 2.006(2) & 13.0(9) & 4.9(3)& 236(12)& 12.5(7)  & 1.17(10)&... & ... \\
	    & 4   & 798.142(12) & 0.77964(3)  & 2.006(2) & 115(11) & 43(4) & 242(22)& 12.9(1.2)&10.6(1.5)&... & ... \\
            & Tot & 798.22(3) & 0.77948(7) & 2.006(2) & 115(11) & 43(4) & 638(8) & 38(4)    & 32(3)	& 24$^9$  & ... \\ 
    \hline
  \end{tabular}
 \end{table*}

\begin{landscape}
 \begin{table}
 \scriptsize
  \caption{Properties of the HI 21~cm and OH 18~cm Line Observations. Columns list the source name and redshift of known intervening 21~cm absorbers or redshift of the host galaxy in the case of intrinsic absorbers with a redshift lying in our spectral coverage, the frequency, ranged searched (in \kms), measured rms noise over the searched region, continuum, and the derived $3\sigma$ upper limit to the \HI\ column density for the 21~cm absorption search. We list the same parameters for the OH 1612, 1667, and 1720~MHz search at known intervening and intrinsic absorbers. The OH continuum flux density and OH column density limit are computed for the 1667~MHz line unless stymied by RFI. Numbers in parentheses show uncertainties in the final digit(s). \newline 
$a$ -- Sources searched for possible intrinsic HI and OH absorption. \newline
$b$ -- The corresponding 21~cm absorbers at the same redshift remain undetected due to the expected 21~cm absorber lying outside the redshift coverage or is irrecoverable due to RFI. \newline
$d$ -- \citet{chengalur99} measures an OH column density of $N_{\rm {OH}}/ T_x=40\times10^{13}$~cm$^{-2}$~K$^{-1}$. \newline
$3$ -- We do not reach sufficient signal to noise to detect the known OH satellite lines at 1612 and 1720~MHz toward 1413+135 at $z_{abs}=0.24671$ \citep{darling04, kanekar04_phrvl}. We do not detect the 1665 or 1667~MHz OH absorption at the same redshift toward this system \citep{kanekar02} with a reported column density of $N_{\rm {OH}}/ T_x = 5.1 \times10^{13}$~cm$^{-2}$~K$^{-1}$.}
  \label{tab:OH}
  \begin{tabular}{lcccccccccccccccccccc}
    \hline
     & & \multicolumn{5}{c}{HI 1420 MHz} & & \multicolumn{3}{c}{OH 1612 MHz} & & \multicolumn{3}{c}{OH 1667 MHz} & & \multicolumn{3}{c}{OH 1720 MHz} & & \\
     \cline{3-7} \cline{9-11} \cline{13-15} \cline{17-19}  \noalign{\smallskip}
    Source & $z_{\rm abs}$ & $\nu$ & $\Delta v$ & $rms$ & Cntm. & $N_{\rm {HI}}/(T_s/f)$ & & $\nu$ & $\Delta v$ & $rms$ & & $\nu$ & $\Delta v$ & $rms$ & & $\nu$ & $\Delta v$ & $rms$ & Cntm. & $N_{\rm {OH}}/ T_x$ \\
    & & (MHz) & (\kms) & (mJy) & (Jy) & ($10^{18} \,(T_s/f)$~cm$^{-2}$~K$^{-1}$) & & (MHz) & (\kms) & (mJy) & & (MHz) & (\kms) & (mJy) & & (MHz) & (\kms) & (mJy) & (Jy) & ($10^{13}$~cm$^{-2}$~K$^{-1}$) \\
    \hline
    0235+164$^b$ & 0.524 &   932.02 & ...  & ...  & ...  & ... &&       1057.9	 & $-$313,	735	&	5.4	&	& 1094.1	& ...	& RFI &	& 1129.0	& ... & RFI	&1.79(17) & $<$55\\ 
0248+430	& 0.39409 &  1018.88(3)   & $-$3654, 2581  & 5.3 & 0.795(3) & 6.7(6)	 && 1156.5 & $-$5118, 1467 & 5.4 && 1196.0	&$-$4386, 2357	& 4.2	&	& 1234.1	&	...	 & ... & 0.801(4) & $<$11.4\\
HB89 0312$-$034$^a$ & 1.072	& 685.52	& $-$5456, 3465	& 8.1	& 2.741(5)	 & $<$0.52 && 778.1 & $-$1656, 1757 & 5.4 & & 804.71 & $-$5017, 2719	& 4.6 && 830.4 & $-$7708, 3704 & 2.4 & 2.328(4) & $<$3.7\\
3C124$^a$ & 1.083	&	681.90	& ...	& RFI &...	&... && 773.99 & $-$275, 1874 & 4.7 && 800.46 & $-$5137, 1770	& 4.2	&&  825.99 & $-$2295, 2115 & 2.8 & 1.935(7) & $<$4.0\\
0458$-$020	& 2.03939 &  467.332(3)  & $-$1316, 846  & 33  & 2.45(5)  & 3.8(1.3)&& 530.45 & ... & RFI	&&	548.58	& ... &	RFI	&& 566.08	&... & 	...	&...	&...	\\
SBS 0804+499$^b$	& 1.4073  & 590.04 & ...  & ...  & ...  & ... && 669.73 & ... & RFI	& 		&	692.63 & ... & RFI	& 	& 714.71 & $-$355, 1159 & 8.8	& 0.813(13) & $<$217\\ 
3C216		& 0.66996	& 850.56(5)    & $-$737, 3580 & 8.1 & 6.39(4)  & 1.8(4) && 965.43 & $-$3015, 1649 & 7.1	&&	998.44	& $-$4701, 2073 & 6.2	&	& 1030.3	&	$-$3904, 606 & 9.3	&5.52(4)&$<$2.3\\
4C$-$05.60$^a$ & 1.094(2) &	678.32	& $-$621, 363	& 3.7	& 2.80(14)	&$<$0.23	&& 769.93 & $-$1658, 1713 & 2.3 && 796.26 & $-$5429, 4907	& 2.1	&& 821.65 & $-$3063, 544 & 2.2 & 2.41(11)& $<$1.6\\
1127$-$145	& 0.31272 &1082.04(5)   & $-$2734, 1722 & 7.6 & 5.548(3) & 6.2(2) && 1228.2 &  $-$3427, 1187 & 4.9 && 1270.2	&$-$2064, 2633 & 6.1	&	&1310.7	&  $-$1885, 2507 & 7.8	 & 5.574(3) & $<$2.3	\\
1243$-$072	& 0.436734 & 988.635(5)   & $-$909, 3347 & 7.9 & 0.64(2)  &  2.7(3)	 && 1122.2 & ... & RFI	 &&	1160.5	& $-$9759, 2149	& 7.1	&	& 1197.5	& $-$2922, 2085 & 5.5	& 0.60(3) & $<$27\\
1413+135$^{b,e}$ & 0.24671&  1139.3 & ... & ... & ... & ... & &  1293.2 & $-$8024, 8807 & 6.1 && 1337.4 	& $-$8113, 2445 &	2.2 & & 1380.1 & 	$-$6374, 4075& 5.6	& 1.1240(6)	&$<$3.8	\\
HB89 1437+624$^a$ &	1.0935(17) & 678.48	& ... & RFI & ... & ... && 770.11 & $-$1577, 1306 & 8.9 && 796.45 & $-$1883, 280	& 4.6	&& 821.84 & $-$1060, 576 & 4.3 & 2.59(12) & $<$3.3\\
3C356$^a$ & 1.079	&	683.22	& ...	& RFI & ...	& ... && 775.48 & $-$2362, 2193 & 6.1 && 802.00 & $-$6344, 4516	& 6.1 && 827.58 & $-$1195, 1960 & 6.6 & 2.644(6) & $<$4.3\\
PKS 1830$-$211	& 0.8851& 753.5(3)     & $-$3585, 3447 & 15  & 9.95(4)  & 37(4)	&& 855.25	&  $-$740, 5258 & 10 &&	884.49	& $-$1629, 309 & 9.5	&& 912.70 	 & ...	 & ... & 9.76(4) & 60(3)$^d$	\\ 
2351+456	& 0.77948	& 798.22(3) & $-$4191, 6957 & 10 & 2.006(2)  & 32(3) && 906.02 &   ... & RFI	 &&	937.0	&	...	&	...	&& 966.88	&	...	 & ... & ... & ...\\
2355$-$106$^b$ & 1.1726 &  653.78 & ... & ... & ... & ... &&	742.07&  $-$13560, 5573 & 5.5 && 767.45 &	... & RFI	&& 791.92 & $-$3929, 4960 & 4.4 & 0.54(5) & $<$147\\
    \hline
  \end{tabular}
 \end{table}
\end{landscape}

\begin{table*}
  \caption{$f(N_{\rm HI},X)$ and $\Omega_{\rm HI}$ Measurements. Columns list the 
(1) $\Delta \log N_{\textrm {HI}}$ bin, 
(2) number of detections made in each $\Delta \log~N_{\rm HI}$ bin, 
(3) total number of observational samples in each column density sensitivity bin, 
(4) total $\Delta X$ searched for which a system with that column density could have been detected, 
(5) calculated column density frequency distribution for the entire sample in each bin, and
(6) measured $\Omega_{\rm HI}$ for each $\log N_{\textrm {HI}}=0.5$~dex bin for our redshift bin below/above the redshift cut at $z=0.69$. We repeat all calculations for $T_s/f$ values of 100, 250, 500, and 1000~K. Numbers in parentheses indicate uncertainties in the final digit(s) of listed quantities, when available.}
  \label{tab:fN}
  \begin{tabular}{lcccccc}
    \hline
    $\log N_{\textrm {HI}}$ & Detections & Samples & $\Delta X$ & log~$f(N_{\rm HI},X)$ & \multicolumn{2}{c}{$\Omega_{\rm HI}\times10^{-3}$} \\
    (cm$^{-2}$) & & & & & $0<z<0.69$ & $0.69<z<2.74$ \\
    \hline
    \hline 
\multicolumn{7}{c}{$T_s/f = 100$~K} \vspace{3pt} \\
\hline 
18.25	&	0	&	9		&	0.740	&	$<-$17.73		&	...		& ...\\	
18.75	&	0	&	84	&	7.23		&	$<-$19.22	&	...		& ...\\	
19.25	&	0	&	484	&	56.03	&	$<-$20.61	&	...		& ...\\	
19.75	&	2	&	1013	&	121.2		&	$-$21.6(7)	&	...		& ...\\	
20.25	&	1	&	506	&	155.2	&	$-$22.5(1.0)	&	...		& ...\\	
20.75	&	4	&	48	&	159.1	&	$-$22.4(5)	&	...		& ...\\	
21.25	&	0	&	1		&	159.2	&	$<-$23.06	&	...		& ...\\	
21.75	&	2	&	2		&	159.5	& 	$-$23.7(9)	&	...		& ...\\	
\hline
Total 	&	9	& 2147	&	159.5	& 	...			&	0.21(10)	& 0.69(45) \vspace{3pt}\\
\hline 
\hline 
\multicolumn{7}{c}{$T_s/f = 250$~K} \vspace{3pt} \\
\hline
18.75	&	0	&	15		&	1.03		&	$<-$18.37	&	...		& ...\\	
19.25	&	0	&	113	&	10.6		&	$<-$19.88	&	...		& ...\\	
19.75	&	0	&	652	&	73.0		&	$<-$21.22	&	...		& ...\\	
20.25	&	2	&	977	&	132.1		&	$-$21.2(7)	&	...		& ...\\	
20.75	&	2	&	364	&	156.9	&	$-$22.7(7)	&	...		& ... \\ 
21.25	&	3	&	24	&	159.2	&	$-$23.1(6)	&	...		& ... \\ 
21.75	&	2	&	2		&	159.5	&	$-$23.7(7)	&	...		& ... \\ 
\hline
Total 	&	9	& 2147	&	159.5	& 	...		&	0.53(16)	& 1.7(9) \vspace{3pt} \\
\hline 
\hline 
\multicolumn{7}{c}{$T_s/f = 500$~K} \vspace{3pt} \\
\hline
18.75	&	0	&	2 		&	0.085	&	$<-$17.29	&	...		& ...\\	
19.25	&	0	&	37		&	3.37		&	$<-$19.39	&	...		& ...\\	
19.75	&	0	&	263	&	28.2		&	$<-$20.81	&	...		& ...\\	
20.25	&	0	&	909	&	101.4	&	$<-$21.86	&	...		& ...\\	
20.25	&	2	&	777	&	147.9	&	$-$22.7(7)	&	...		& ...\\	
21.25	&	4	&	155	&	159.1	&	$-$22.9(5)	&	...		& ...\\	
21.75	&	1	&	2		&	159.2	&	$-$24.0(1.0)	&	...		& ...\\	
22.25	&	2	&	2		&	159.5	&	$-$24.2(7)	&	...		& ...\\	
\hline
Total 	&	9	& 2147	&	159.5	& 	...		&	1.05(33)	& 3.4(1.6) \vspace{3pt} \\
\hline 
\hline 
\multicolumn{7}{c}{$T_s/f = 1000$~K} \vspace{3pt} \\
\hline
19.25	&	0	&	9		&	0.740	&	$<-$18.73	&	...		& ...\\	
19.75	&	0	&	84	&	7.23		&	$<-$20.22	&	...		& ...\\	
20.25	&	0	&	484	&	56.03	&	$<-$21.61		&	...		& ...\\	
20.75	&	2	&	1013	&	121.2		&	$-$22.6(7)	&	...		& ...\\	
21.25	&	1	&	506	&	155.2	&	$-$23.5(1.0)	&	...		& ...\\	
21.75	&	4	&	48	&	159.1	&	$-$23.4(5)	&	...		& ...\\	
22.25	&	0	&	1		&	159.2	&	$<-$24.1		&	...		& ...\\	
22.75	&	2	&	2		&	159.5	&	$-$24.7(8)	&	...		& ...\\	
\hline
Total 	&	9	& 2147	&	159.5	& 	...		&	2.1(5)	& 6.9(2.7) \vspace{3pt}   \\
    \hline
  \end{tabular}
 \end{table*}




\bibliographystyle{mnras}








\bsp	
\label{lastpage}
\end{document}